\documentclass[aip,jcp,amsmath,amssymb,graphicx,preprint]{revtex4-1}
\usepackage{graphicx, url, float, color, soul}
\graphicspath{{Images/}}

\draft 

\usepackage{etoolbox}
\providetoggle{comments}\settoggle{comments}{true}
\iftoggle{comments}{
    \usepackage{todonotes}
    \newcommand\ed[1]{\todo[color=red!10,inline]{ED:~#1}}
    \newcommand\sa[1]{\todo[color=orange!10,inline]{SA:~#1}}
}{
    \newcommand\ed[1]{}
    \newcommand\sa[1]{}
}

\begin{document}

\preprint{AIP}

\title{Investigating the Role of Non-Covalent Interactions in Conformation and Assembly of Triazine-Based Sequence-Defined Polymers}


\author{Surl-Hee Ahn}
\email{sahn1@stanford.edu}
\affiliation{Chemistry Department, Stanford University, Stanford, California 94305, USA}

\author{Jay W. Grate}
\email{jwgrate@pnnl.gov}
\affiliation{Pacific Northwest National Laboratory, Richland, Washington 99352, USA}

\author{Eric F. Darve}
\email{darve@stanford.edu}
\affiliation{Mechanical Engineering Department, Stanford University, Stanford, California 94305, USA}


\date{\today}

\begin{abstract}
Grate and co-workers at Pacific Northwest National Laboratory recently developed high information content triazine-based sequence-defined polymers that are robust by not having hydrolyzable bonds and can encode structure and functionality by having various side chains. Through molecular dynamics (MD) simulations, the triazine polymers have been shown to form particular sequential stacks, have stable backbone-backbone interactions through hydrogen bonding and $\pi$-$\pi$ interactions, and conserve their \emph{cis/trans} conformations throughout the simulation. However, we do not know the effects of having different side chains and backbone structures on the entire conformation and whether the \emph{cis} or \emph{trans} conformation is more stable for the triazine polymers. For this reason, we investigate the role of non-covalent interactions for different side chains and backbone structures on the conformation and assembly of triazine polymers in MD simulations. Since there is a high energy barrier associated to the \emph{cis}-\emph{trans} isomerization, we use replica exchange molecular dynamics (REMD) to sample various conformations of triazine hexamers. To obtain rates and intermediate conformations, we use the recently developed concurrent adaptive sampling (CAS) algorithm for dimer of triazine trimers. We found that the hydrogen bonding ability of the backbone structure is critical for the triazine polymers to self-assemble into nanorod-like structures, rather than that of the side chains, which can help researchers design more robust materials.
\end{abstract}

\pacs{Valid PACS appear here}
\maketitle

\section{\label{sec:intro} Introduction}
Molecular dynamics (MD) simulations are becoming as indispensable as experiments, since they can give us insight into mechanisms of how bio-molecules change their conformations with fine resolution. We can also test the effects of different experimental conditions, discern favorable conformations, and discover major pathways and associated rates for the bio-molecules. In short, MD simulations are truly becoming ``microscopes" that can elucidate important biophysical phenomena such as protein folding. 

However, MD simulations by themselves are limited in predictive power, since the bio-molecules routinely get ``stuck" in metastable states and do not change their conformations for a long period of time. Moreover, most interesting biological processes have timescales in milliseconds and longer, whereas MD simulations have to be run using femtosecond time steps, since they are limited by the fastest motions in the system such as vibrations of water. The long timescale is usually due to the presence of energy barriers or the process having a long diffusion time.

As a result, many enhanced sampling methods have been developed by various researchers over the past few decades to overcome this timescale barrier. The enhanced sampling methods can be broadly divided into two classes: (1) thermodynamic methods that bias the system to shorten the long time-correlation of trajectories (due to energy barriers) and efficiently sample free energy landscapes and (2) state-based methods that divide up the conformational space and sample both thermodynamic and kinetic properties. Some of the popular thermodynamic methods include replica exchange molecular dynamics (REMD), umbrella sampling, metadynamics, and adaptive biasing force method\cite{torrie1977, sugita1999, laio2008, darve2008}. On the other hand, some of the popular state-based methods include building a Markov state model (MSM) that reconstructs global kinetic properties from short simulations, and the weighted ensemble (WE) method that samples full trajectories going over barriers\cite{bowman2009progress, huber1996}. Depending on the system and what properties we are interested in uncovering, we can use the appropriate enhanced sampling method for the system.

In this paper, we are interested in using MD simulations to uncover properties of triazine-based sequence-defined polymers that were recently developed by Grate and co-workers\cite{grate2016}. These new sequence-defined polymers encode structure and functionality by having various side chains and are robust by not having susceptible peptide bonds that can be cut by proteases. In Ref.~\onlinecite{grate2016}, MD simulations have shown that the triazine polymers can have multiple backbone-backbone interactions through hydrogen bonding and $\pi$-$\pi$ interactions. In particular, the all \emph{cis} triazine polymers self-assemble and form a nanorod-like structure, which has motifs resembling those of DNA, $\alpha$-helices and $\beta$-sheets (not experimentally verified). Overall, the biomimetic triazine polymers have great potential to become useful building blocks for new macromolecules and materials with desired functions.

However, there are many questions that need to be answered regarding the triazine polymers. For instance, we do not know how different side chains other than S-ethyl and different backbone structures may modulate the self-assembly process. In addition, even though the triazine polymers conserve their \emph{cis/trans} conformations throughout the simulation, the \emph{cis}-\emph{trans} isomerization has a high energy barrier of $\Delta G^{\neq} = 15$ kcal/mol (experimentally measured), which is close to the rotational barrier for peptides that is 16--20 kcal/mol\cite{archer2002, amm1998, birkett2000}. This is due to the bond having a partial double bond character from the delocalization of the triazine $\pi$ electrons and the nitrogen lone pair. Nonetheless, the \emph{cis}-\emph{trans} isomerization is a process that can happen in seconds in the laboratory, so the all \emph{cis} and/or all \emph{trans} may not be the most stable conformations and each triazine polymer will most likely transition into its most stable conformation after seconds pass. In general, the \emph{cis}-\emph{trans} isomerization is known to play an important role in protein folding, cellular signaling, and ion channel gating\cite{craveur2013}. 

Unfortunately, MD simulations need to be run for an intractable period of time to even show a single \emph{cis}-\emph{trans} isomerization. Hence, MD simulations show the triazine polymers conserving their \emph{cis/trans} conformations throughout the limited runs, even though this may not be true in reality after much longer timescales have been reached. As previously mentioned, it is essential to use enhanced sampling methods to overcome the timescale gap between simulations and biological processes, observe the isomerizations, and discover the most stable conformations for the triazine polymers. In investigating the role of different side chains and backbone structure on the conformations of a single triazine hexamer in implicit solvent in MD simulations, we use replica exchange molecular dynamics (REMD)\cite{sugita1999}. This thermodynamic enhanced sampling method is suitable in overcoming the high free energy barrier associated with rotating the bond between the linker nitrogen and the triazine ring and observing folded hexamer conformations, as done in Ref.~\onlinecite{grate2016}, which followed Ref.~\onlinecite{voelz2011}. 

To investigate the role of different side chains and backbone structure on the self-assembly of dimers of triazine trimers in explicit solvent in MD simulations, we use the recently developed concurrent adaptive sampling (CAS) algorithm\cite{ahn2017}. REMD was not used in this case due to its high computational cost when simulating explicit solvent systems. Instead, this state-based enhanced sampling method is used because it does not suffer from having high computational cost when simulating explicit solvent systems, can easily handle more than two reaction coordinates, and can consider reaction coordinates that have integer values, i.e., have a reaction coordinate that equals the total number of hydrogen bonds and $\pi$-$\pi$ interactions for the self-assembly of triazine polymers. Hydrogen bonds are strong driving forces of self-assembly due to their complementarity, directionality, and strength\cite{archer2002, rehm2008}. $\pi$-$\pi$ interactions are weaker but can help stabilize self-assembled molecules~\cite{rehm2008}.

The next sections detail the methods used and results from our studies. By using REMD and the CAS algorithm, we found that the hydrogen bonding ability of the backbone structure is essential for the triazine polymers to self-assemble into nanorod-like structures, rather than that of the side chains. This main result and other results in this paper can help researchers design materials with desired properties without having to test various starting structures beforehand.

\section{\label{sec:methods} Methods}
\subsection{\label{sec:CAS_algorithm} CAS Algorithm}
The concurrent adaptive sampling (CAS) algorithm is a state-based enhanced sampling method that is based on the weighted ensemble (WE) method\cite{huber1996, zhang2010, bhatt2010, zhang2007, suarez2014, suarez2016, abdul2012, costaouec2013, badi2014, trott2016, zwier2016}. The WE method replaces a single long simulation with many simulations that are resampled at frequent intervals and carry probabilistic weights. This way, the WE method is able to observe rare trajectories and obtain overall better sampling statistics. The system's kinetics are not altered, so we can directly obtain both thermodynamic and kinetic properties from the method. If we were only interested in sampling thermodynamic properties, however, then other methods like umbrella sampling and replica exchange molecular dynamics (REMD) would work as well\cite{torrie1977, sugita1999}. We can also obtain a statistical model in terms of state transitions and in this sense, the WE method bares similarity to building Markov state models (MSMs)\cite{bowman2009progress, bowman2009using, bowman2013, lane2011, chodera2014}. However for MSMs, macrostates, or small regions of conformational space, are constructed such that transitions between them are Markovian, and controlling the Markovian error may be difficult or even practically impossible\cite{suarez2016}. Additionally, we are not able to obtain long trajectories from MSMs.

The WE method, on the other hand, does not require the Markovian assumption and thus overcomes its associated limitations. To start, the conformational space is partitioned into discrete sets or macrostates and reaction coordinates (e.g., dihedral angles, number of non-covalent bonds) to keep track of during the simulation are chosen beforehand. The reaction coordinates' values determine which macrostate each short simulation, or ``walker," belongs to after the simulation finishes running. Note that non-differentiable, discrete reaction coordinates can be considered for the WE method, in contrast to several biasing force enhanced sampling methods like metadynamics\cite{laio2008}. This can be useful for sampling systems that self-assemble, where the number of non-covalent bonds can be an important reaction coordinate.

\begin{figure}
\centering
\includegraphics[width=80mm]{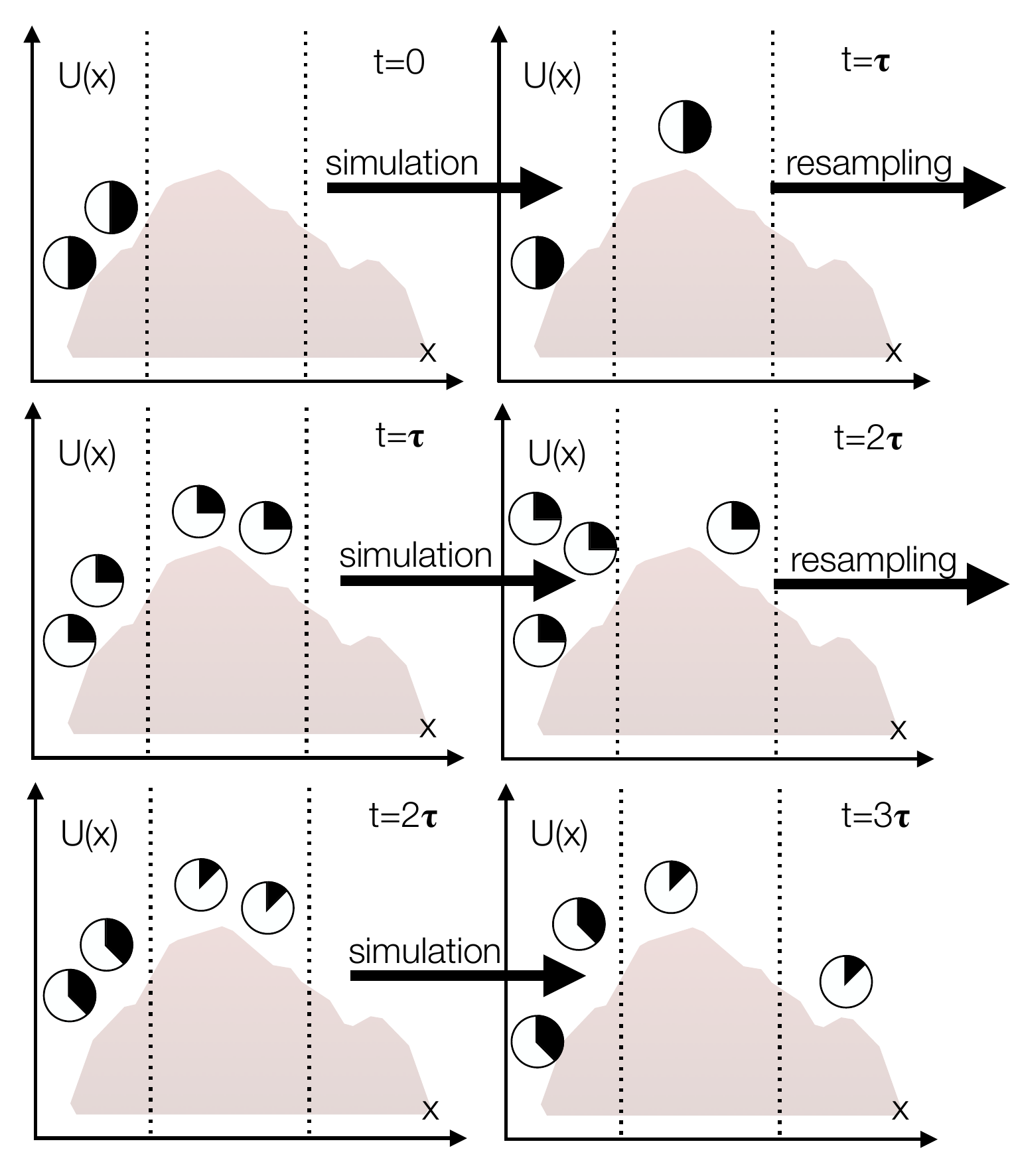} 
\caption{\label{fig:resampling} Diagram of how resampling is carried out in the CAS algorithm. $U(x)$ denotes the energy in terms of the reaction coordinate $x$. The simulation time is set to $\tau$ and the target number of walkers per macrostate is set to 2. The walkers, represented as circles, carry weights that are represented by the black portion in the circle. The walkers are first split into two walkers for each of the visited macrostates at $t = \tau$ and are merged where there are three walkers and split where there is one walker at $t = 2\tau$. The WE method resampling diagram from Ref.~\onlinecite{donovan2013} was modified to incorporate the resampling method from the CAS algorithm.}
\end{figure}

Within each macrostate, the WE method maintains a fixed target number of walkers that carry probabilities or ``weights." A fixed target number of walkers is maintained by merging or splitting walkers in a statistically correct way, which is called ``resampling." Fig.~\ref{fig:resampling} shows how the walkers, which are represented as circles, are resampled in a simple simulation. The walkers' weights are represented by the black portion in the circle, e.g., the two walkers in the initial stage of the simulation both have weights of 0.5, so half of the circles are filled with black. This technique is used to maintain a constant stream of walkers, irrespective of the energy barrier height. Note that the walkers' weights are always equal to the mean weight of the macrostate in Fig.~\ref{fig:resampling}, which is different from the original WE method, where the walkers' weights range from the mean weight to two times the mean weight. It has been proven that having equal weights is optimal since it minimizes variance and statistical errors (see Chapter 7 by Darve and Ryu in Schlick\cite{schlick2012}, and Darve\cite{darve2013}). Without resampling, then the walkers would be depleted in macrostates near an energy barrier or overcrowded in macrostates at low energy. Consequently, walkers without resampling would not be able to overcome energy barriers and sample rare pathways and intermediates. These same steps are repeated until the walkers' weights converge to steady-state probabilities. Exact fluxes and pathways with probabilities can also be obtained with no bias if the reactant and the product are specified.

Unlike MSMs, there is no need to adjust the simulation time and the macrostate decomposition to control accuracy for the WE method because the Markovian assumption is not required. The only thing to note is that we need to pick a simulation time that is long enough to observe desired transitions but not too long so that we inadvertently miss those transitions. After we pick the simulation time and run the WE method for some number of steps with resampling, the distribution of walkers will relax within each macrostate and will be closer to converging to the exact distribution. Although the WE method is guaranteed to converge to the exact distribution, it loses efficiency if macrostates are not correctly defined. In addition, finding the right partition becomes significantly more difficult for high-dimensional spaces. Nonetheless, the WE method results are not sensitive to the simulation time, and the macrostate decomposition only controls the efficiency of the method. 

Moreover, convergence is easy to monitor for the WE method, whereas MSMs may need to be reconstructed again if the memory effects are found to be big. To determine convergence of the WE method, we monitor the macrostates' probabilities and determine whether the error is small or not. To obtain accurate fluxes, we monitor the error of the fluxes and run the simulation until it is small enough\cite{schlick2012, darve2013}. In other words, we can run the WE method for longer until the errors are small enough without losing the data that we have already gathered. However, both methods have difficulty in getting initial data and early trajectories with zero information, unless good reaction coordinates are provided to move the system forward. Despite this, the WE method is preferable to MSMs in many cases given that it is more robust and that MSMs have uncontrollable errors unless macrostates are chosen carefully.

Considering these shortcomings of the vanilla WE method, the CAS algorithm was recently developed\cite{ahn2017, izaguirre2015}. The method can be used for systems in which the reaction coordinates are largely unknown because it can easily have more than two reaction coordinates, making the true reaction coordinates a function of those selected. In addition, the macrostates can be adaptively constructed as the CAS algorithm simulation proceeds so that we can sample conformations and pathways without having to define an intractable number of macrostates in high-dimensional space. Furthermore, the macrostates can be constructed in an optimal way by using the committor function, which makes the simulation focus sampling the slowest process and control computational cost as a result. Other related WE methods that can have many reaction coordinates and adaptive macrostates include the WE-based string method and the WExplore method\cite{zhang2010, adelman2013, dickson2014, dickson2016, dickson2017}. In this paper, however, we did not use these features from the CAS algorithm since there was no need for the systems that we studied. Finally, since the CAS algorithm is an extension of the WE method, it can be coupled with any MD simulation program to run walkers simultaneously, achieving computational efficiency proportional to available computational resources, similar to the WE method implementations like WESTPA and Workqueue\cite{abdul2012, zwier2015}.

However, note that the CAS algorithm only addresses a few shortcomings of the vanilla WE method, i.e., macrostate partitioning and limitation to one or two reaction coordinates. The CAS algorithm's efficiency depends on the initialization data and the choice of reaction coordinates. Additionally, trial and error is necessary to get optimal parameters (simulation time $\tau$ and target number of walkers per macrostate) to use the CAS algorithm in practice.

\subsection{\label{sec:reaction_coordinates} Reaction Coordinates}
\begin{figure}
\centering
\includegraphics[width=80mm]{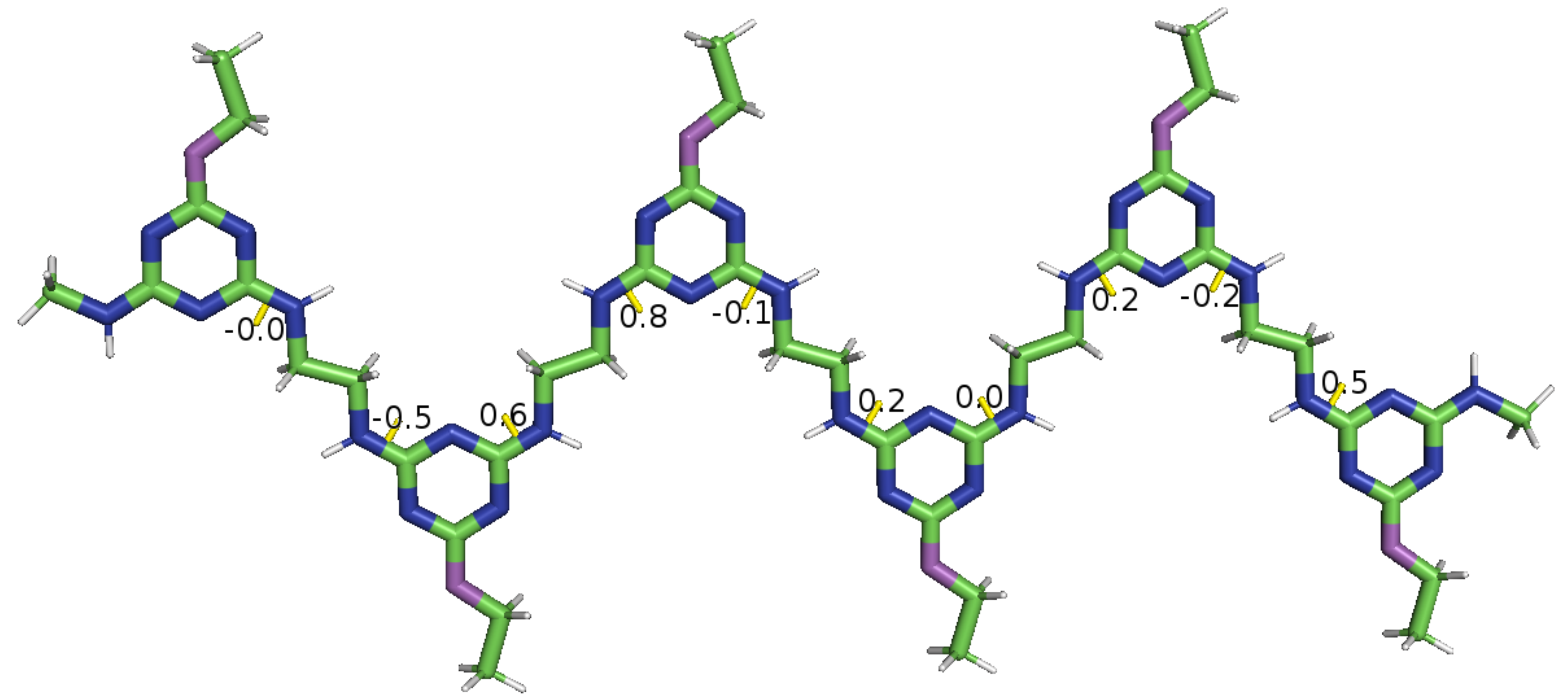} 
\caption{\label{fig:dihedrals} Dihedrals of an all \emph{cis} triazine hexamer with S-ethyl side chains and amino backbone. Since the hexamer is all \emph{cis}, all of the ten dihedrals are around $0^{\circ}$.}
\end{figure}

To study the self-assembly of dimers of triazine trimers and conformational changes of a single triazine hexamer, we chose the reaction coordinates to be the total number of non-covalent interactions, i.e., hydrogen bonds and $\pi$-$\pi$ interactions and the dihedrals that are associated with bonds that can be \emph{cis} or \emph{trans}. We kept track of hydrogen bonds that were 2.5 $\AA$ or shorter between the backbone and the triazine rings and $\pi$-$\pi$ interactions that were 4.2 $\AA$ or shorter in distance between centers of mass of two triazine rings. Note that the dihedrals are not exactly the same as the conventional $\omega$ dihedrals, which determine the \emph{cis/trans} conformation in peptide bonds. But like the regular $\omega$ dihedrals, the molecule is \emph{cis} when the dihedrals are all equal to $0^{\circ}$ and \emph{trans} when they are all equal to $180^{\circ}$. Fig.~\ref{fig:dihedrals} shows the dihedrals that indicate \emph{cis/trans} bonds for a single triazine hexamer.

Specifically, for the dimers of triazine trimers with sulfur backbone (which cannot hydrogen bond), we kept track of three reaction coordinates, i.e., the total number non-covalent interactions (ranging from 0 to 4 or 0 to 8), the total number of \emph{trans} bonds for the first trimer (ranging from 0 to 4, where Eq.~(\ref{eq:sigmoid}) is used to indicate the degree of isomerizing to \emph{trans} for each dihedral angle), and the total number of \emph{trans} bonds for the second trimer (ranging from 0 to 4, where Eq.~(\ref{eq:sigmoid}) is used again for each dihedral). For the dimers of triazine trimers with amino backbone (which can hydrogen bond), we only kept track of one reaction coordinate, i.e., the total number of non-covalent interactions (ranging from 0 to 11, with the maximum number varying for different trimers). Although Ref.~\onlinecite{ahn2017} showed that the CAS algorithm was able to sample all four \emph{cis}-to-\emph{trans} isomerizations for a single triazine trimer by keeping track of all four dihedrals, we needed better reaction coordinates that would sample \emph{cis}-to-\emph{trans} isomerizations and vice versa more easily and have macrostates converge to steady-state quickly. Hence, we focused our effort in sampling the free energy landscape and fluxes between the initial and final states for all \emph{cis} trimers and all \emph{trans} trimers separately.

\begin{equation}
\mbox{\# of \emph{trans} bonds} = \begin{cases}
0 & \mbox{if $\frac{1}{1+\exp(-0.05(\omega-90)} \leq 0.02$}\\
1 & \mbox{if $\frac{1}{1+\exp(-0.05(\omega-90)} \geq 0.98$}\\
\frac{1}{1+\exp(-0.05(\omega-90))} & \mbox{if $0.02 < \frac{1}{1+\exp(-0.05(\omega-90))} < 0.98$}
 \end{cases}
\label{eq:sigmoid}
\end{equation}

\bigskip

For the single triazine hexamer, we kept track of two reaction coordinates, i.e., the total number of hydrogen bonds (ranging from 0 to 11, where the maximum number slightly varies between different hexamers) and the total number of \emph{trans} bonds for the hexamer (ranging from 0 to 10, where Eq.~(\ref{eq:sigmoid}) is used again for each dihedral). The number of $\pi$-$\pi$ interactions was not kept track of due to the difficulty of differentiating true $\pi$-$\pi$ interactions from triazine rings that were just close together by distance. This way, the dimer of triazine trimers case was reduced to a one-dimensional or three-dimensional space sampling problem and the single triazine hexamer case was reduced to a two-dimensional space sampling problem.

\begin{figure}
\centering
\begin{tabular}{cc}
\includegraphics[width=80mm]{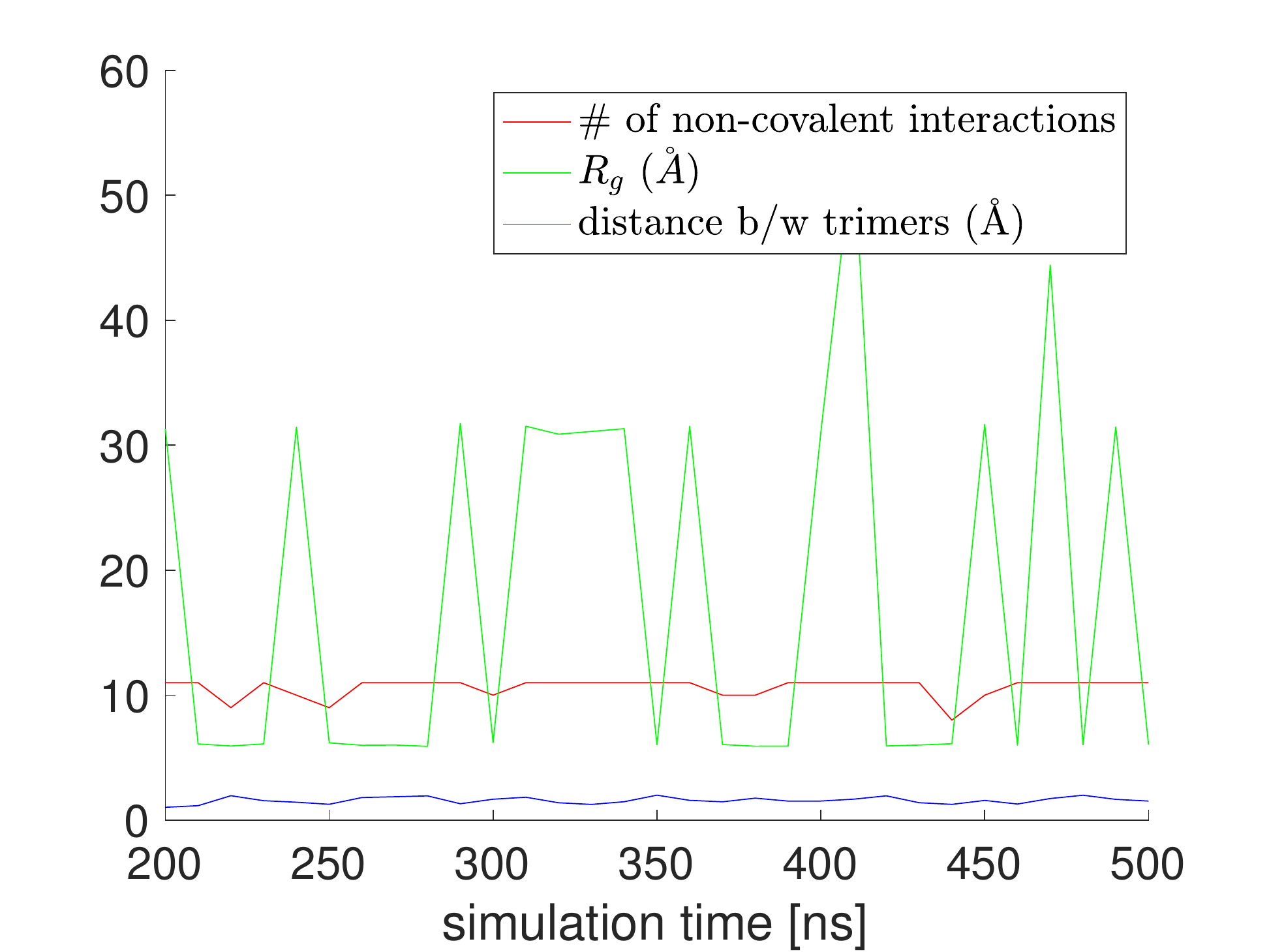} & \includegraphics[width=80mm]{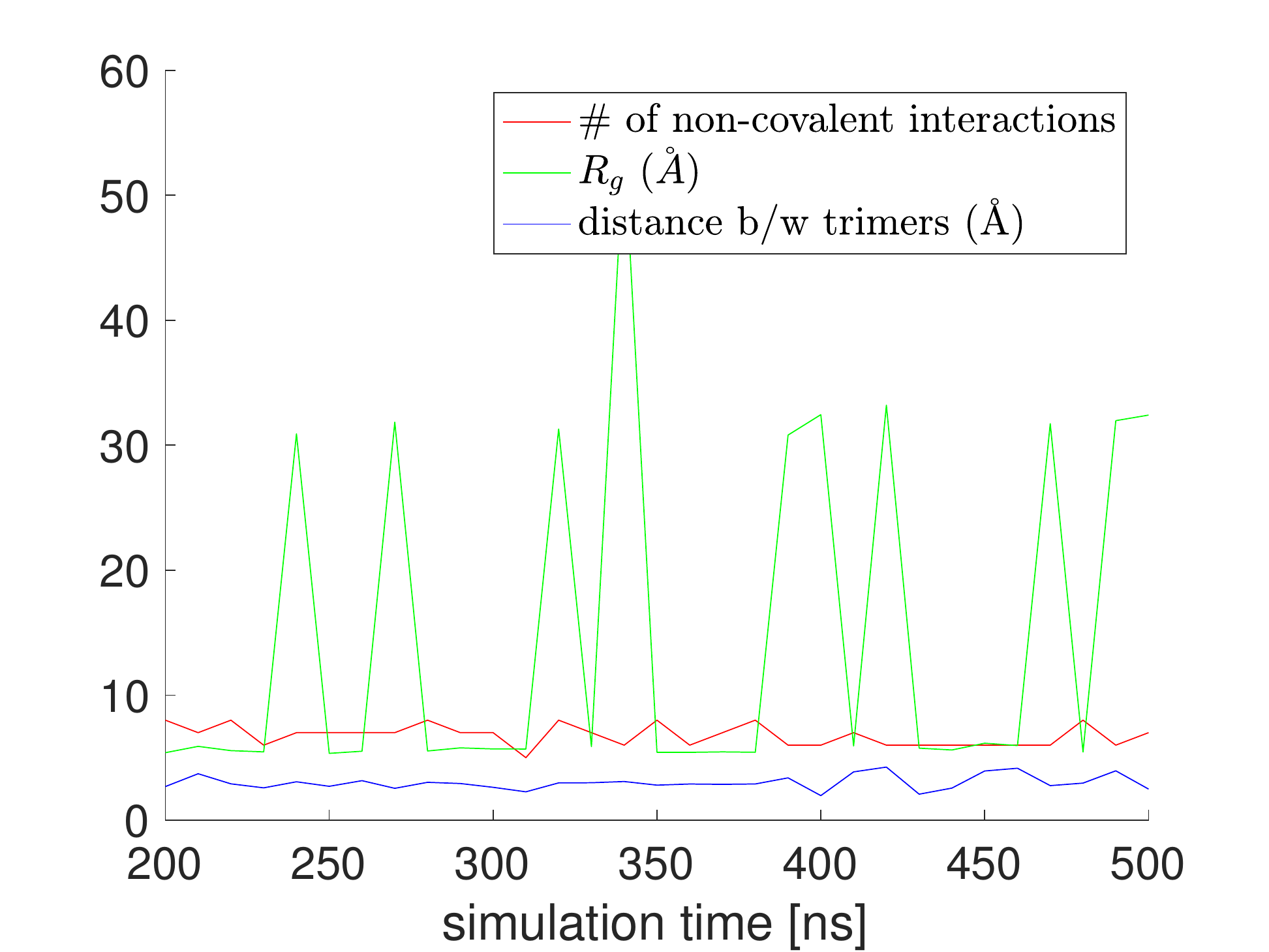} \\
(a) Plot of reaction coordinates (\emph{cis}). & (b) Plot of reaction coordinates (\emph{trans}). \\[6pt]
\end{tabular}
\caption{\label{fig:col_var_plot} Plots of reaction coordinates for a dimer of triazine trimers (all \emph{cis} and all \emph{trans}) with S-ethyl side chains and amino backbone from brute force MD simulations.}
\end{figure}

The total number of non-covalent interactions is a non-continuous reaction coordinate, which would not be suitable for most biasing force enhanced sampling methods that require differentiable reaction coordinates. Fortunately, the CAS algorithm can have discrete reaction coordinates. To test whether the total number of non-covalent interactions is indeed a good choice for a reaction coordinate, we tested other choices such as the radius of gyration ($R_g$) and the distance between the centers of mass of triazine trimers. First, we plotted how each reaction coordinate, i.e., the total number of non-covalent interactions, radius of gyration, and distance between the centers of mass of triazine trimers, changes over time from 200 to 500 ns of brute force MD simulation data, for all \emph{cis} and all \emph{trans} cases for the original dimer of triazine trimers in Ref.~\onlinecite{grate2016}. We picked 200 ns as a starting point since that is when the dimer stays around its most stable state (nanorod for all \emph{cis} and intertwined for all \emph{trans}). From Fig.~\ref{fig:col_var_plot}, we can see that while the total number of non-covalent interactions and distance between trimers change only slightly, the radius of gyration fluctuates dramatically and takes on different values for the same structure throughout the simulation. Hence, we concluded that the radius of gyration is not a suitable reaction coordinate to describe the conformations of the dimer of triazine trimers. 

\begin{figure}
\centering
\begin{tabular}{cc}
\includegraphics[width=80mm]{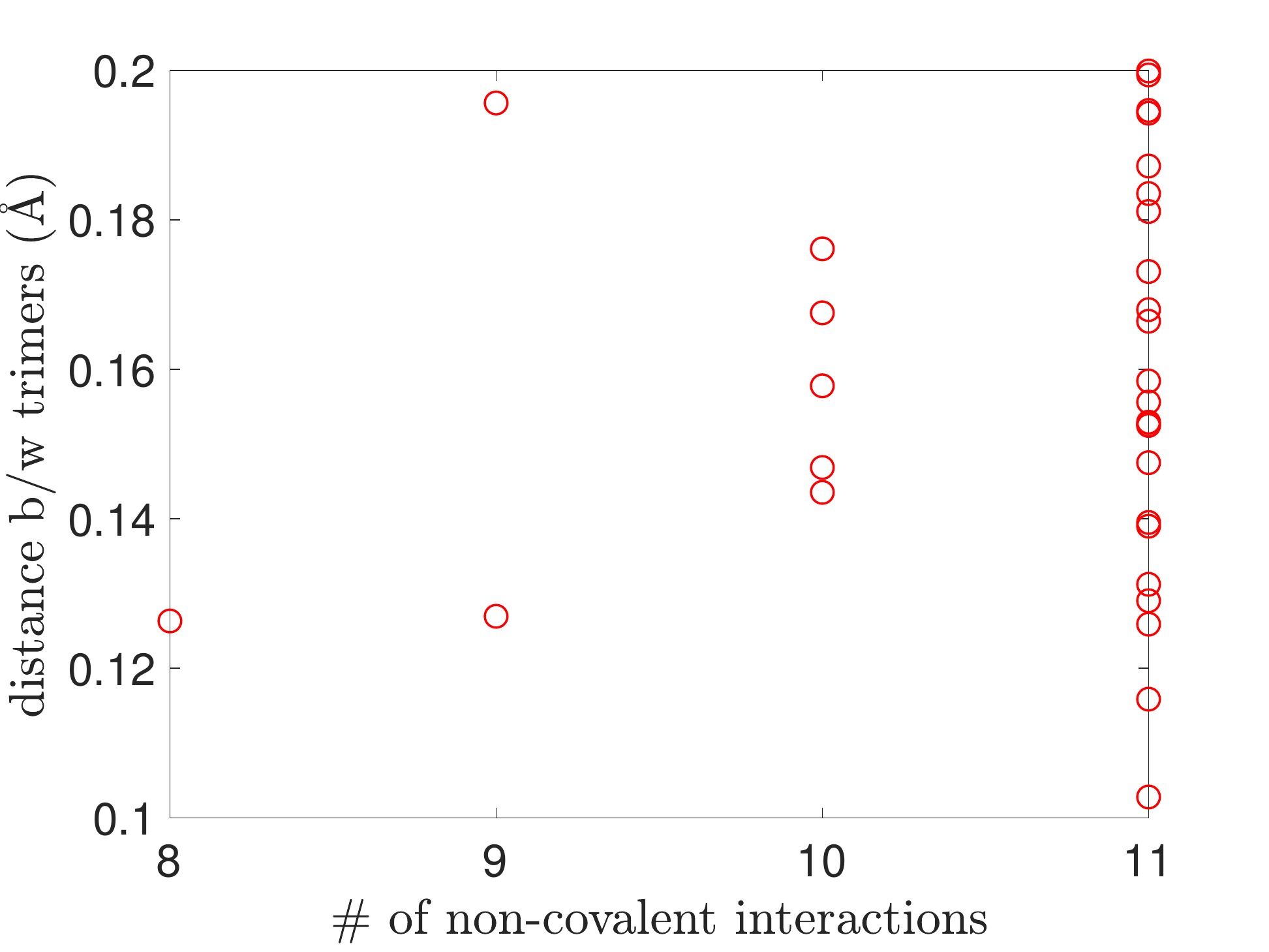} & \includegraphics[width=80mm]{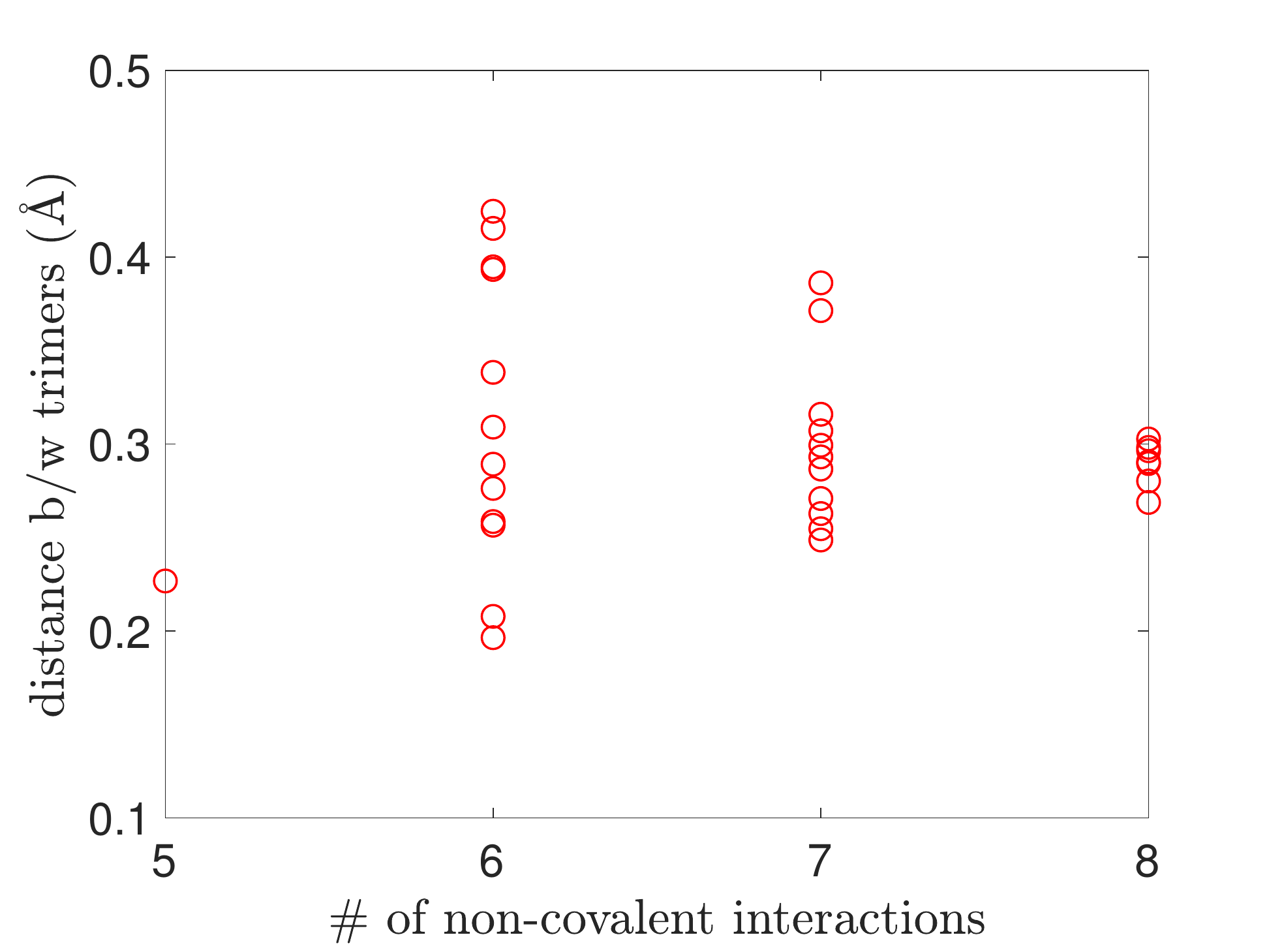} \\
(a) Plot of distance vs. non-covalent interactions (\emph{cis}). & (b) Plot of distance vs. non-covalent interactions (\emph{trans}). \\[6pt]
\end{tabular}
\caption{\label{fig:col_var_plot_noncov_dist} Plots of distance between trimers vs. \# of non-covalent interactions for a dimer of triazine trimers (all \emph{cis} and all \emph{trans}) with S-ethyl side chains and amino backbone from brute force MD simulations.}
\end{figure}

To further test whether the distance between trimers is a suitable reaction coordinate, we plotted the distance between trimers for each total number of non-covalent interactions from 200 to 500 ns of brute force MD simulation data. We wanted to see whether there was a correlation between the two reaction coordinates. From Fig.~\ref{fig:col_var_plot_noncov_dist}, we can see that the two reaction coordinates are not significantly correlated with each other, since for the same total number of non-covalent interactions, the distance between trimers can take many different values and vice versa. Hence, in order to precisely differentiate between different conformations, we concluded that the total number of non-covalent interactions is a more suitable reaction coordinate compared to the distance between trimers. 

\subsection{\label{sec:triazine_polymers} Triazine Polymers}
In Ref.~\onlinecite{grate2016}, in which the first MD simulations of the triazine polymers were published, the dimer of triazine trimers, in all \emph{cis} and in all \emph{trans} conformations for separate simulations, had alkane thiol (specifically S-ethyl) side chains that could not participate in hydrogen bonding and an amino backbone that could participate in hydrogen bonding. It was shown that from brute force MD simulations a dimer of all \emph{cis} triazine trimers forms a nanorod-like structure (which represented 100\% of the total ensemble) that is stabilized by hydrogen bonds between the two backbones and three pairs of parallel $\pi$-$\pi$ interactions between the two trimers' triazine rings. Similarly, from replica exchange molecular dynamics (REMD) simulations a single hexamer with the same side chains and backbone folds onto itself and forms a nanorod-like structure (which represented 19\% of the total ensemble) that is also stabilized by hydrogen bonds and $\pi$-$\pi$ interactions. Although the MD simulation results have yet to be experimentally verified, the MD simulations showed that the triazine polymers have potential to be useful building blocks for new materials, especially if they can form the nanorod.

To investigate the effects of having different side chains and backbone structures, we made several variants with the generalized Amber force field (GAFF) and the programs ACPYPE and Ante-chamber\cite{da2012, wang2004, wang2006}. Specifically, four different variants were made for the dimer of triazine trimers case for all \emph{cis} and all \emph{trans} separately, and thus eight different systems in total: 
\begin{enumerate}
    \item alkyl amino (amino-ethyl) side chains and amino backbone
    \item amino side chains and amino backbone
    \item amino side chains and sulfur backbone
    \item alkane thiol (S-ethyl) side chains and sulfur backbone 
\end{enumerate}
Note that the relative strength of the hydrogen bonding ability of the side chains is measured by the electronegativity of the nitrogen atom in the side chains. Hence, the amino-ethyl side chains have weaker hydrogen bonding compared to the amino side chains. The dimer of triazine trimers' variants' chemical structures and properties, along with the original dimer with S-ethyl side chains and amino backbone, are listed in Table~\ref{tab:trimers_summary}. All ten systems, including the \emph{cis} and \emph{trans} conformations of the original molecule, were simulated with the CAS algorithm. Similarly, three different variants were made for the single triazine hexamer case for all \emph{cis} and all \emph{trans} separately, and thus six different systems in total: 
\begin{enumerate}
	\item alkyl amino with protecting group (amino-ethyl-sulfide) side chains and amino backbone
    \item alkyl amino (amino-ethyl) side chains and amino backbone
    \item amino side chains and amino backbone
\end{enumerate}
Again, the relative strength of the hydrogen bonding ability of the side chains is measured by the electronegativity of the nitrogen atom in the side chains. Hence, when we rank the side chains from weakest to strongest in terms of hydrogen bonding ability, we get amino-ethyl-sulfide, amino-ethyl, and amino side chains. The single triazine hexamer's variants' chemical structures and properties, along with the original hexamer with S-ethyl side chains and amino backbone, are listed in Table~\ref{tab:hexamers_summary}. All eight systems, including the \emph{cis} and \emph{trans} conformations of the original molecule, were simulated with REMD.

The goal was to find out whether the side chain and/or the backbone having hydrogen bonding abilities was essential for the nanorod to form. The hydrogen bonding side chains also differed from each other in terms of strength so that we could investigate whether the difference in hydrogen bonding strength affected the thermodynamic and kinetic properties of the triazine polymer.

\begin{table}
\caption{\label{tab:trimers_summary} Dimer of triazine trimers' chemical structures and properties. The following systems were simulated with the CAS algorithm.}
\begin{ruledtabular}
\begin{tabular}{ccc}
Chemical structure & Side chain & Backbone\\
\hline
& & \\
\includegraphics[width=60mm]{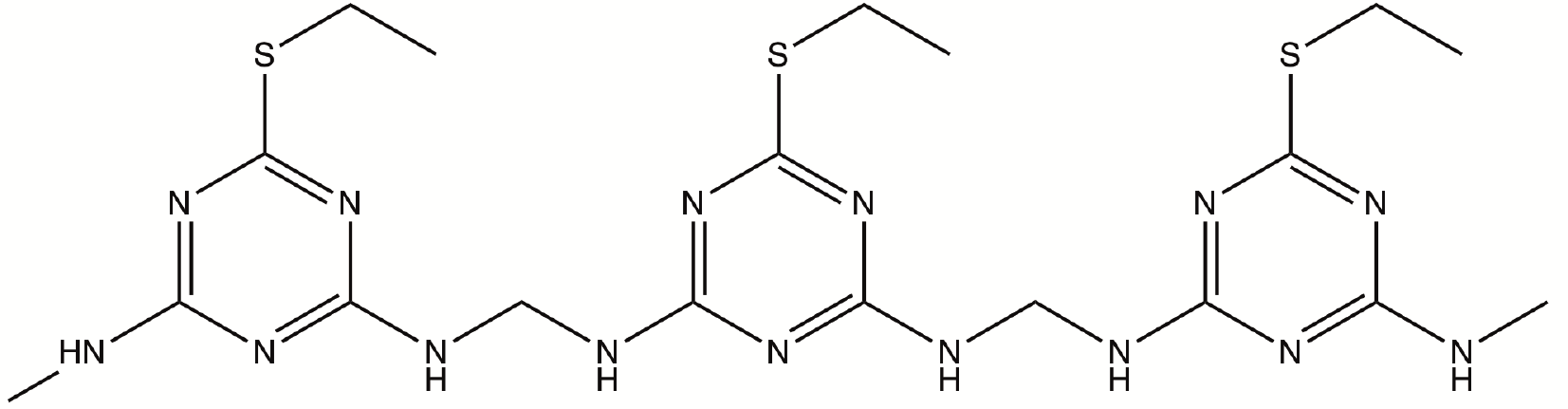} & Cannot hydrogen bond & Can hydrogen bond\\
\includegraphics[width=60mm]{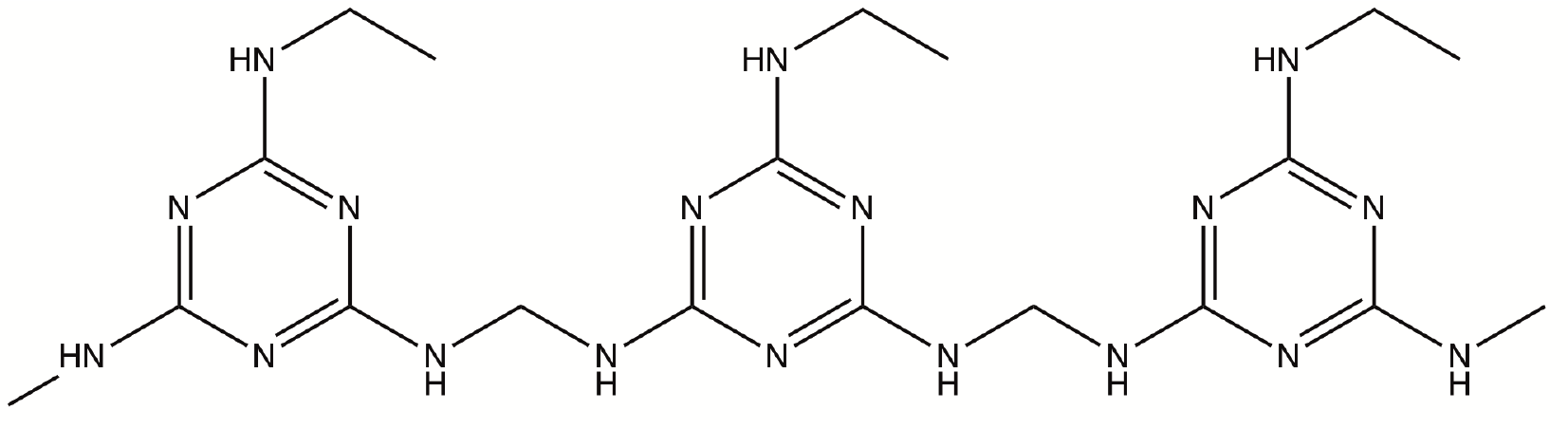} & Can hydrogen bond & Can hydrogen bond\\
\includegraphics[width=60mm]{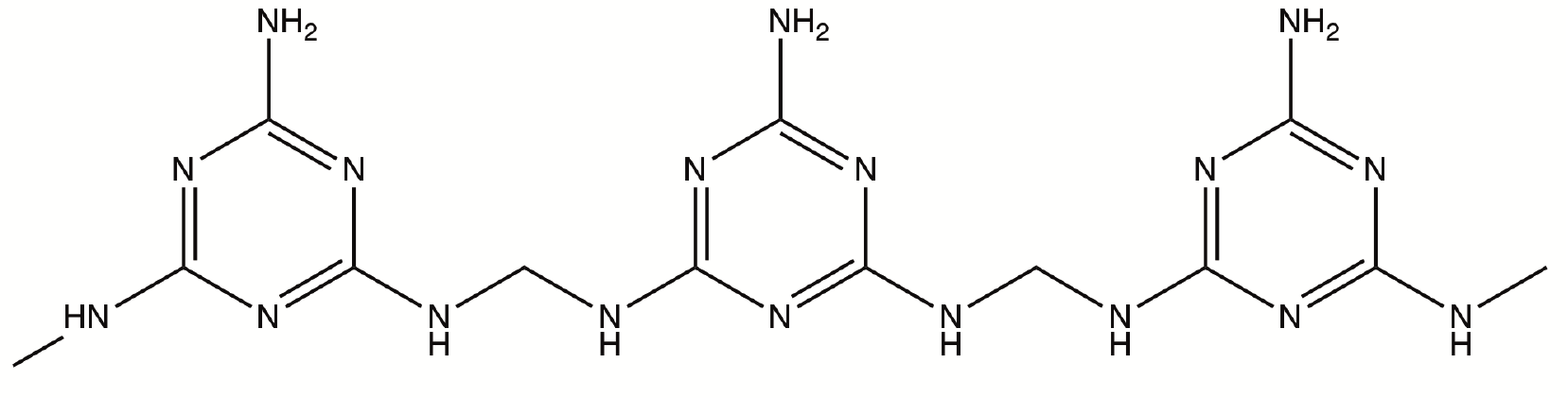} & Can hydrogen bond (stronger) & Can hydrogen bond\\
\includegraphics[width=60mm]{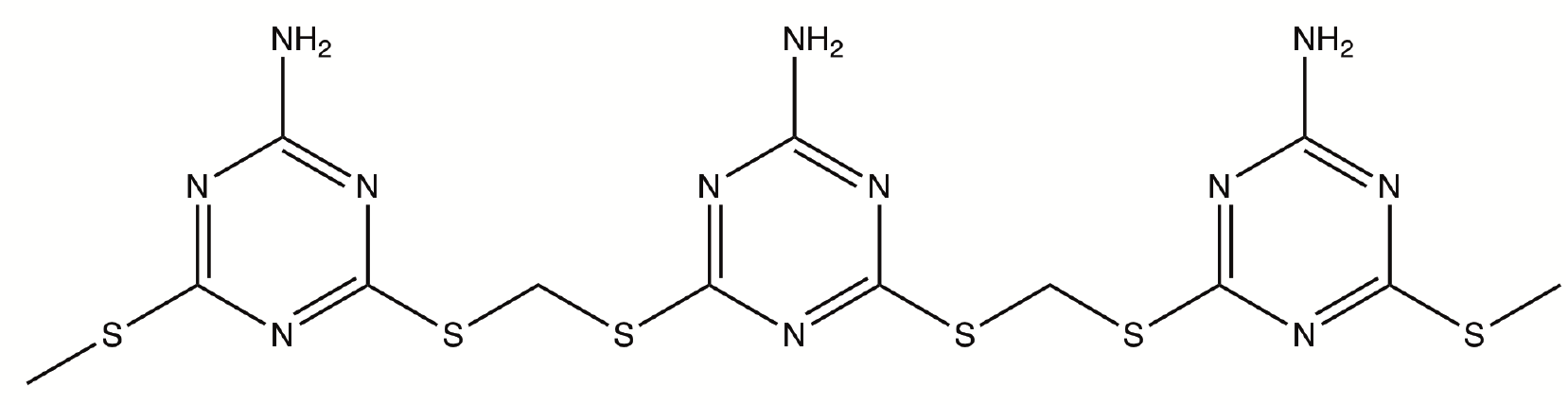} & Can hydrogen bond (stronger) & Cannot hydrogen bond\\
\includegraphics[width=60mm]{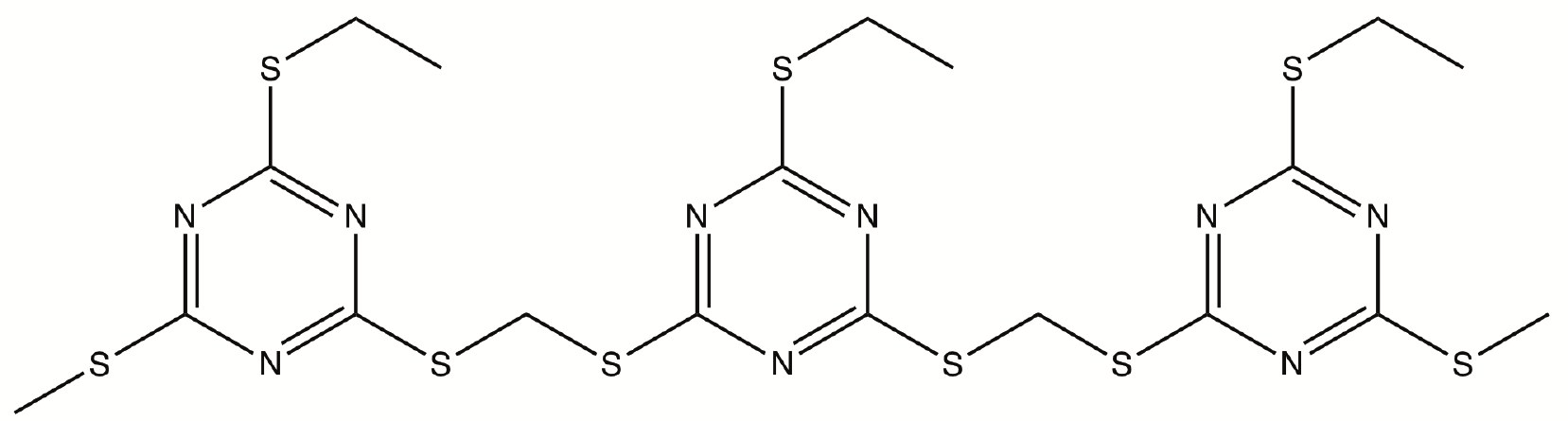} & Cannot hydrogen bond & Cannot hydrogen bond\\
\end{tabular}
\end{ruledtabular}
\end{table}

\begin{table}
\caption{\label{tab:hexamers_summary} Single triazine hexamer's chemical structures and properties. The following systems were simulated with REMD.}
\begin{ruledtabular}
\begin{tabular}{ccc}
Chemical structure & Side chain & Backbone\\
\hline
& & \\
\includegraphics[width=60mm]{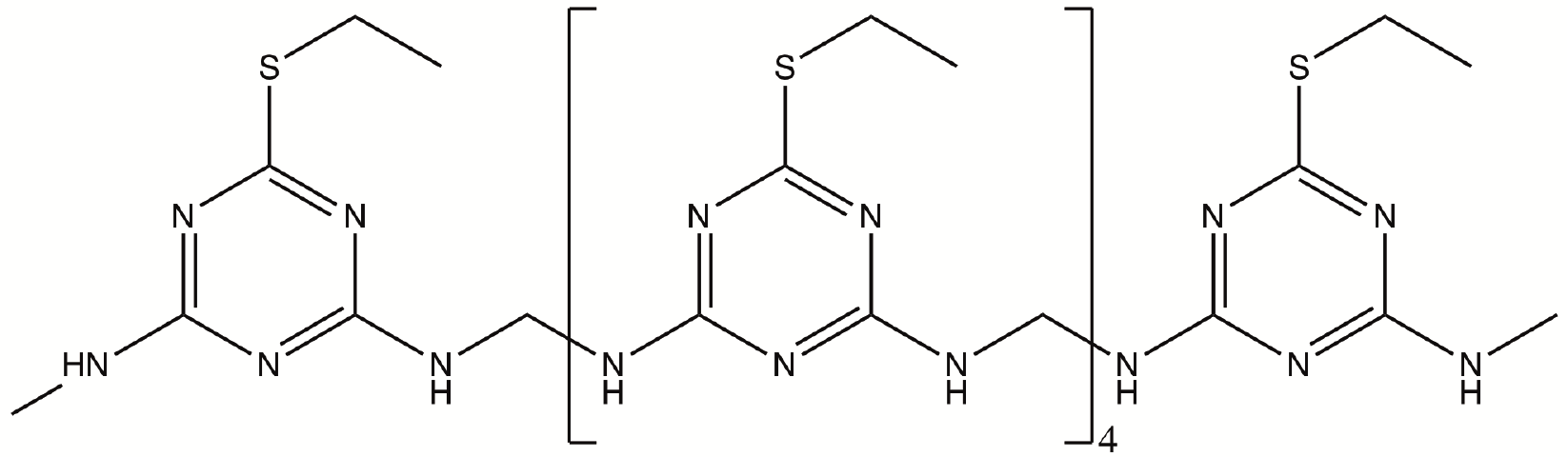} & Cannot hydrogen bond & Can hydrogen bond\\
\includegraphics[width=60mm]{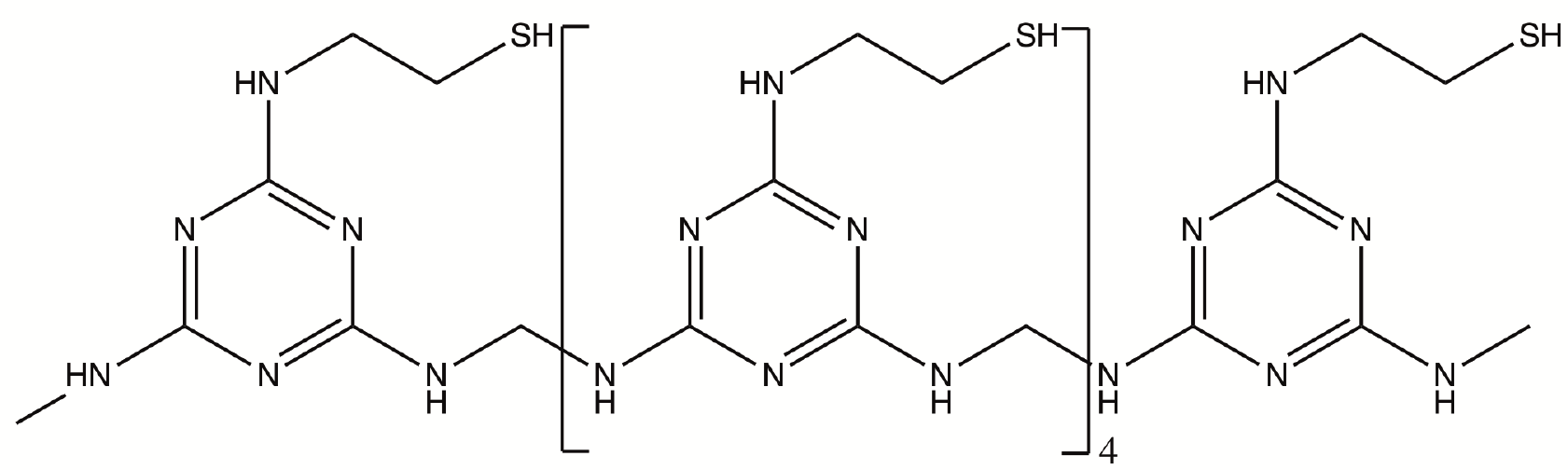} & Can hydrogen bond & Can hydrogen bond\\
\includegraphics[width=60mm]{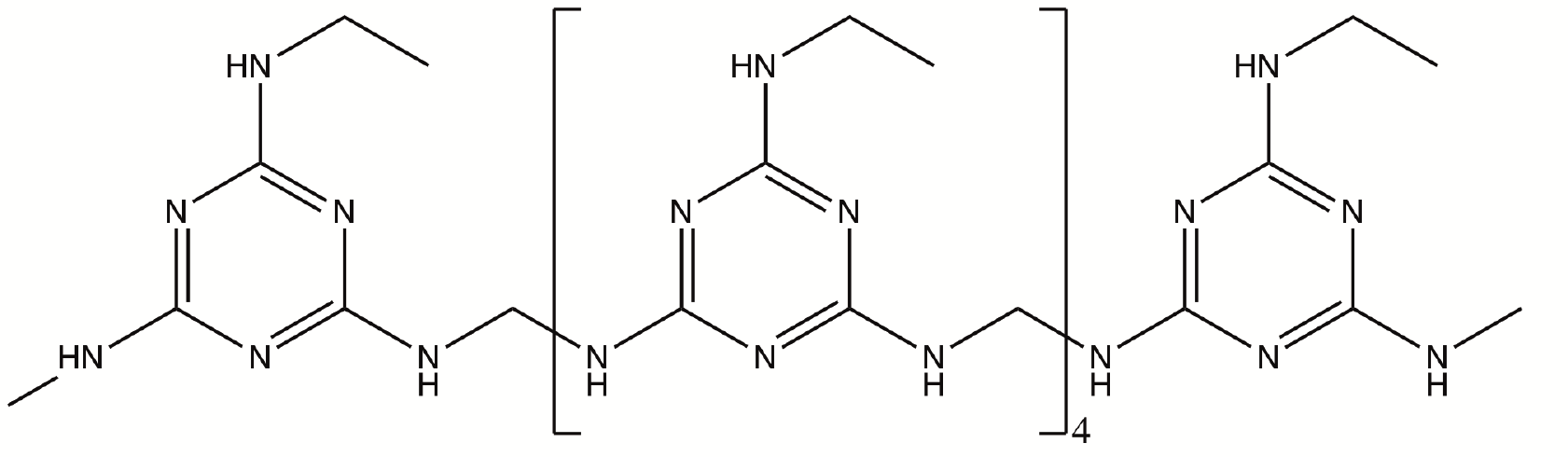} & Can hydrogen bond (stronger) & Can hydrogen bond\\
\includegraphics[width=60mm]{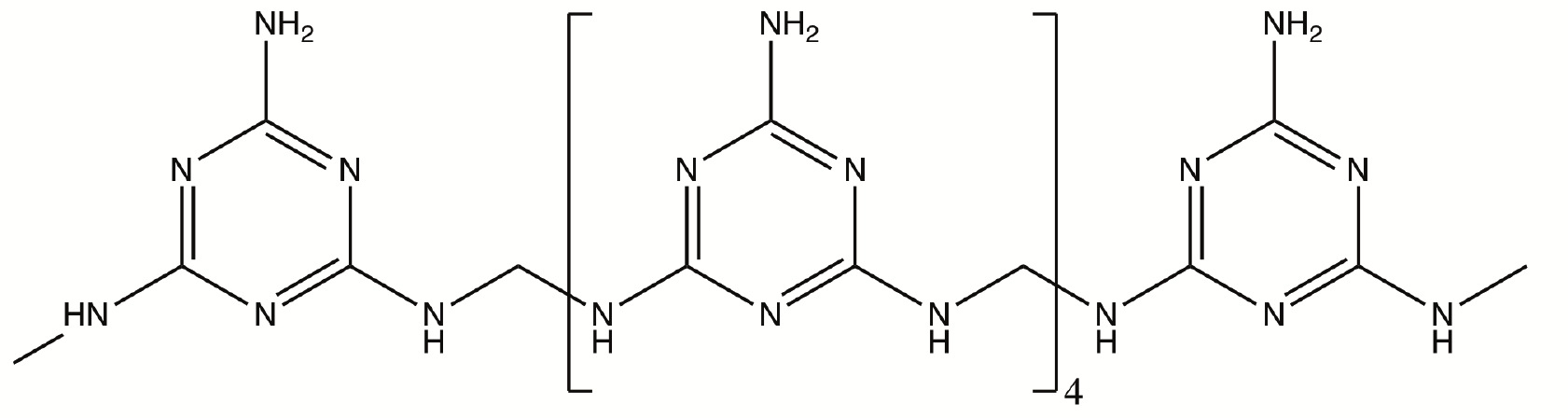} & Can hydrogen bond (strongest) & Can hydrogen bond\\
\end{tabular}
\end{ruledtabular}
\end{table}

\subsection{\label{sec:simulation_protocol} Simulation Protocol}
The triazine polymers were all simulated with GROMACS 4.6.4 at temperature $T = 300\ \mbox{K}$ with time step $\Delta\mbox{t} = 2\ \mbox{fs}$\cite{pronk2013}. Most simulation parameters were identical to the ones in Ref.~\onlinecite{grate2016}, including the force field that was generated using generalized Amber force field (GAFF), the explicit solvent with SPC water molecules and 0.115 M KCl for the dimer of triazine trimers case, and the Generalized Born/Surface Area (GB/SA) implicit solvent for the single triazine hexamer case\cite{wang2004, voelz2011}. 

For the dimer of triazine trimers simulations, 500 ns of brute force MD simulations in the NPT ensemble with a Parrinello-Rahman barostat with a 1 ps coupling time at 1 bar and a Nose-Hoover thermostat with a 0.2 ps coupling time at 300 K were initially run to initialize the CAS algorithm simulations, since having good initial conditions speeds up convergence\cite{parrinello1981, hoover1985}. The CAS algorithm simulations  were run for 2 $\mu$s for the dimer of triazine trimers with amino backbone and for 2.3 $\mu$s for the dimer of triazine trimers with sulfur backbone. The total simulation time is calculated by the cumulative number of macrostates $\times$ target number of walkers $\times$ simulation time $\tau$. The target number of walkers per macrostate $n_w$ was set to 20, and the simulation time $\tau$ was set to $100\ \mbox{ps}$. As previously mentioned, REMD was not used for the dimer of triazine trimers simulations due to its high computational cost when simulating explicit solvent systems. 

For the single triazine hexamer simulations, 500 ns of REMD simulations in the NVT ensemble with velocity-rescale temperature coupling (0.2 ps coupling time) were run\cite{bussi2007}. The GB/SA implicit solvent with the Onufriev/Bashford/Case algorithm for calculating Born radii, solvent dielectric constant of 78.3, and infinite van der Waals and Coulomb cutoffs were used\cite{onufriev2004}. The hydrophobic solvent accessible surface area is calculated using an analytical continuum electrostatics (ACE)-type approximation and the internal dielectric constant is set to 1, which are the default settings on GROMACS\cite{schaefer1996, calimet2001}. The REMD simulation parameters were identical to the ones in Ref.~\onlinecite{grate2016}, which followed Ref.~\onlinecite{voelz2011} as a reference. Specifically, 16 replicas that uniformly span from 300 K to 800 K were used for each simulation, with exchanges occurring every 1000 steps (2 ps) and a Metropolis acceptance rate of about 50\%. The replicas were first equilibrated for 200 ps at each temperature before the 500 ns production runs, which saved conformations and potential energies every 2 ps. REMD was used for the single triazine hexamer simulations to observe the hexamers folding onto itself, since there is a high energy barrier associated with rotating the bond between the linker nitrogen and the triazine ring as previously mentioned.

\section{\label{sec:results} Results}
\subsection{\label{sec:remd_results} Replica Exchange Molecular Dynamics}
\begin{figure}
\centering
\includegraphics[width=80mm]{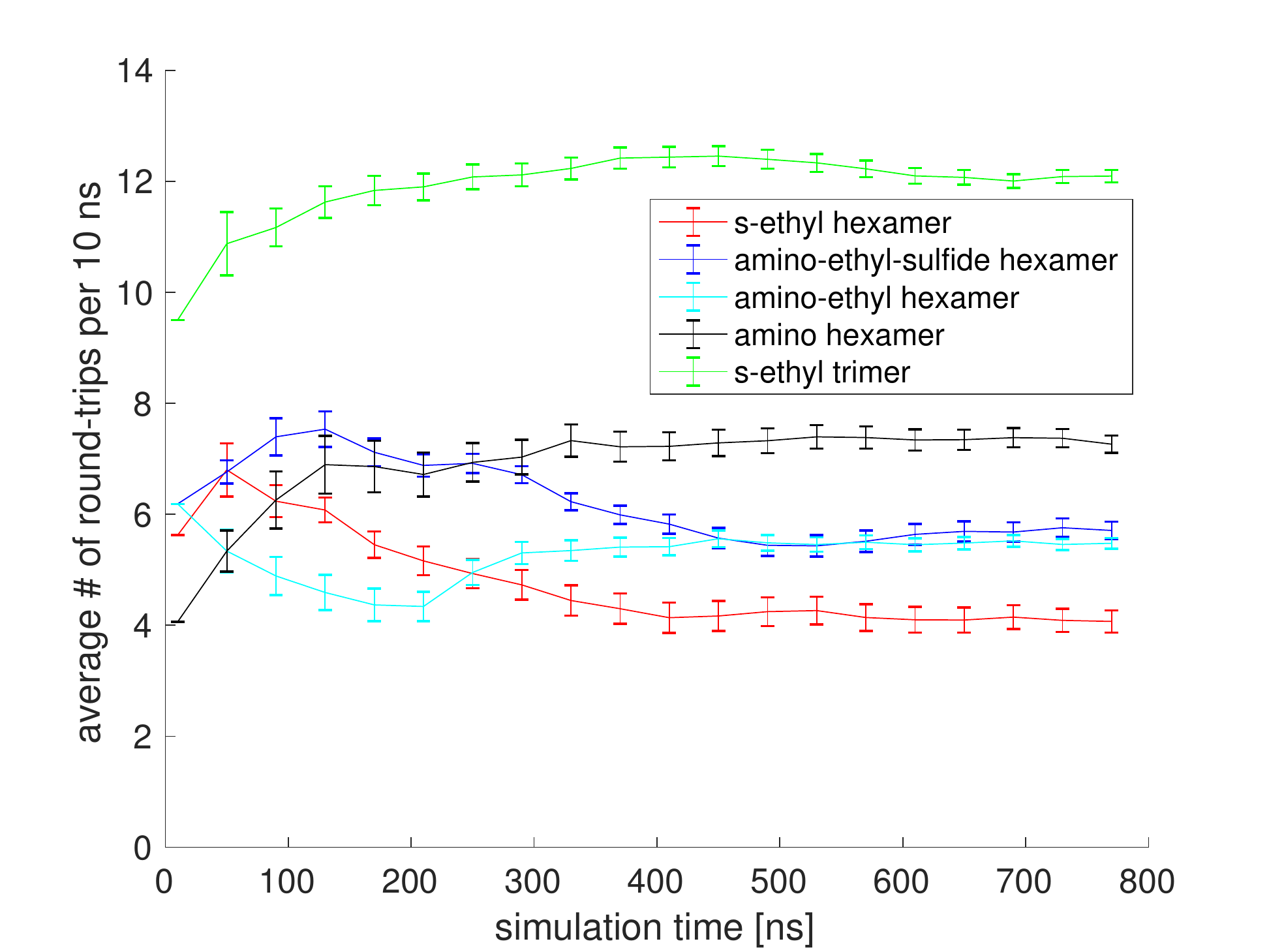} 
\caption{\label{fig:remd_convergence} Plot of average number of round-trips in a given observation time $\tau = 10$ ns vs. simulation time. The legend indicates the side chains of the particular hexamer tested and one triazine trimer with S-ethyl side chains. The average number of round-trips converges to a stable value with small error bars for each of the four hexamers and the single trimer.}
\end{figure}

To obtain the free energy landscape and the most stable conformation for each hexamer listed in Table~\ref{tab:hexamers_summary}, we simulated each hexamer, all \emph{cis} and all \emph{trans} separately as starting conformations, with replica exchange molecular dynamics (REMD) for 500 ns, as done in Ref.~\onlinecite{grate2016} and Ref.~\onlinecite{voelz2011}. We also ran a REMD simulation of a single triazine trimer with S-ethyl side chains to investigate the free energy landscape of a much simpler molecule for reference and followed the same simulation protocol as the hexamers. We used the last 400 ns of the 500 ns simulation run and since both all \emph{cis} and all \emph{trans} simulations were used for each trimer/hexamer, we ended up using 800 ns of simulation data to calculate the free energy landscape for each trimer/hexamer. The multistate Bennett acceptance ratio (MBAR), specifically the Python implementation by Shirts and Chodera, was used to calculate the free energy landscapes\cite{shirts2008}. To test for convergence of all REMD simulations, we followed Ref.~\onlinecite{sindhikara2010} and Ref.~\onlinecite{zheng2007} and calculated the average number of round-trips, i.e., how many times a replica visits both the lowest and the highest temperatures for each hexamer in a given observation time $\tau$. We set $\tau = 10$ ns and measured how the average number of round-trips changes as we increase the simulation time. Fig.~\ref{fig:remd_convergence} shows how the average number of round-trips converges to a stable value with small error bars as the simulation time increases for each of the four hexamers and the single trimer, indicating that the REMD simulations have converged. The standard deviation of the number of round trips was multiplied by 2, which approximately represents 95\% confidence interval, for error bars. 

\begin{figure}[htbp]
\centering
\begin{tabular}{cc}
\includegraphics[width=60mm]{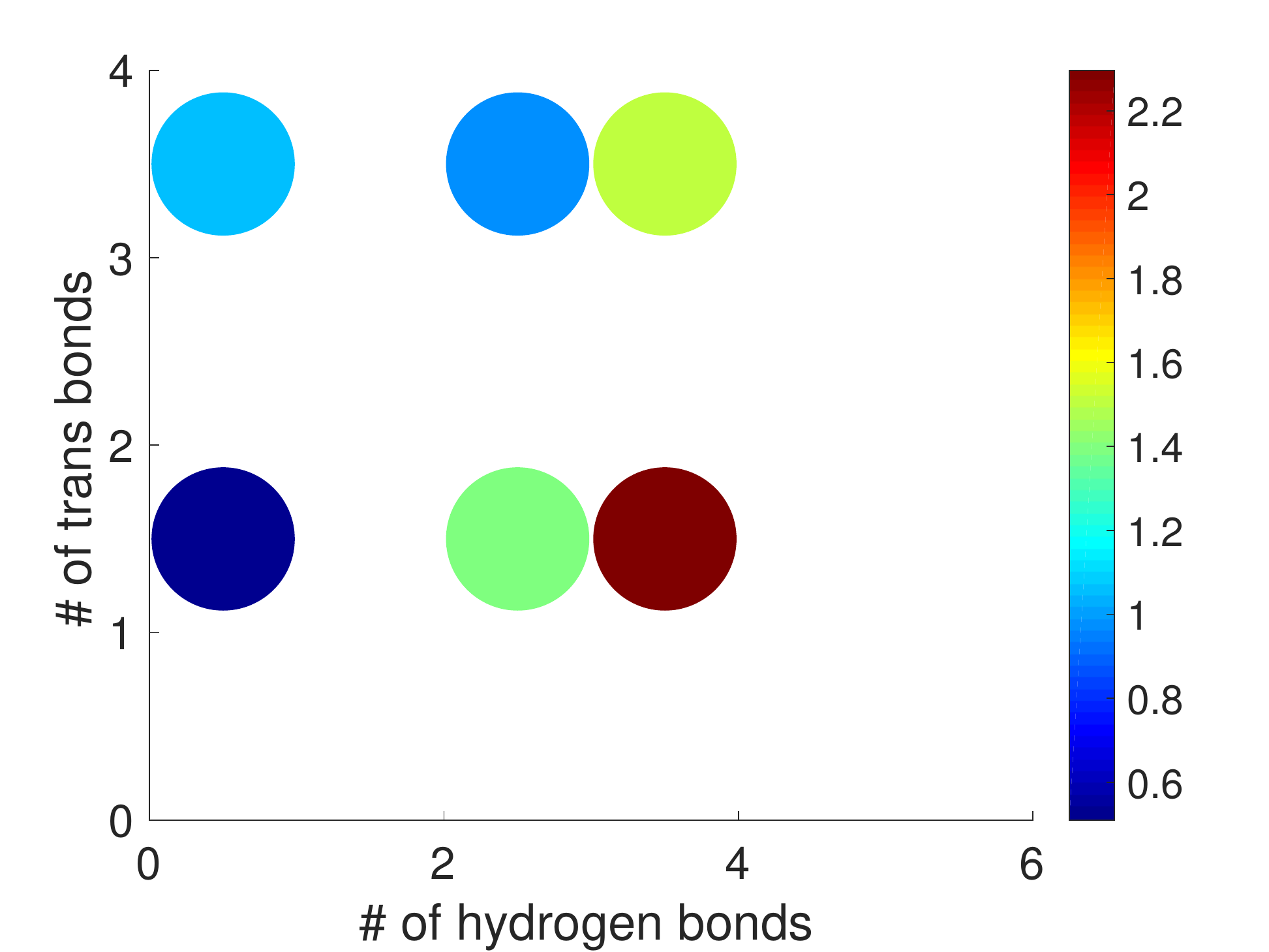} & \includegraphics[width=50mm]{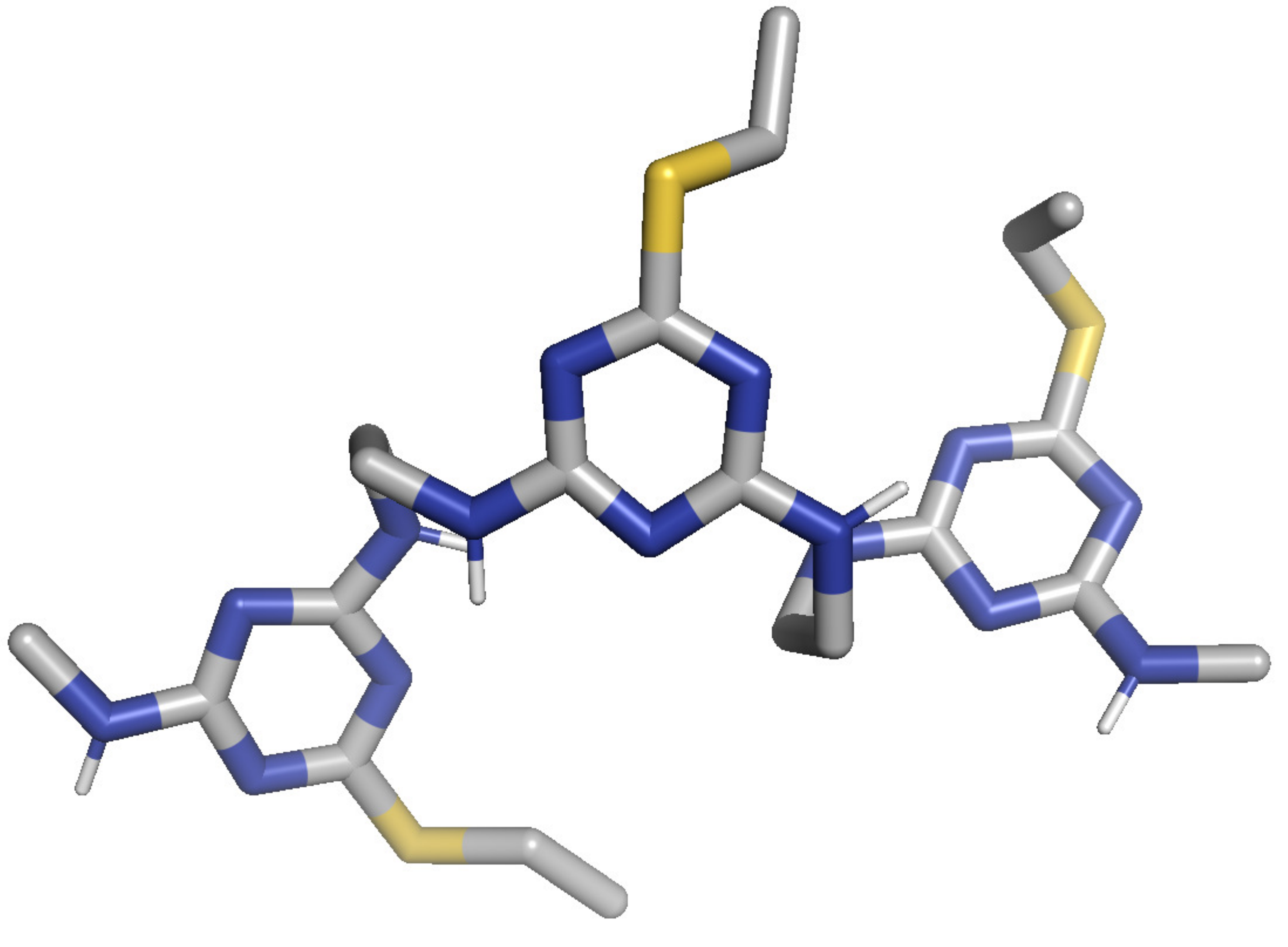} \\
(a) Free energy landscape. & (b) Most stable conformation. \\[6pt]
\end{tabular}
\caption{\label{fig:original_trimer_remd_free_energy_conformations} Free energy landscape and most stable conformation of a single triazine trimer with S-ethyl side chains and amino backbone from REMD simulations. Figure (a) plots the free energies in log scale or $-k_BT\ln P$ (kcal/mol), where $P$ denotes the weight, truncated at 5 kcal/mol, and the color bar indicates which colors correspond to which free energies in log scale (kcal/mol). Figure (b) shows the most stable conformation (0 hydrogen bonds and 1 \emph{trans} bond).}
\end{figure}

\begin{figure}[htbp]
\centering
\begin{tabular}{cc}
\includegraphics[width=60mm]{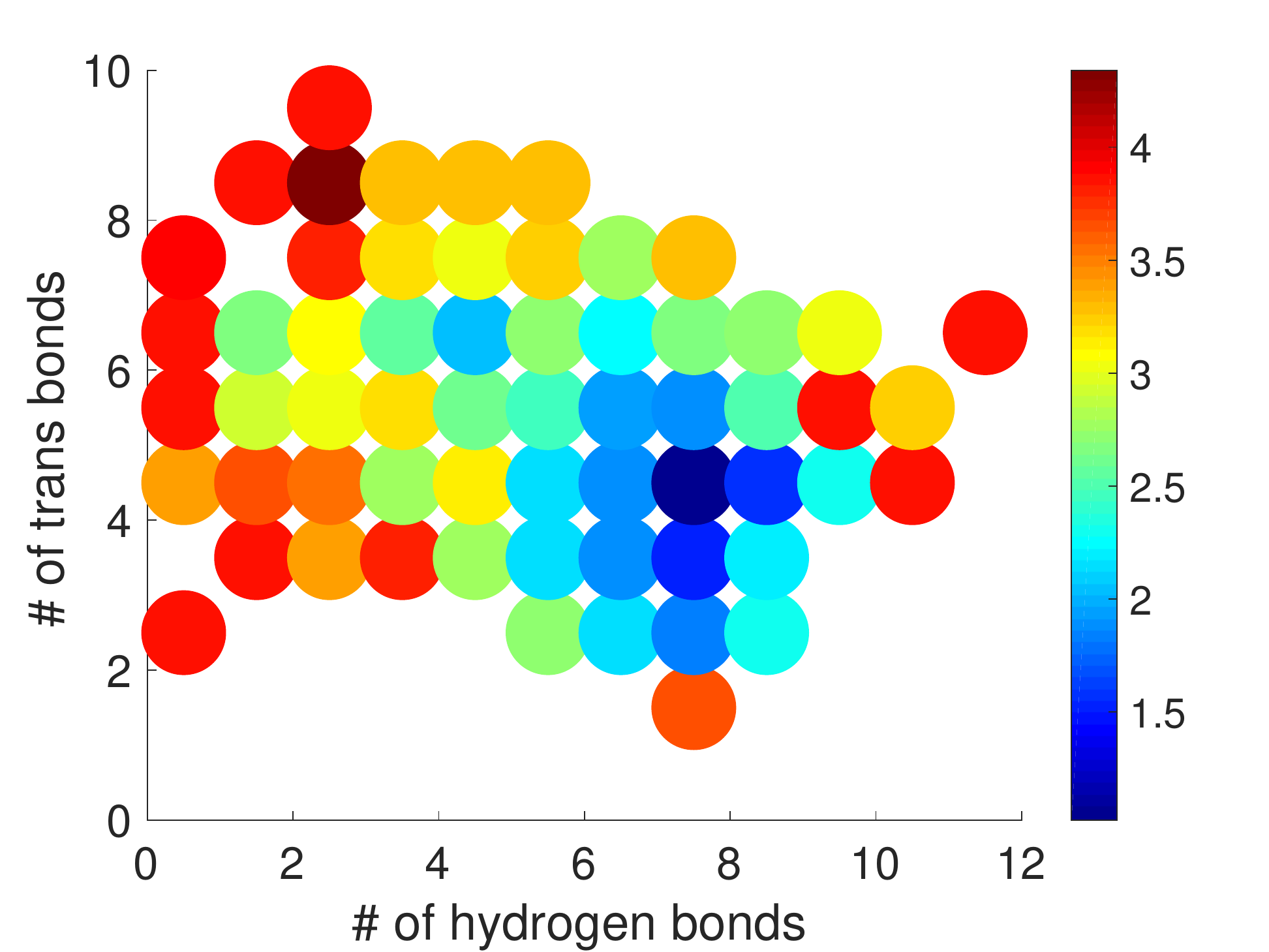} & \includegraphics[width=50mm]{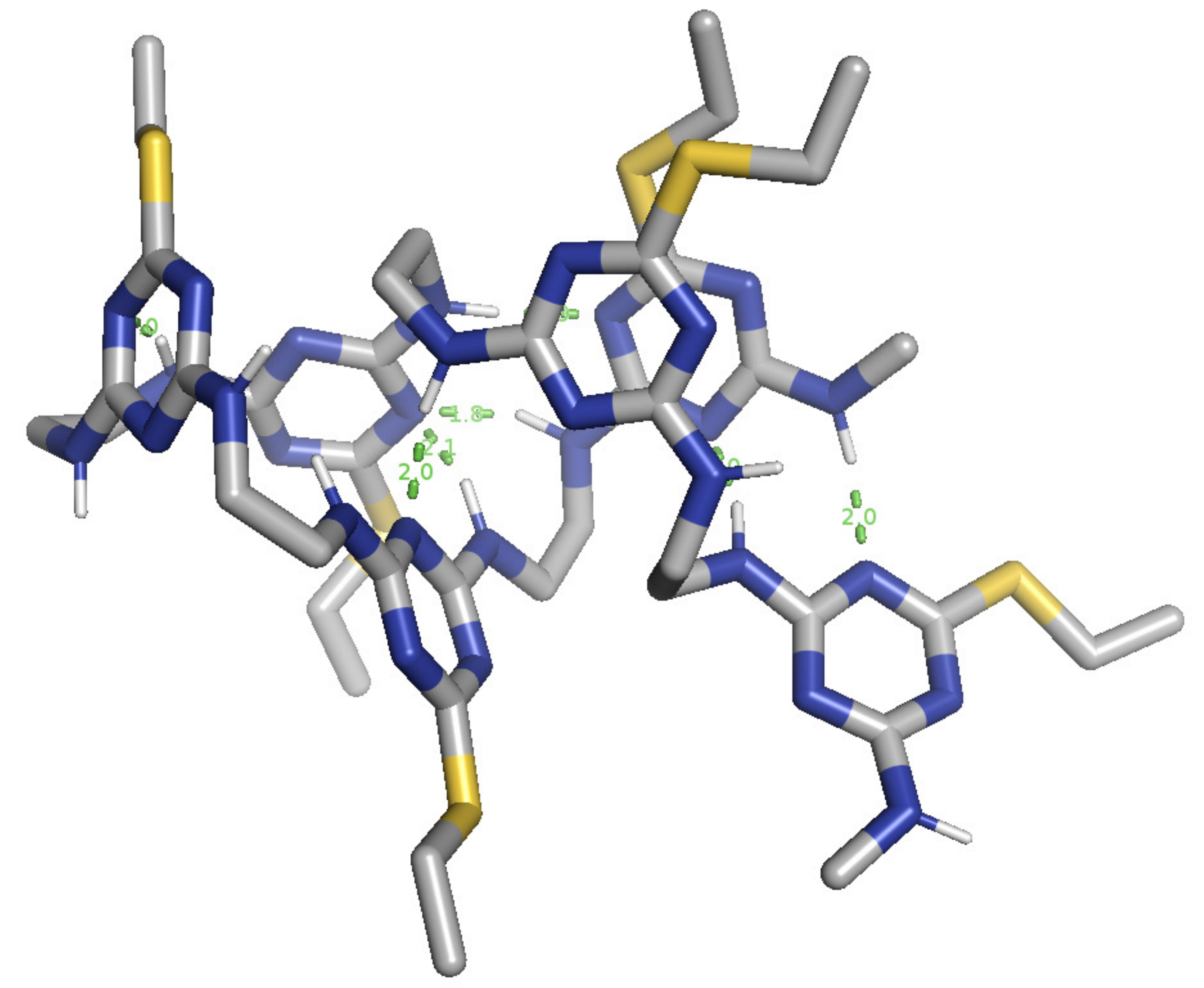} \\
(a) Free energy landscape. & (b) Most stable conformation. \\[6pt]
\multicolumn{2}{c}{\includegraphics[width=60mm]{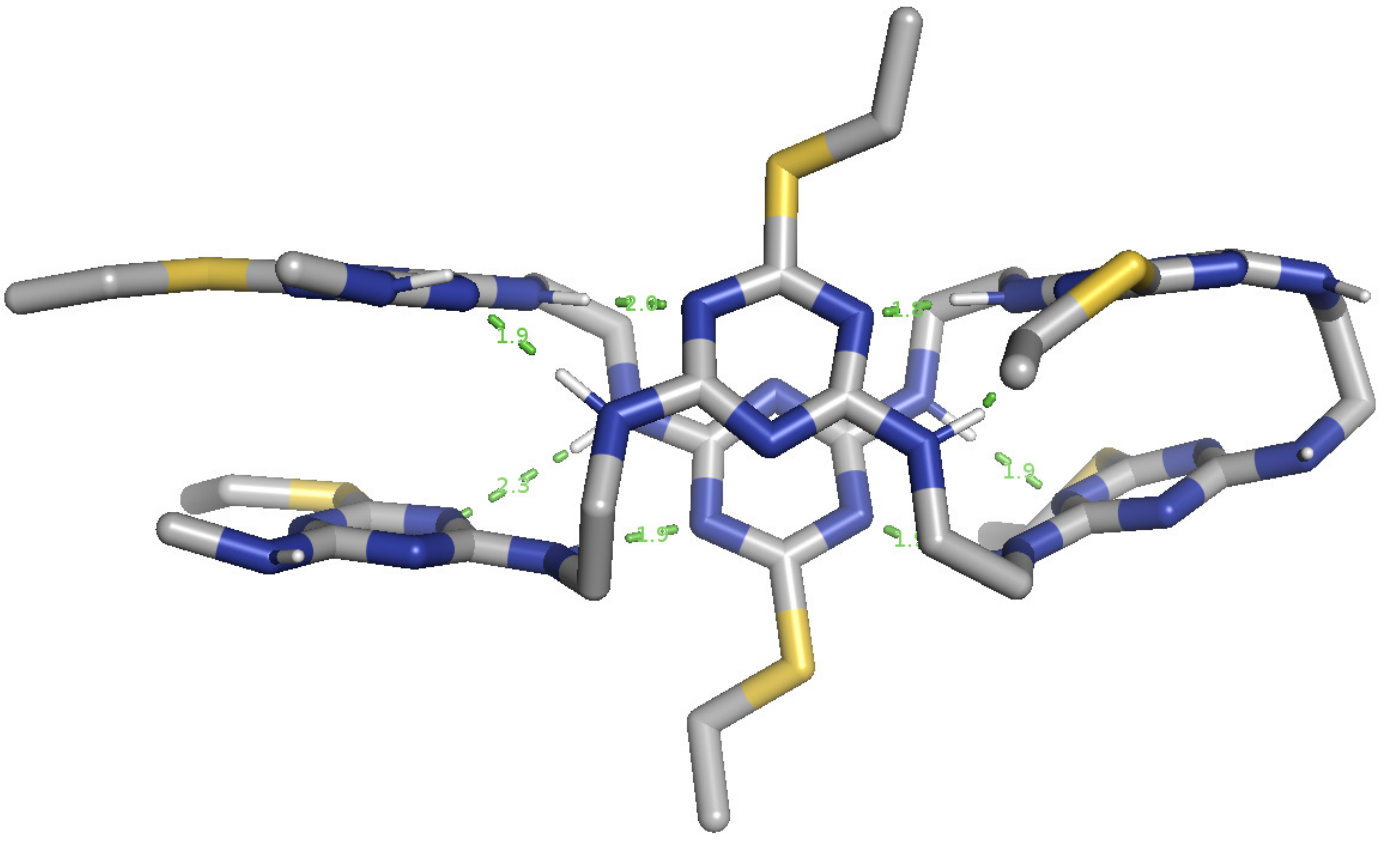}}\\
\multicolumn{2}{c}{(c) Nanorod structure.}
\end{tabular}
\caption{\label{fig:original_hexamer_remd_free_energy_conformations} Same as Fig.~\ref{fig:original_trimer_remd_free_energy_conformations} but for a single triazine hexamer with S-ethyl side chains.}
\end{figure}

\begin{figure}[htbp]
\centering
\begin{tabular}{cc}
\includegraphics[width=60mm]{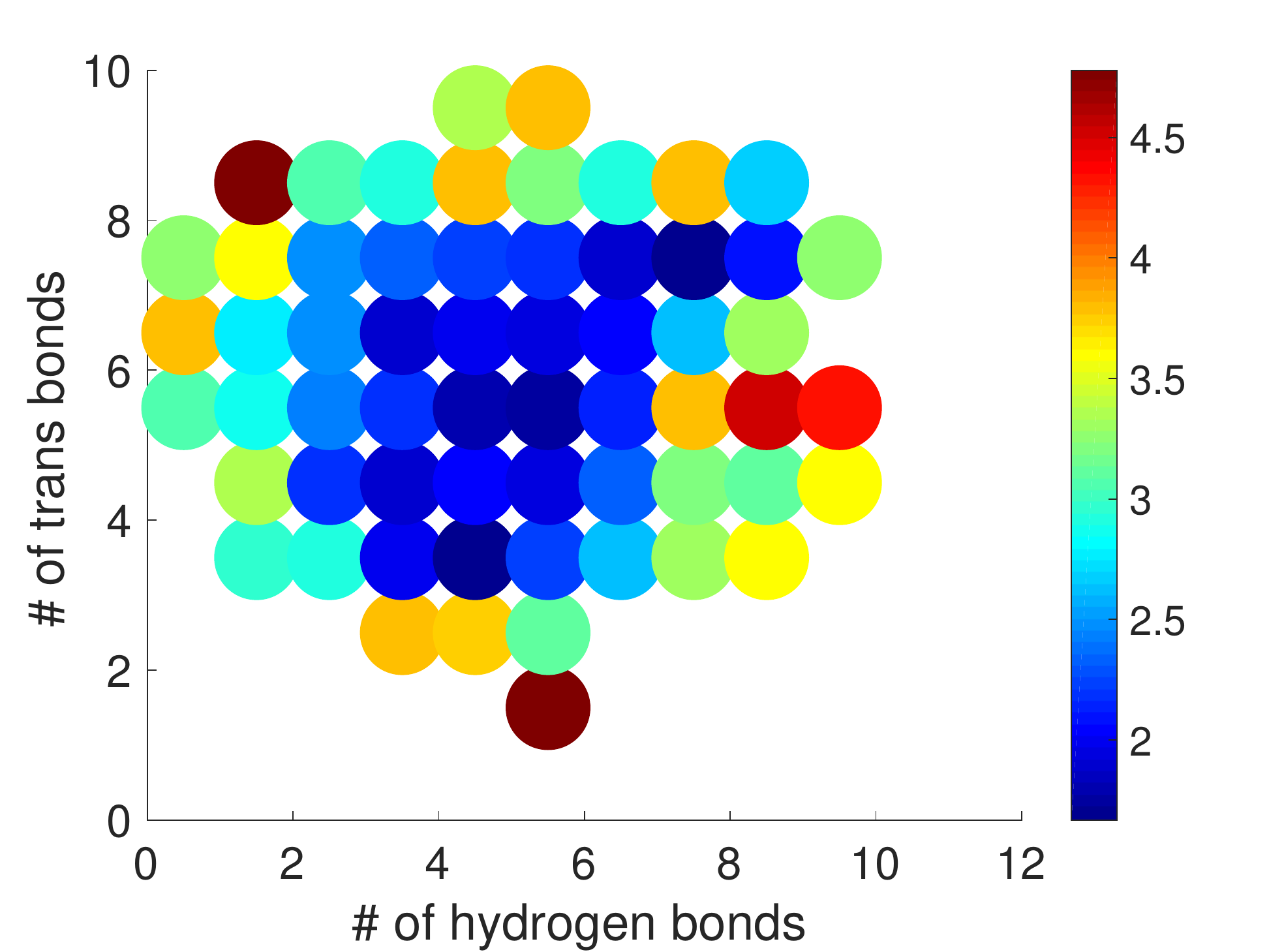} & \includegraphics[width=50mm]{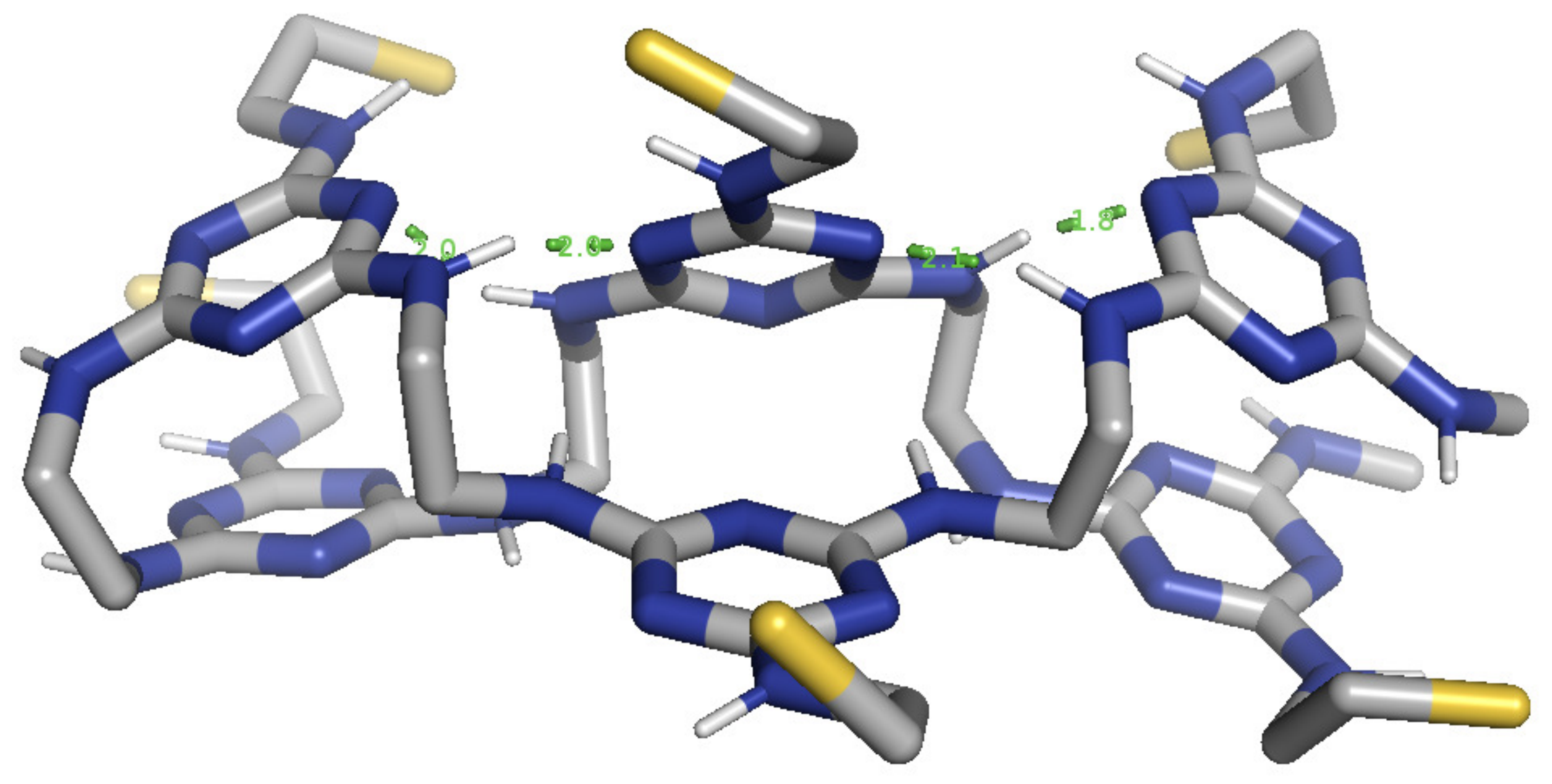} \\
(a) Free energy landscape. & (b) Most stable conformation/nanorod structure. \\[6pt]
\end{tabular}
\caption{\label{fig:tns_hexamer_remd_free_energy_conformations} Same as Fig.~\ref{fig:original_trimer_remd_free_energy_conformations} but for a single triazine hexamer with amino-ethyl-sulfide side chains.}
\end{figure}

\begin{figure}[htbp]
\centering
\begin{tabular}{cc}
\includegraphics[width=60mm]{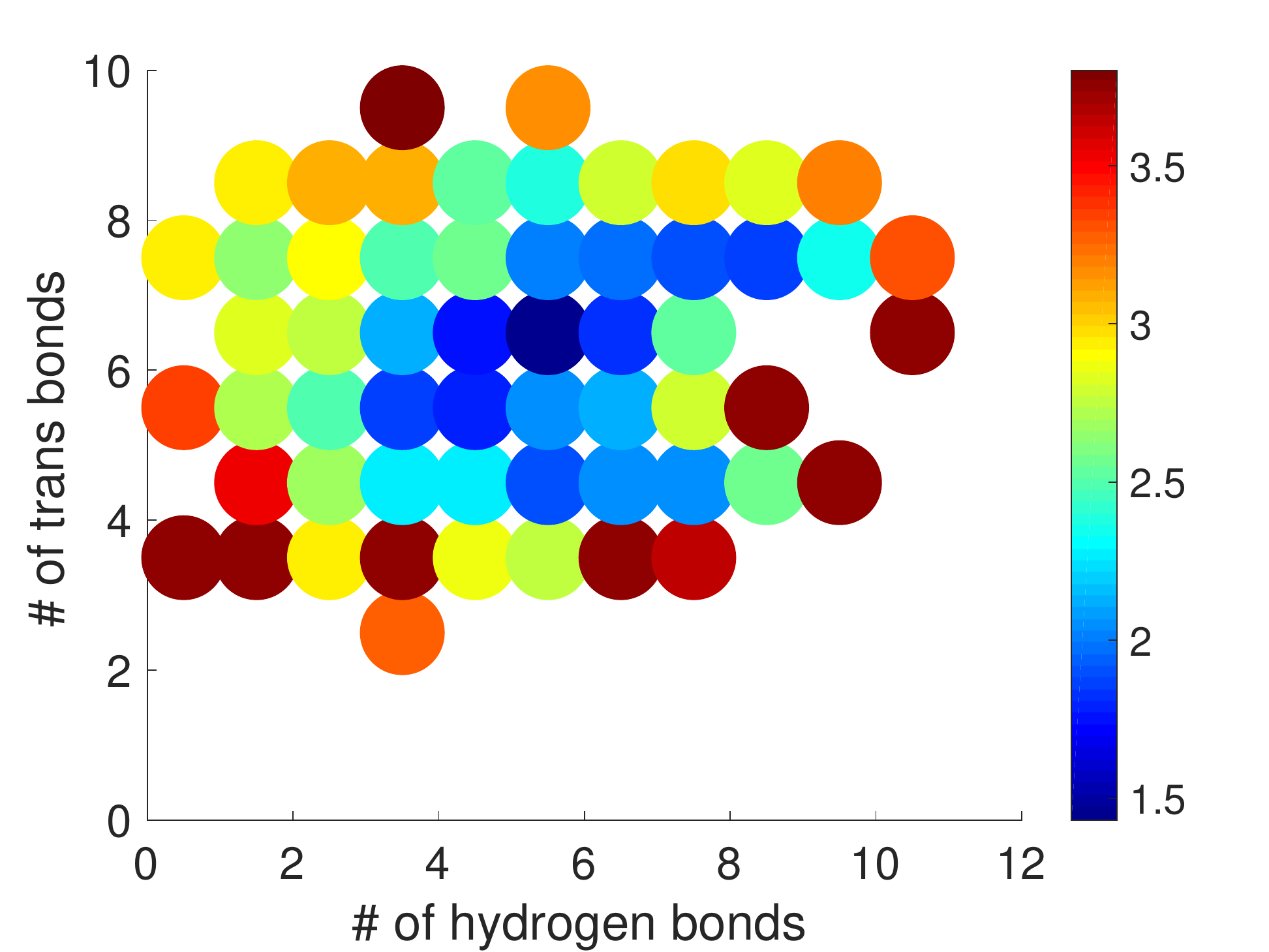} & \includegraphics[width=50mm]{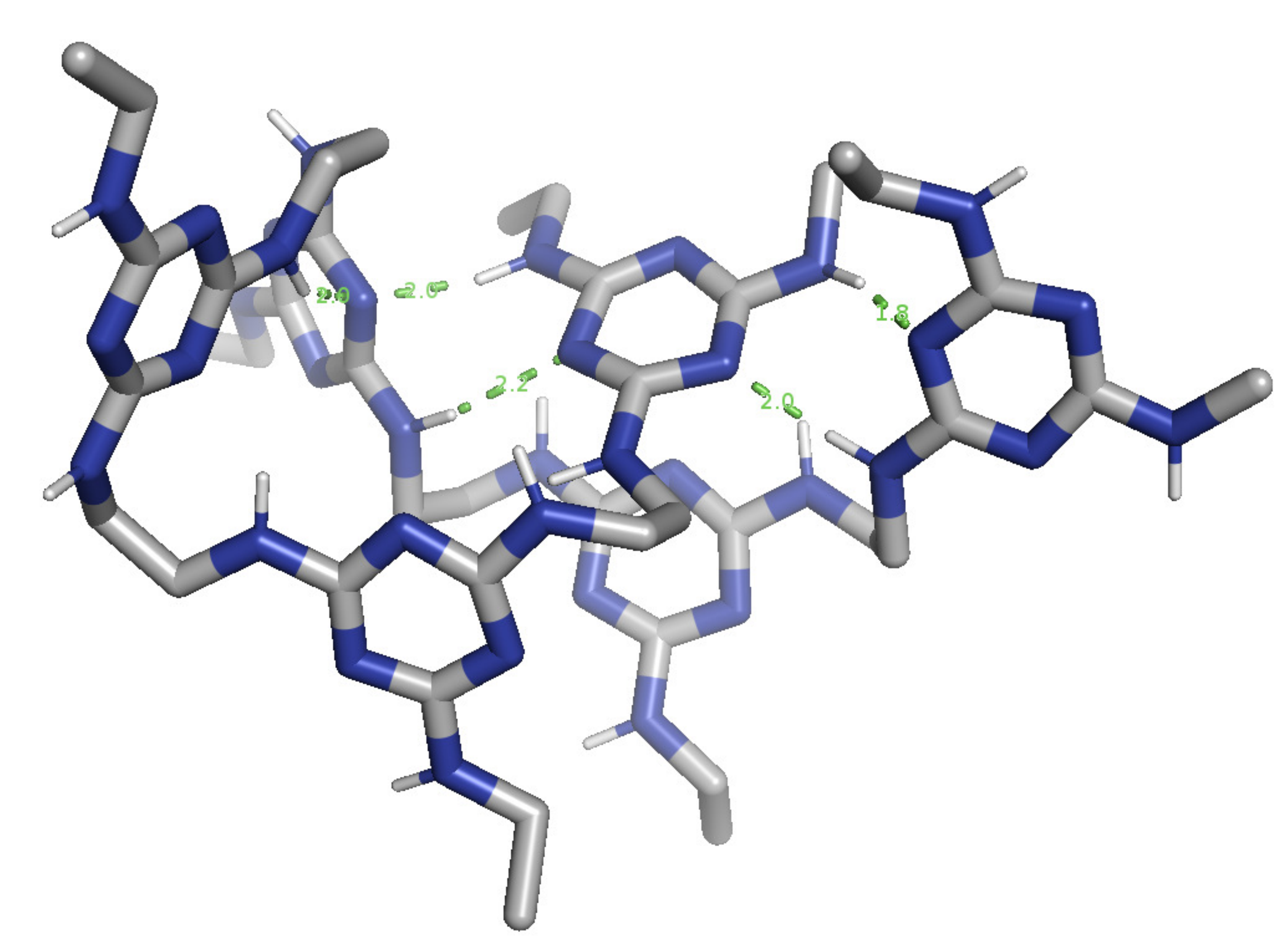} \\
(a) Free energy landscape. & (b) Most stable conformation. \\[6pt]
\multicolumn{2}{c}{\includegraphics[width=60mm]{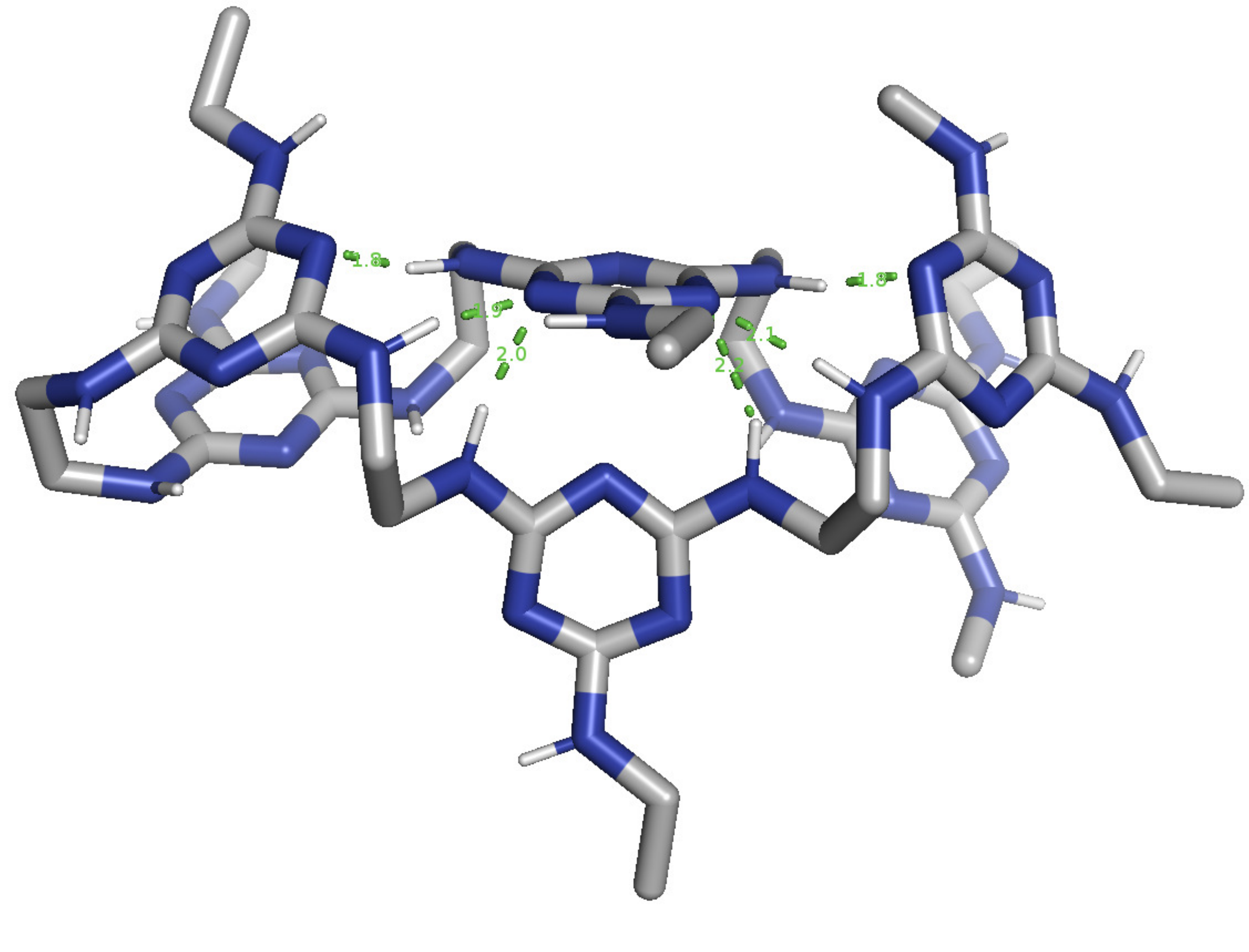}}\\
\multicolumn{2}{c}{(c) Nanorod structure.}
\end{tabular}
\caption{\label{fig:hn_hexamer_remd_free_energy_conformations} Same as Fig.~\ref{fig:original_trimer_remd_free_energy_conformations} but for a single triazine hexamer with amino-ethyl side chains.}
\end{figure}

\begin{figure}[htbp]
\centering
\begin{tabular}{cc}
\includegraphics[width=60mm]{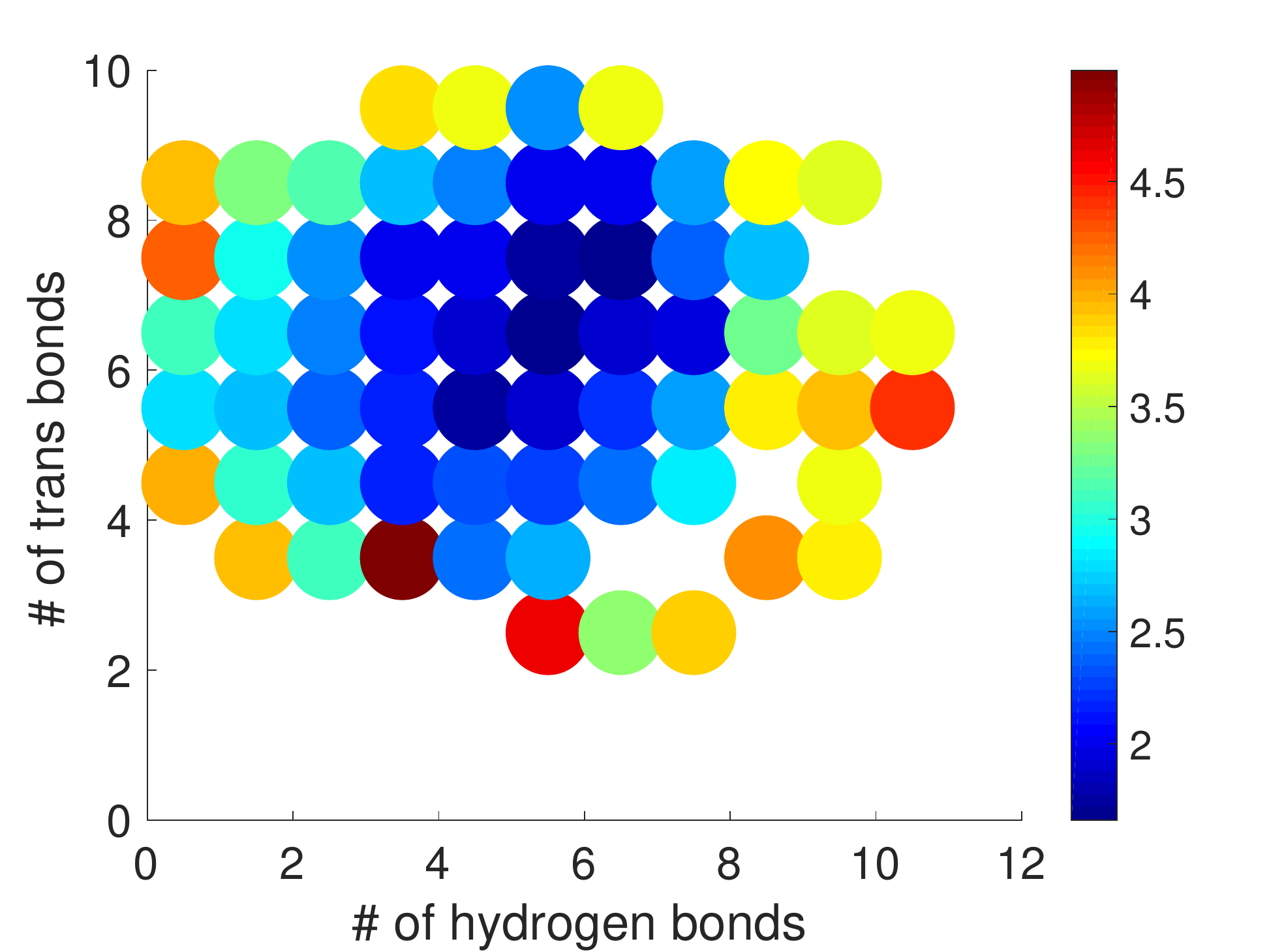} & \includegraphics[width=50mm]{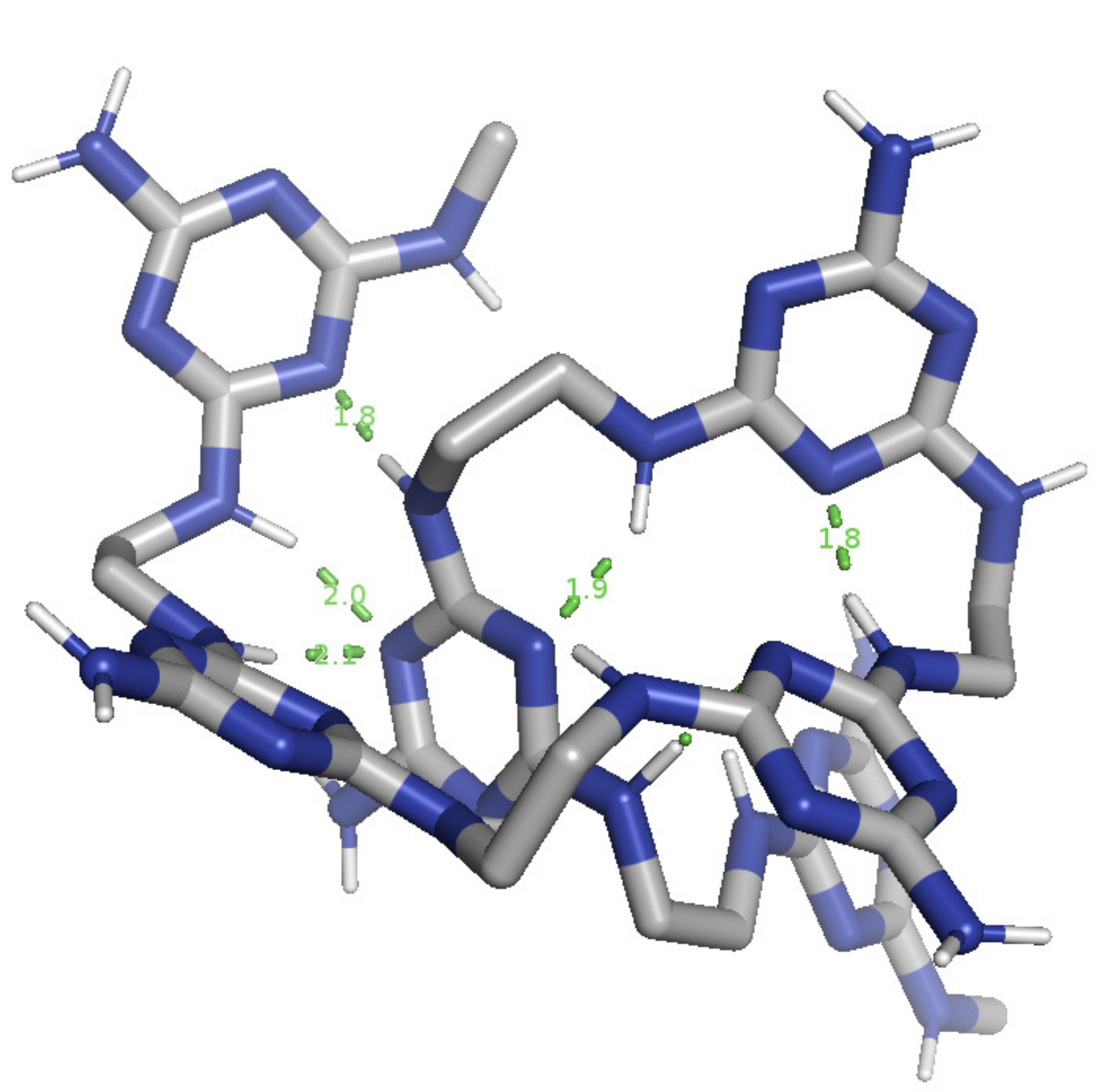} \\
(a) Free energy landscape. & (b) Most stable conformation. \\[6pt]
\multicolumn{2}{c}{\includegraphics[width=60mm]{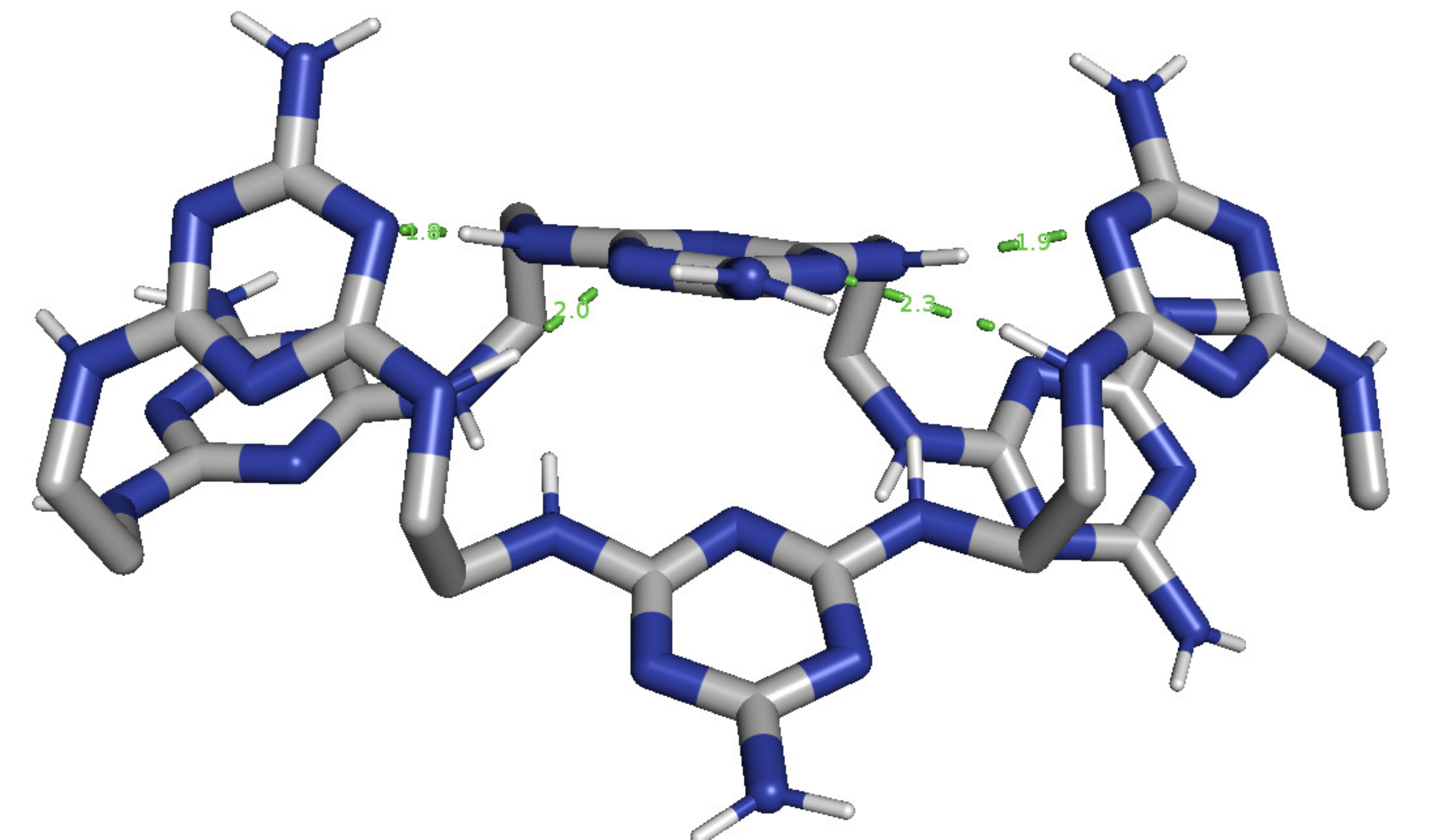}}\\
\multicolumn{2}{c}{(c) Nanorod structure.}
\end{tabular}
\caption{\label{fig:amino_hexamer_remd_free_energy_conformations} Same as Fig.~\ref{fig:original_trimer_remd_free_energy_conformations} but for a single triazine hexamer with amino side chains.}
\end{figure}

With REMD, each triazine trimer/hexamer was able isomerize easily from \emph{cis} to \emph{trans} and vice versa. Fig.~\ref{fig:original_trimer_remd_free_energy_conformations} shows the free energy landscape and the most stable conformation obtained for the trimer with S-ethyl side chains (0 hydrogen bonds and 1 \emph{trans} bond). Fig.~\ref{fig:original_hexamer_remd_free_energy_conformations}, Fig.~\ref{fig:tns_hexamer_remd_free_energy_conformations}, Fig.~\ref{fig:hn_hexamer_remd_free_energy_conformations}, and Fig.~\ref{fig:amino_hexamer_remd_free_energy_conformations} show the free energy landscape and the most stable conformation obtained for the hexamer with S-ethyl side chains (7 hydrogen bonds and 5 \emph{trans} bonds), the hexamer with amino-ethyl-sulfide side chains (4 hydrogen bonds and 3 \emph{trans} bonds), the hexamer with amino-ethyl side chains (5 hydrogen bonds and 6 \emph{trans} bonds), and the hexamer with amino side chains (6 hydrogen bonds and 7 \emph{trans} bonds), respectively.

As previously mentioned, the original triazine polymers in Ref.~\onlinecite{grate2016} had S-ethyl side chains that cannot participate in hydrogen bonding and amino backbone that can participate in hydrogen bonding. A single hexamer with the same side chains and backbone was shown that it can fold onto itself and form a nanorod structure stabilized by hydrogen bonds and $\pi$-$\pi$ interactions from REMD simulations, which comprised 19\% of the total ensemble when k-means was used as a clustering method and root-mean-square distance (RMSD) to the starting conformation was used as a distance metric\cite{grate2016}. The nanorod structure represented the second of the four ordered clusters. Fig.~\ref{fig:original_hexamer_remd_free_energy_conformations} shows that the nanorod structure indeed appears when the hexamer folds onto itself and has 8 hydrogen bonds and 2-3 \emph{trans} bonds. 

The other hexamers have hydrogen bonding side chains, in contrast to the original triazine hexamer with S-ethyl side chains. The amino side chains have stronger hydrogen bonding than amino-ethyl side chains, which have stronger hydrogen bonding than amino-ethyl-sulfide side chains by not having a protecting group attached. All of the hexamers, including the original hexamer, have an amino backbone that can participate in hydrogen bonding as well. For all of the three hexamers with hydrogen bonding side chains, a nanorod structure with the middle triazine rings interacting in a T-shaped or parallel displaced fashion appears, in contrast to the nanorod structure with all triazine rings interacting in a sandwiched fashion. Fig.~\ref{fig:tns_hexamer_remd_free_energy_conformations}, Fig.~\ref{fig:hn_hexamer_remd_free_energy_conformations}, and Fig.~\ref{fig:amino_hexamer_remd_free_energy_conformations} show the nanorod structure obtained for the hexamer with amino-ethyl-sulfide side chains (4 hydrogen bonds and 3 \emph{trans} bonds), the hexamer with amino-ethyl side chains (6 hydrogen bonds and 6 \emph{trans} bonds), and the hexamer with amino side chains (4 hydrogen bonds and 3 \emph{trans} bonds), respectively. Interestingly, for the hexamer with amino-ethyl-sulfide side chains, the most stable conformation is the nanorod, unlike the other hexamers that do not have the nanorod as their most stable conformation.

Overall, REMD is effective at overcoming \emph{cis}-to-\emph{trans} barriers and vice versa and mapping out the entire free energy landscape for the triazine trimers/hexamers. Although rates can be approximately obtained after post-processing the data and constructing a master equation as done in Ref.~\onlinecite{buchete2008}, REMD in general fails to obtain kinetic properties since it alters the real kinetics of the system. In addition, REMD scales poorly for explicit solvent systems like the dimer of triazine trimers. Hence, a state-based method that does not alter the kinetics like the concurrent adaptive sampling (CAS) algorithm needs to be used to efficiently obtain rates and pathways between two states of interest.

\subsection{\label{sec:cas_results} Concurrent Adaptive Sampling Algorithm}
\begin{figure}[p]
\centering
\begin{tabular}{cc}
\includegraphics[width=80mm]{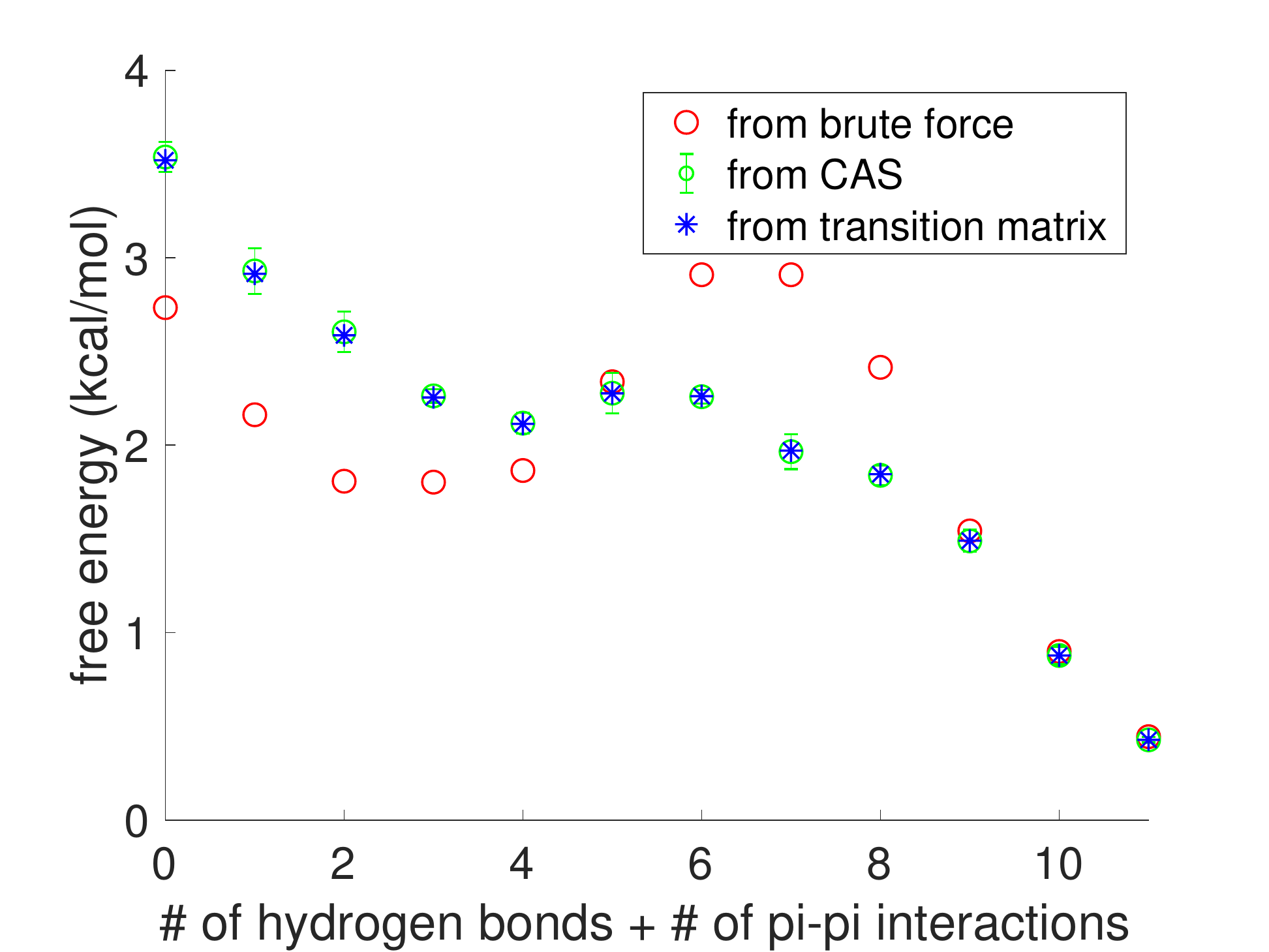} & \includegraphics[width=80mm]{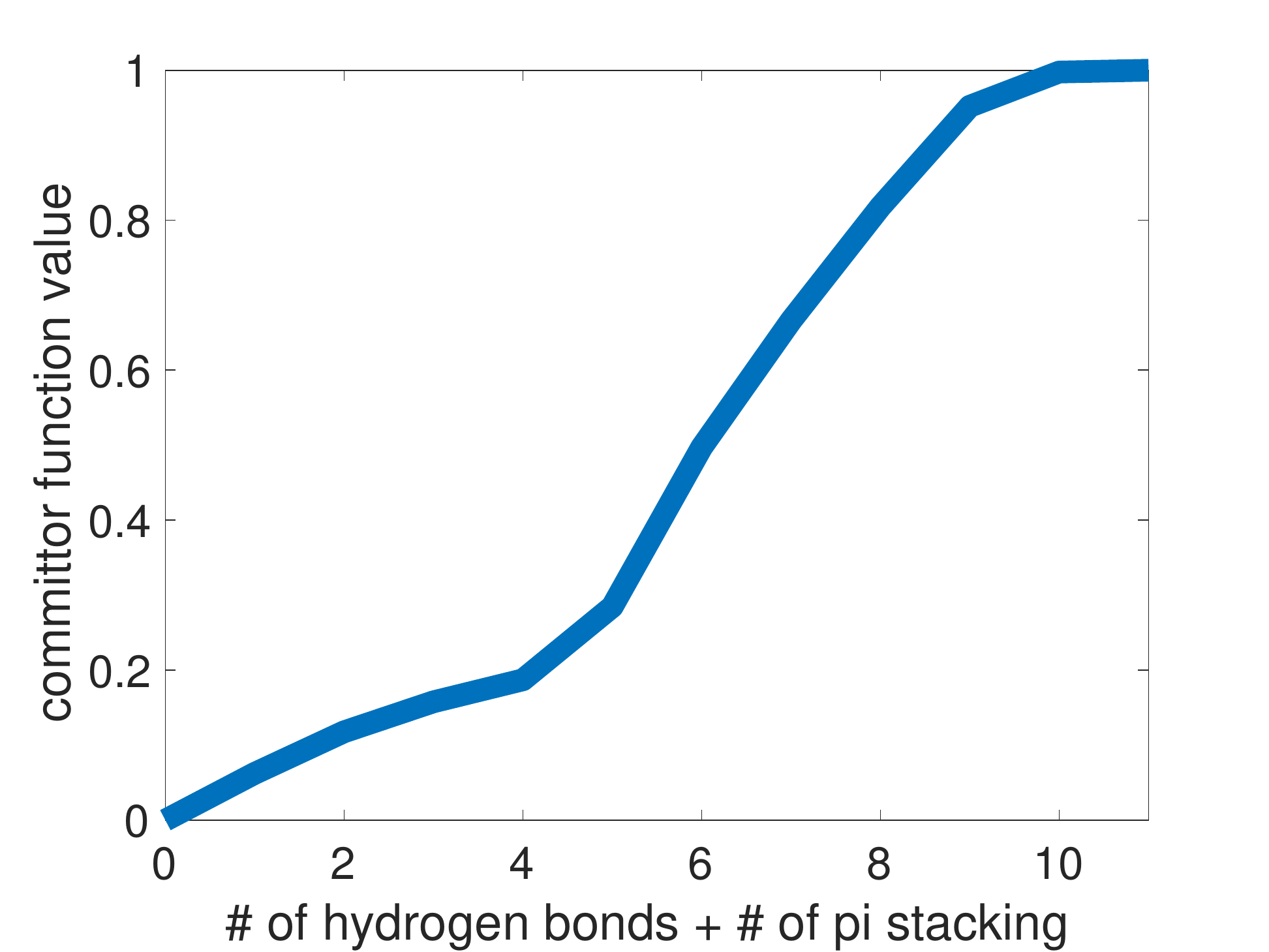} \\
(a) Free energy landscape (\emph{cis}). & (b) Committor function (\emph{cis}). \\[6pt]
\includegraphics[width=80mm]{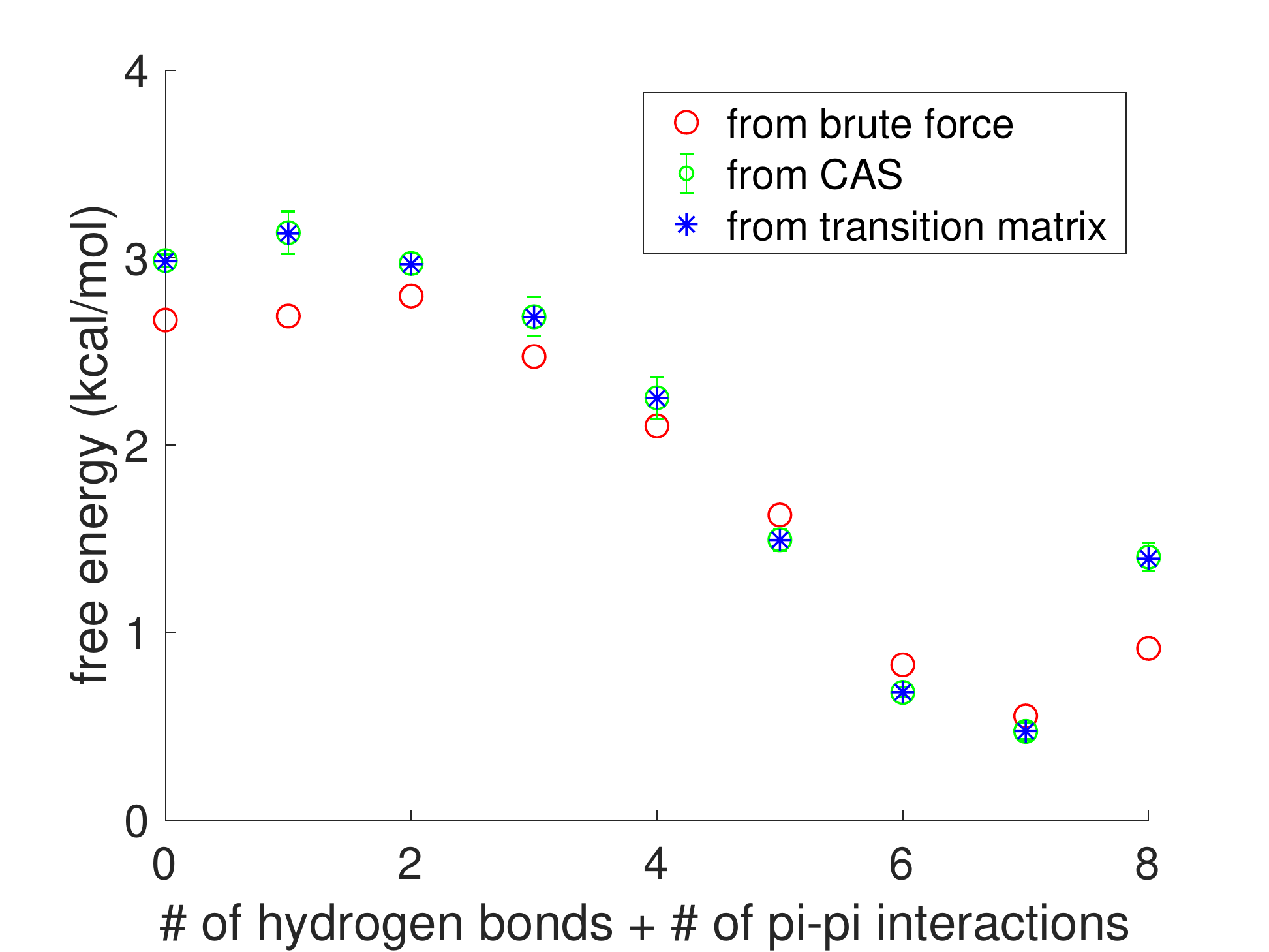} & \includegraphics[width=80mm]{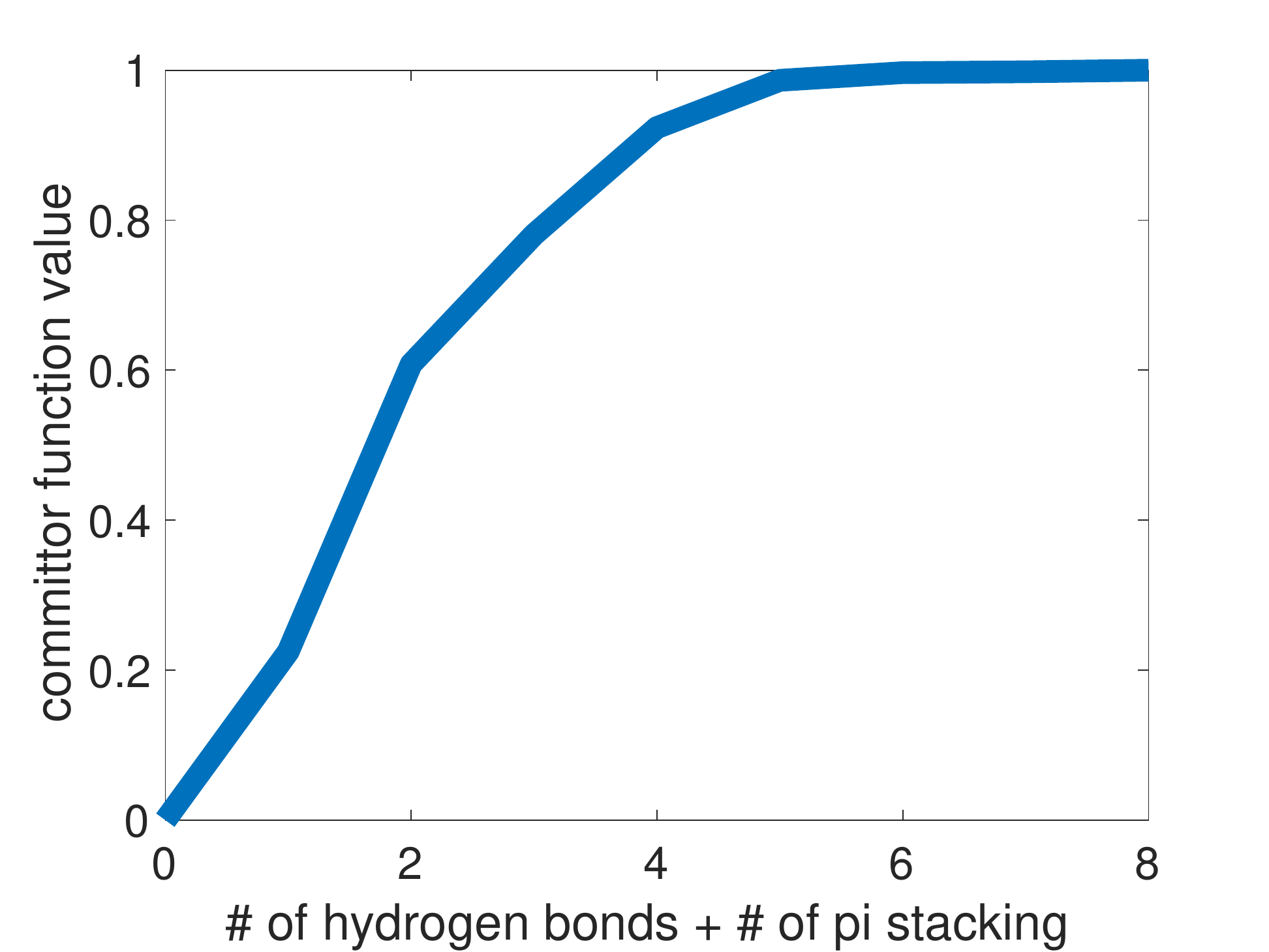} \\
(c) Free energy landscape (\emph{trans}). & (d) Committor function (\emph{trans}). \\[6pt]
\end{tabular}
\caption{\label{fig:original_trimers_free_energy} Free energy landscapes and committor functions of dimer of triazine trimers (all \emph{cis} and all \emph{trans}) with S-ethyl side chains and amino backbone from the CAS algorithm simulations.}
\end{figure}

\begin{figure}[p]
\centering
\begin{tabular}{cc}
\includegraphics[width=80mm]{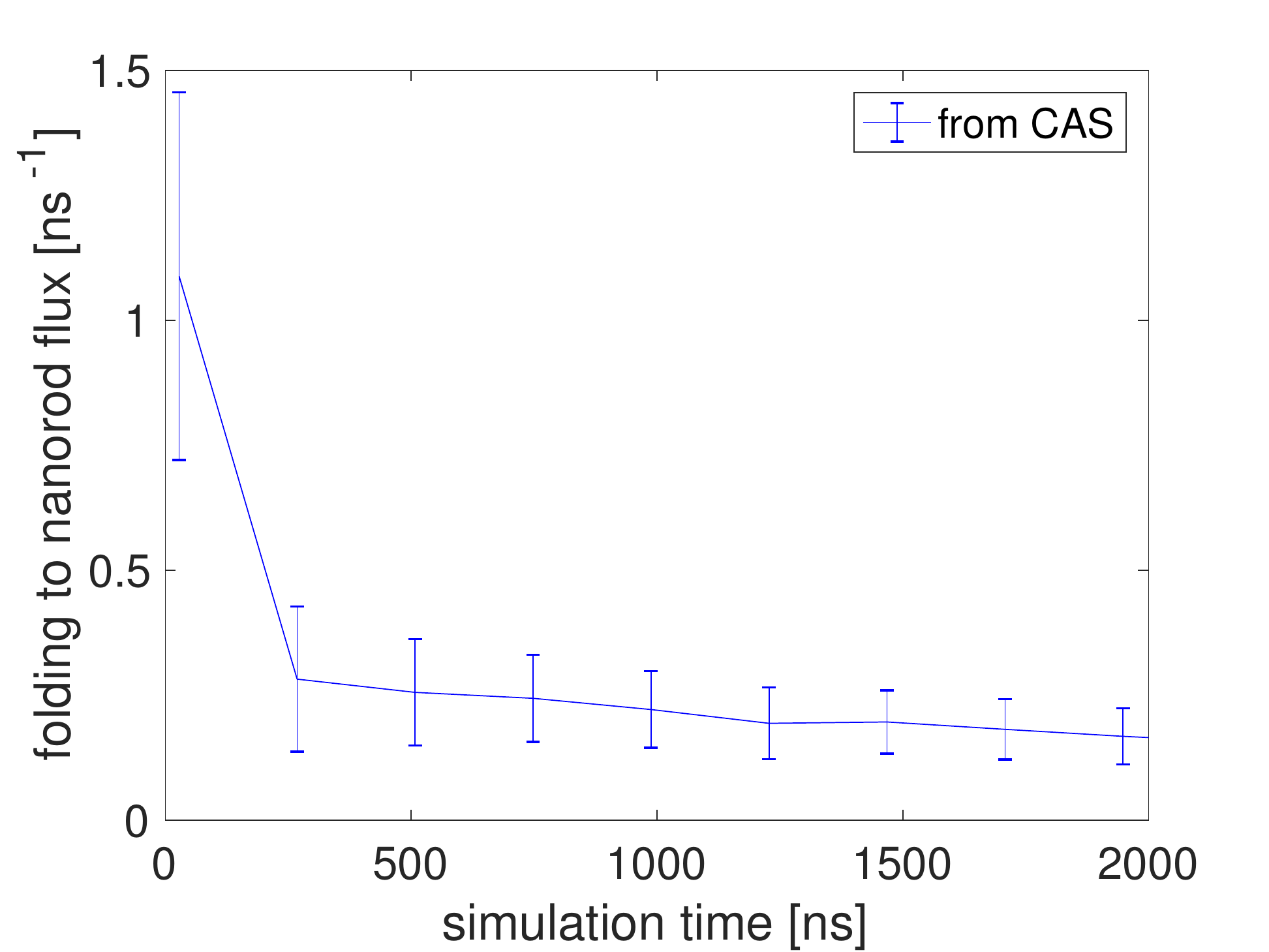} & \includegraphics[width=80mm]{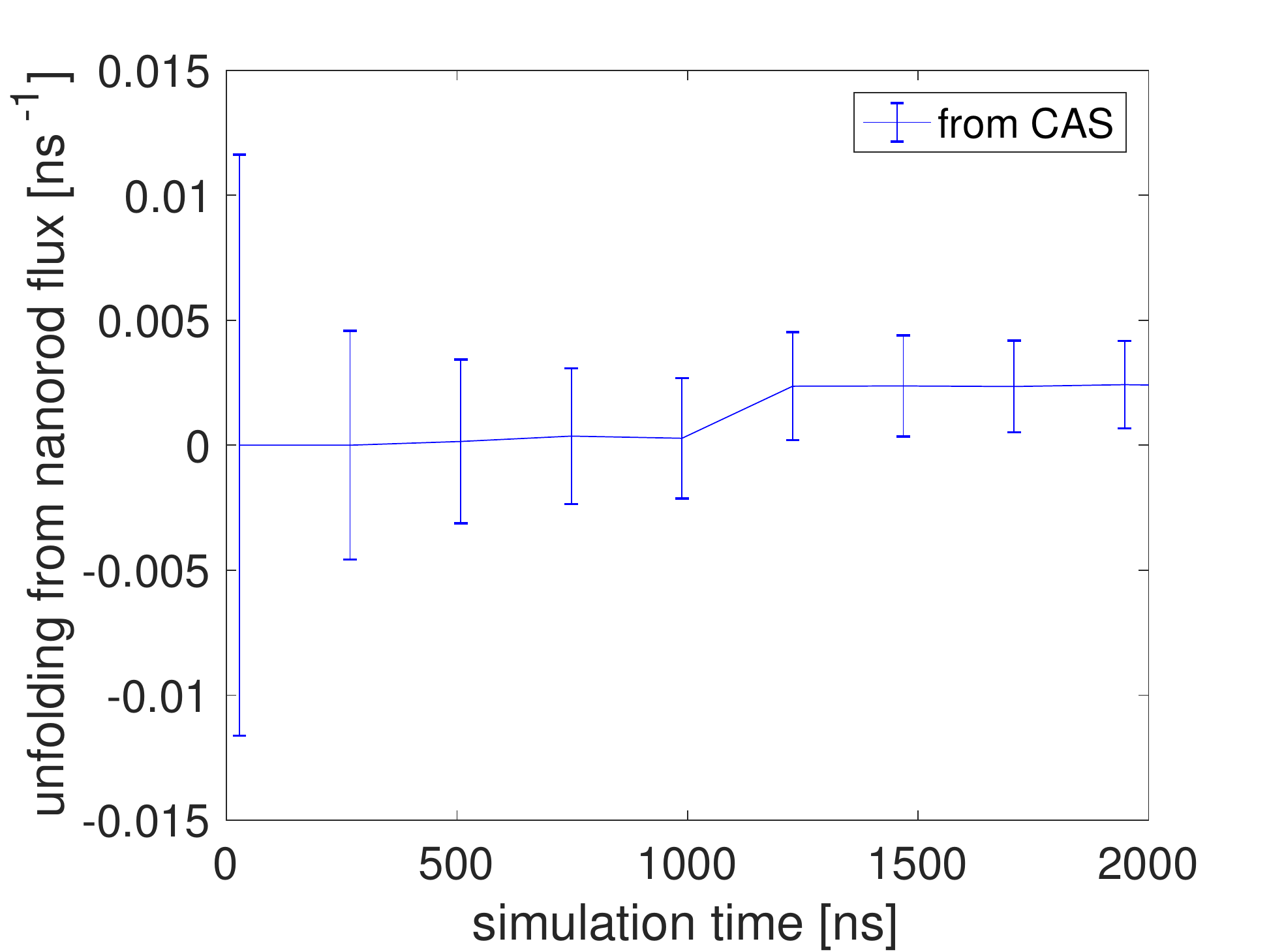} \\
(a) Forward flux (\emph{cis}). & (b) Backward flux (\emph{cis}). \\[6pt]
\includegraphics[width=80mm]{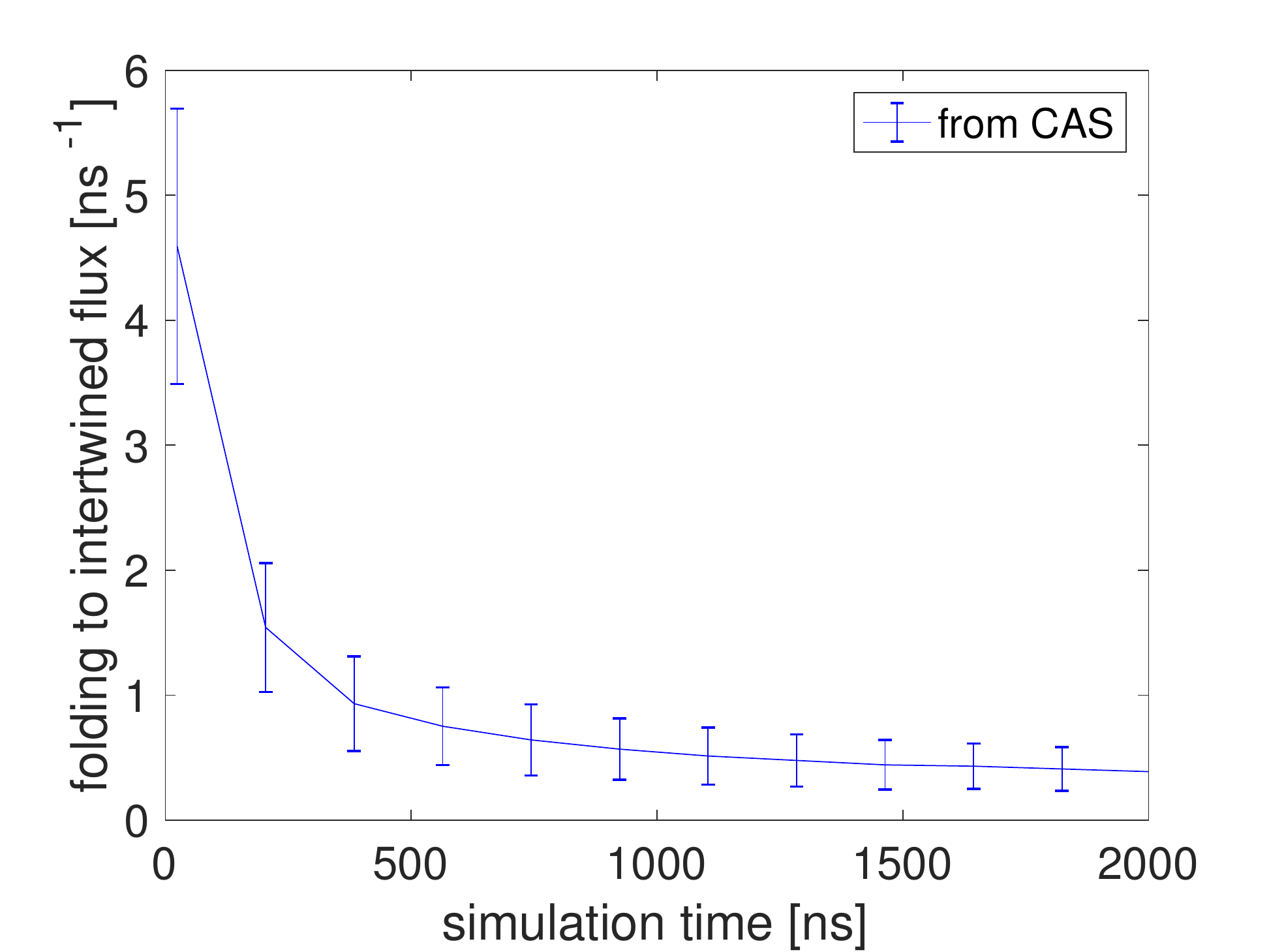} & \includegraphics[width=80mm]{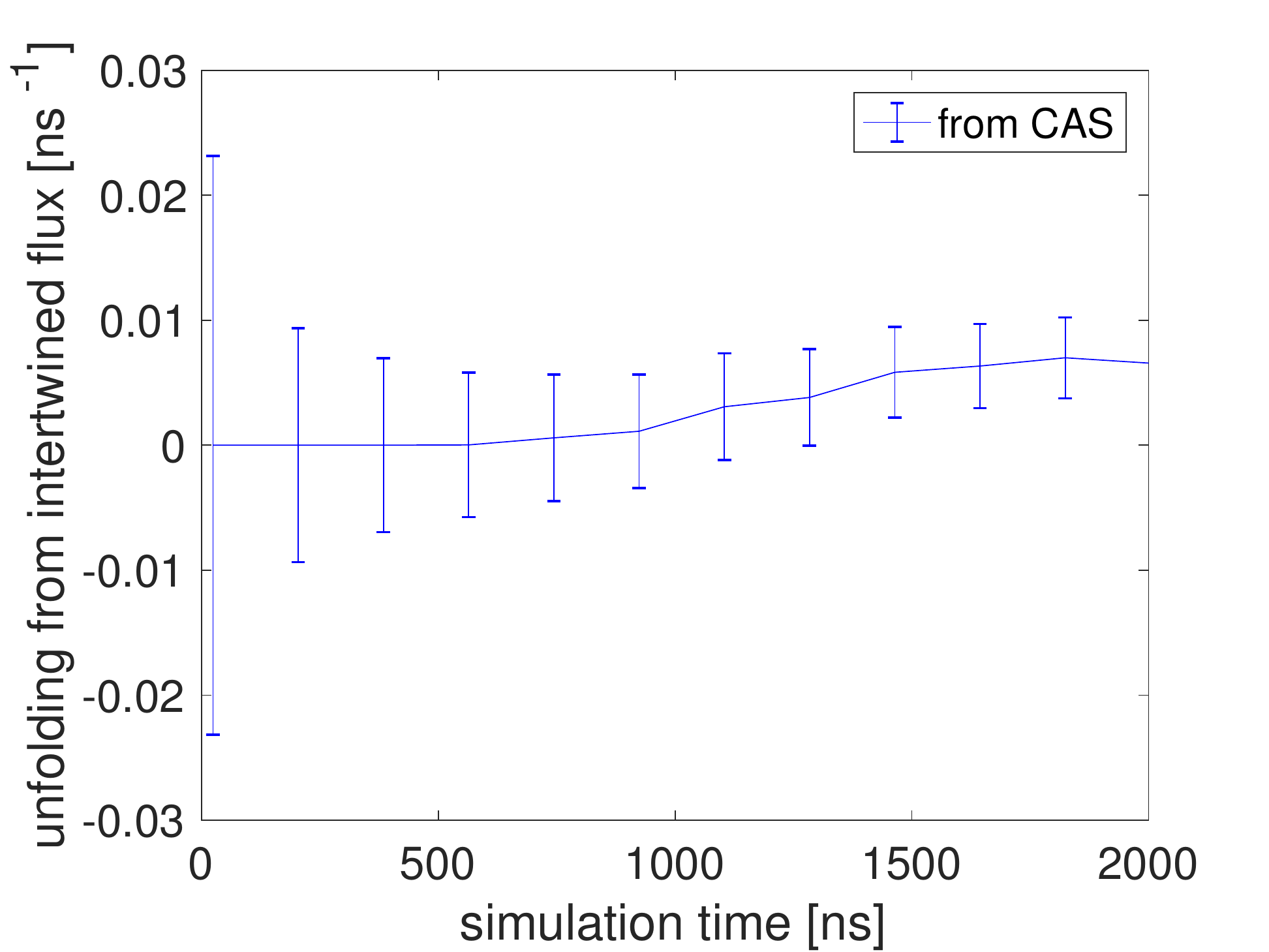} \\
(c) Forward flux (\emph{trans}). & (d) Backward flux (\emph{trans}). \\[6pt]
\end{tabular}
\caption{\label{fig:original_trimers_fluxes} Forward and backward fluxes of dimer of triazine trimers (all \emph{cis} and all \emph{trans}) with S-ethyl side chains and amino backbone from the CAS algorithm simulations.}
\end{figure}

\begin{figure}[p]
\centering
\begin{tabular}{cc}
\includegraphics[width=80mm]{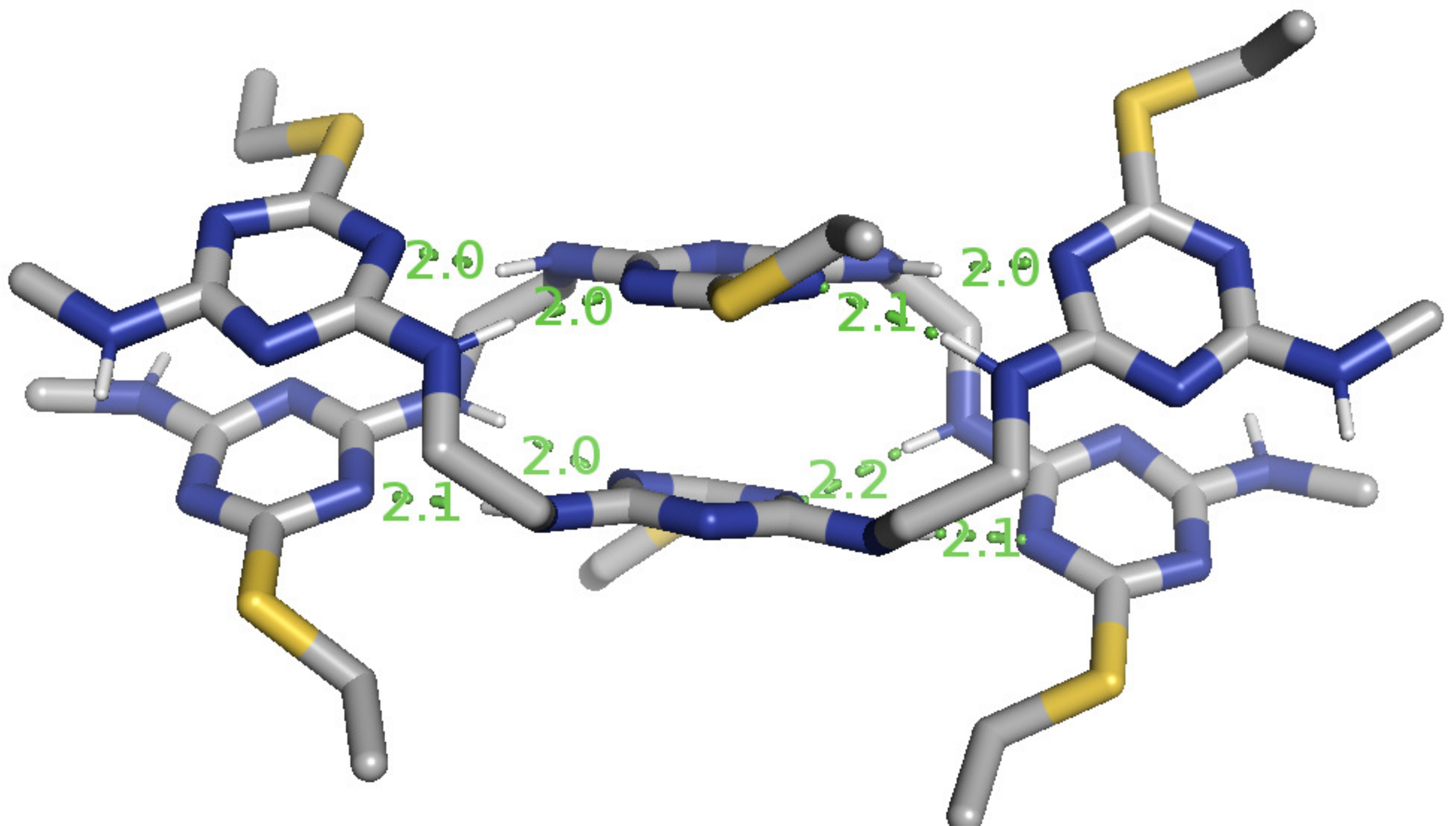} & \includegraphics[width=60mm]{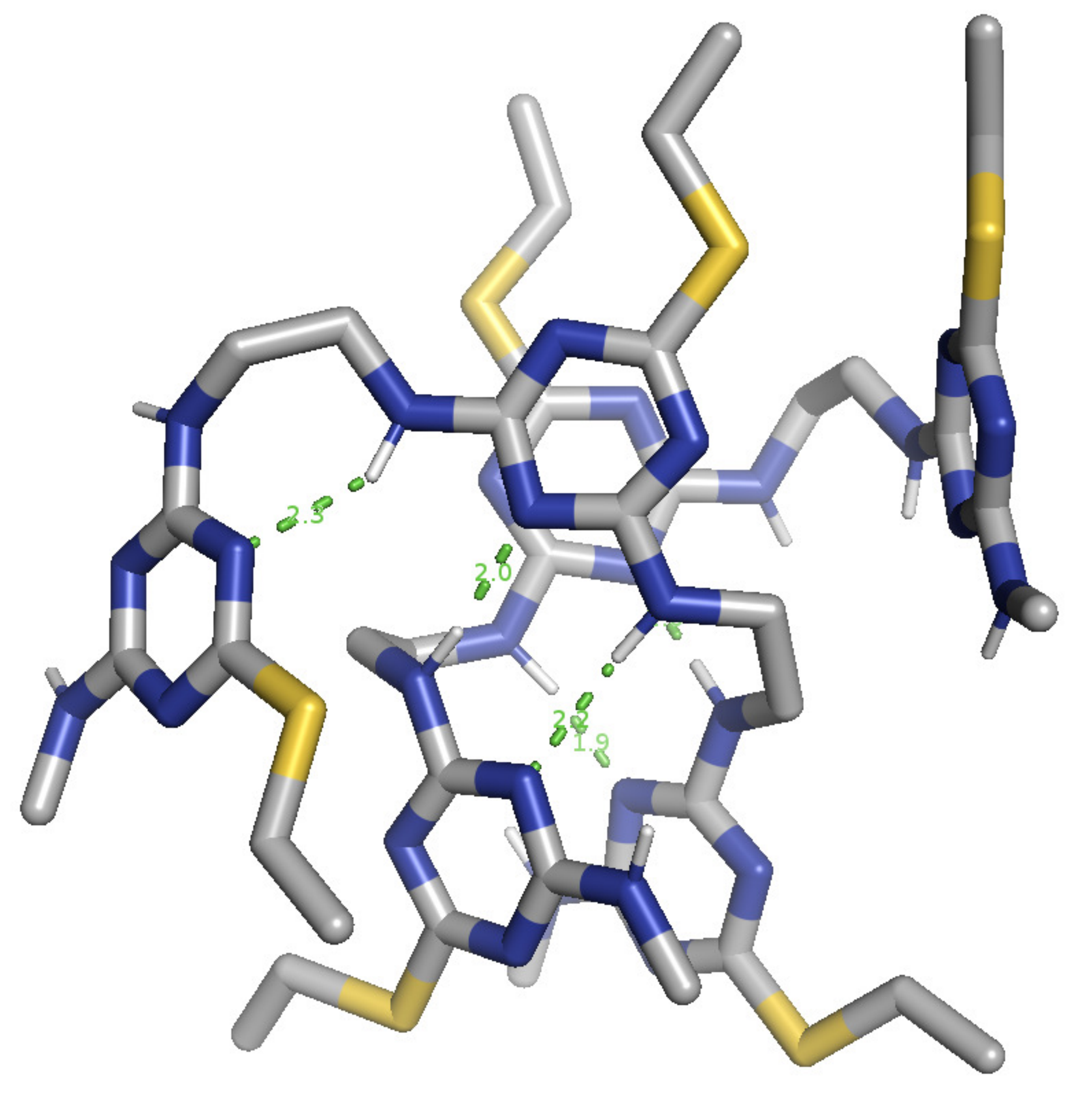} \\
(a) Nanorod structure (\emph{cis}). & (b) Intertwined structure (\emph{trans}). \\[6pt]
\includegraphics[width=80mm]{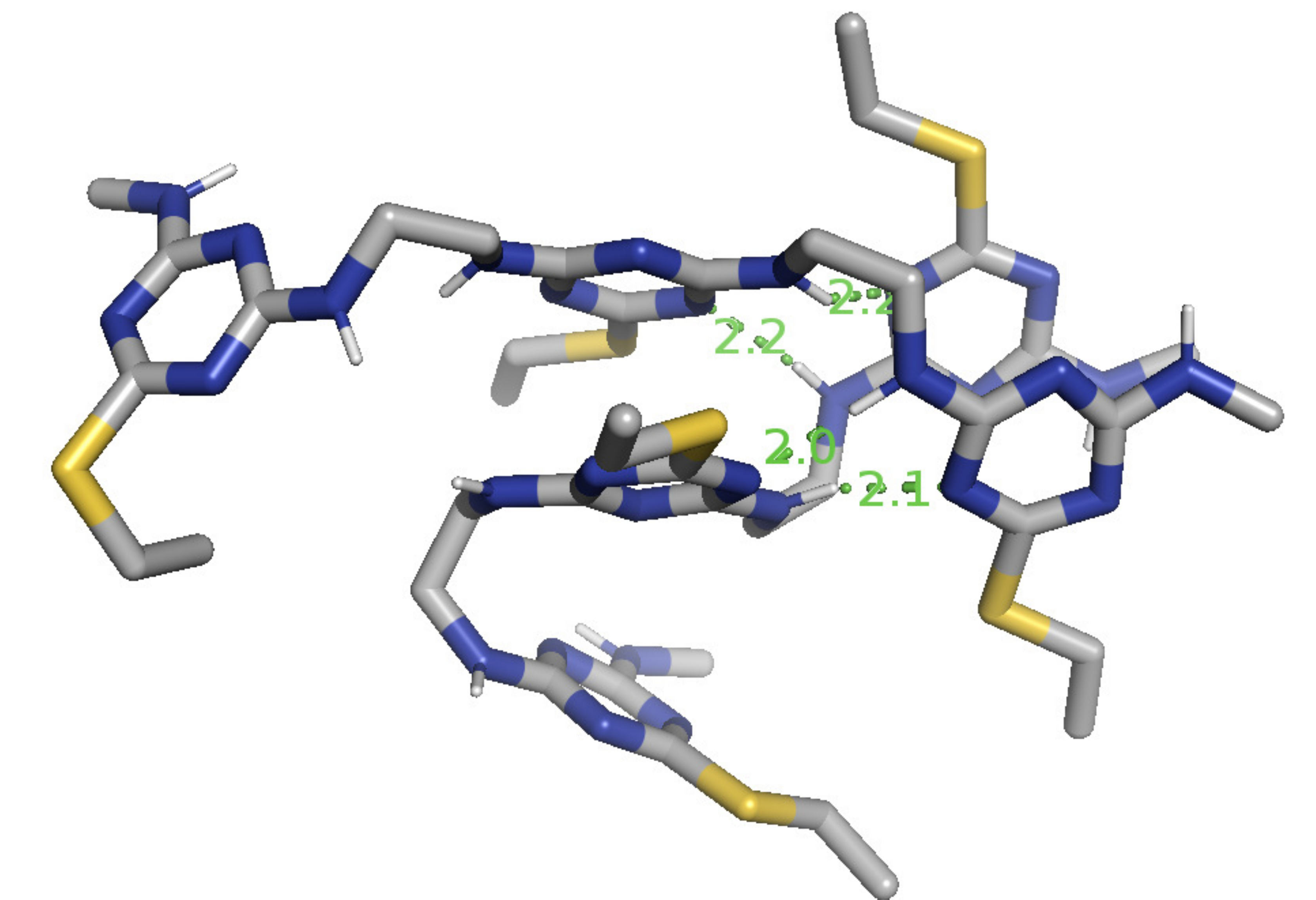} & \includegraphics[width=60mm]{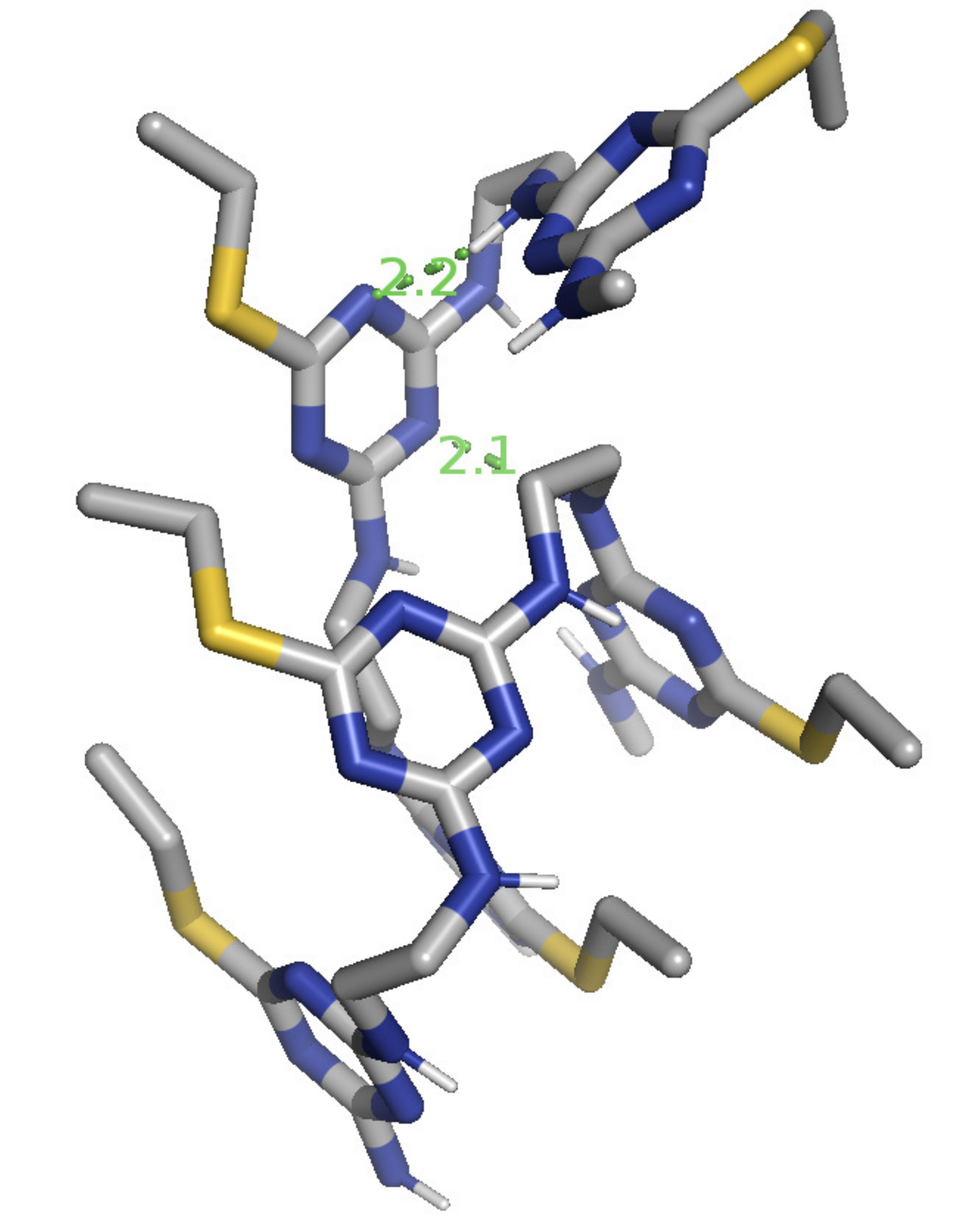} \\
(c) Intermediate conformation (\emph{cis}). & (d) Intermediate conformation (\emph{trans}). \\[6pt]
\end{tabular}
\caption{\label{fig:original_trimers_conformations} Most stable and intermediate conformations for a dimer of triazine trimers (all \emph{cis} and all \emph{trans}) with S-ethyl side chains and amino backbone. Figure (a) shows the nanorod structure that has 11 non-covalent interactions in total (8 hydrogen bonds and 3 $\pi$-$\pi$ interactions). Figure (b) shows the intertwined structure has 7 non-covalent interactions in total (5 hydrogen bonds and 2 $\pi$-$\pi$ interactions). Figures (c) and (d) show the intermediate conformations for all \emph{cis} (6 non-covalent interactions) and all \emph{trans} (2 non-covalent interactions), respectively.}
\end{figure}

\begin{figure}[p]
\centering
\begin{tabular}{cc}
\includegraphics[width=80mm]{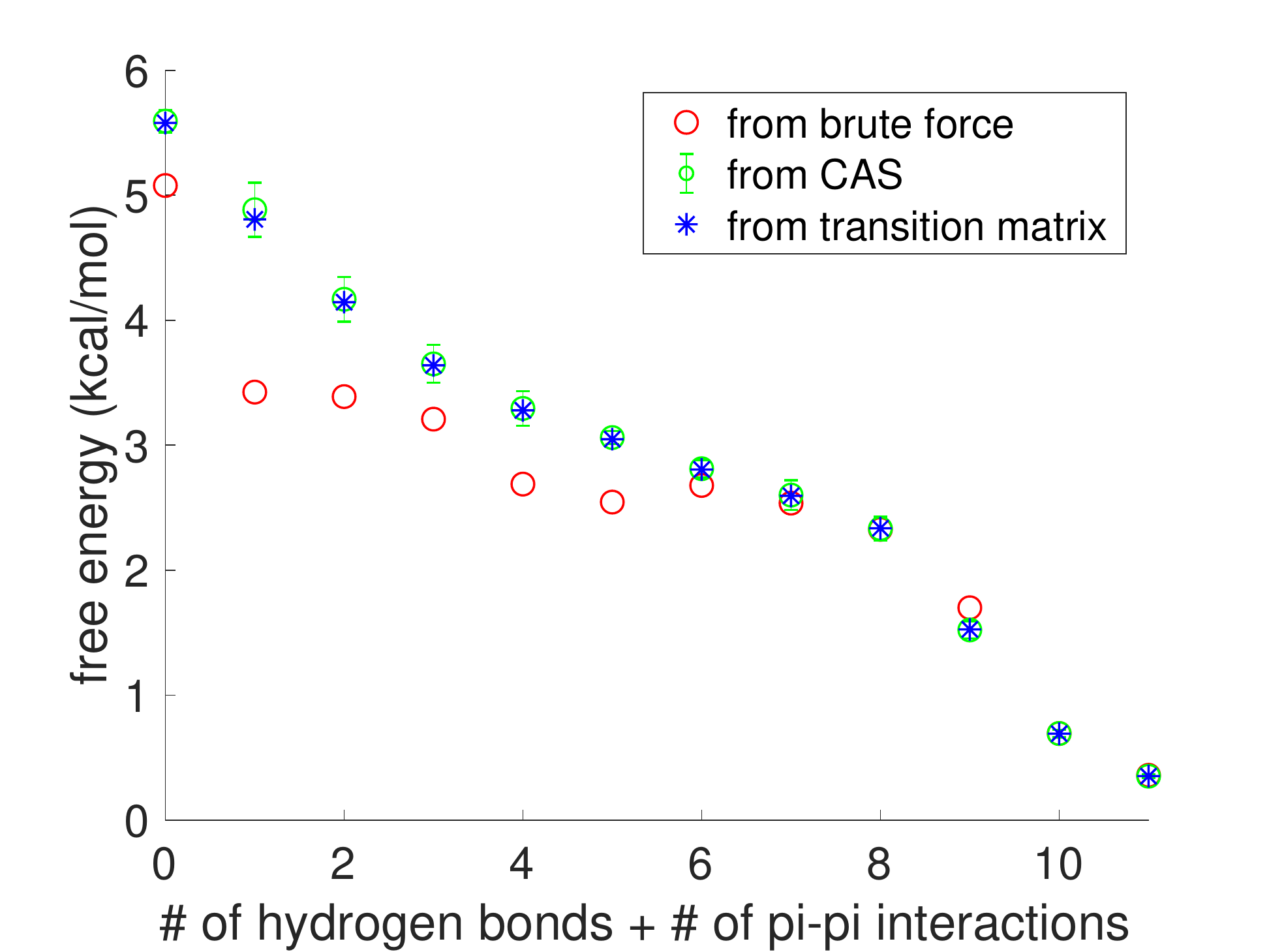} & \includegraphics[width=80mm]{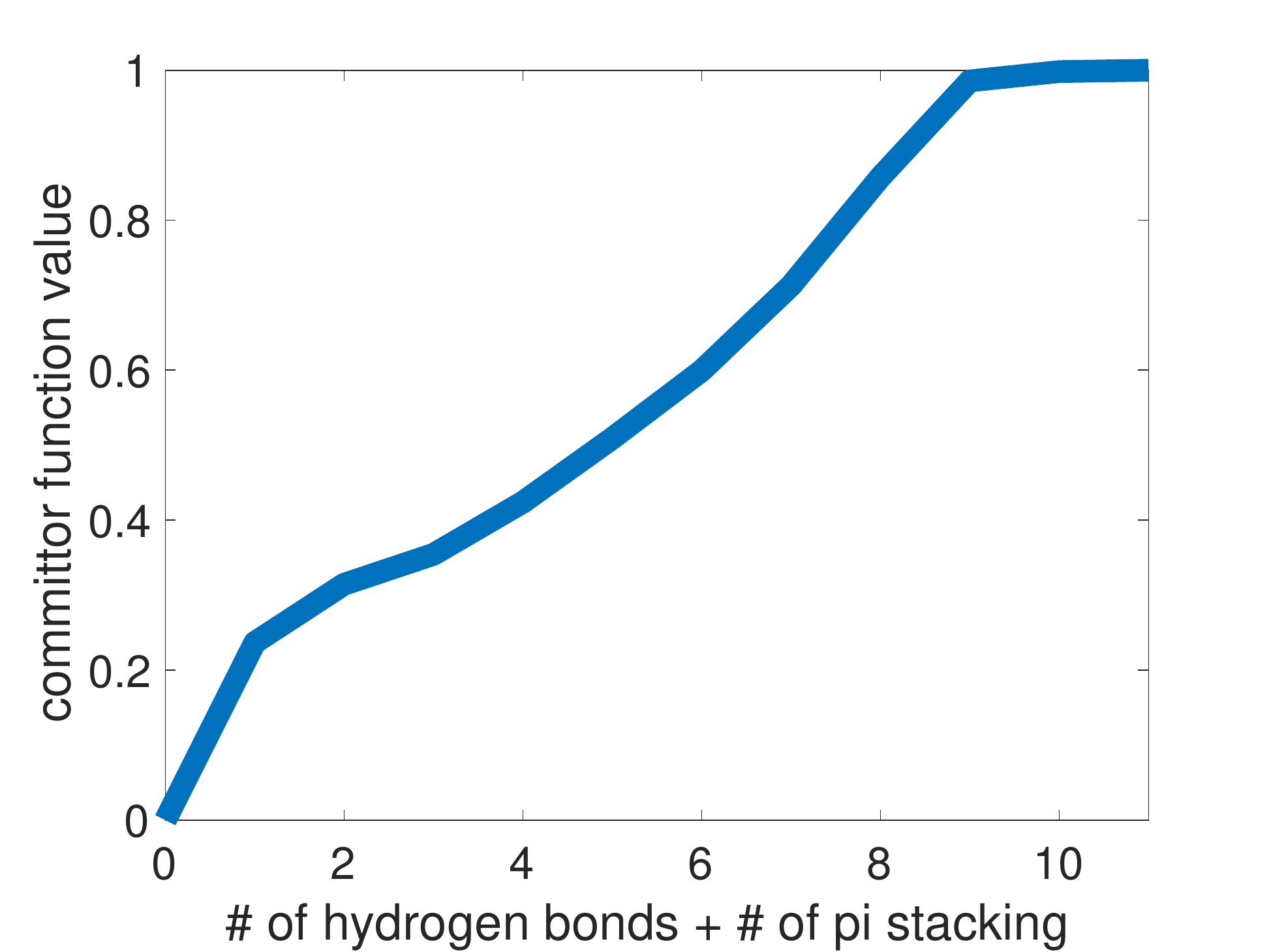} \\
(a) Free energy landscape (\emph{cis}). & (b) Committor function (\emph{cis}). \\[6pt]
\includegraphics[width=80mm]{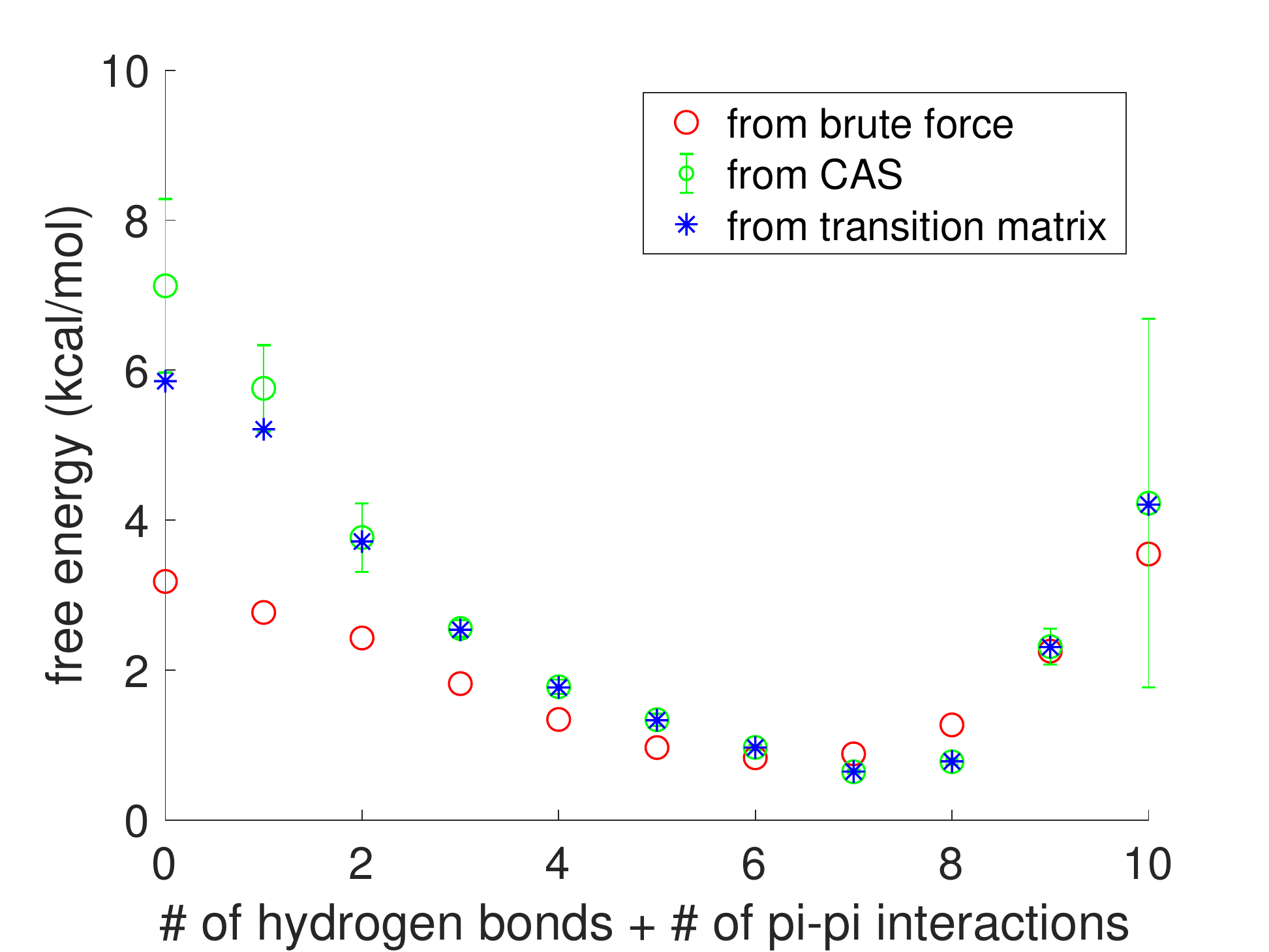} & \includegraphics[width=80mm]{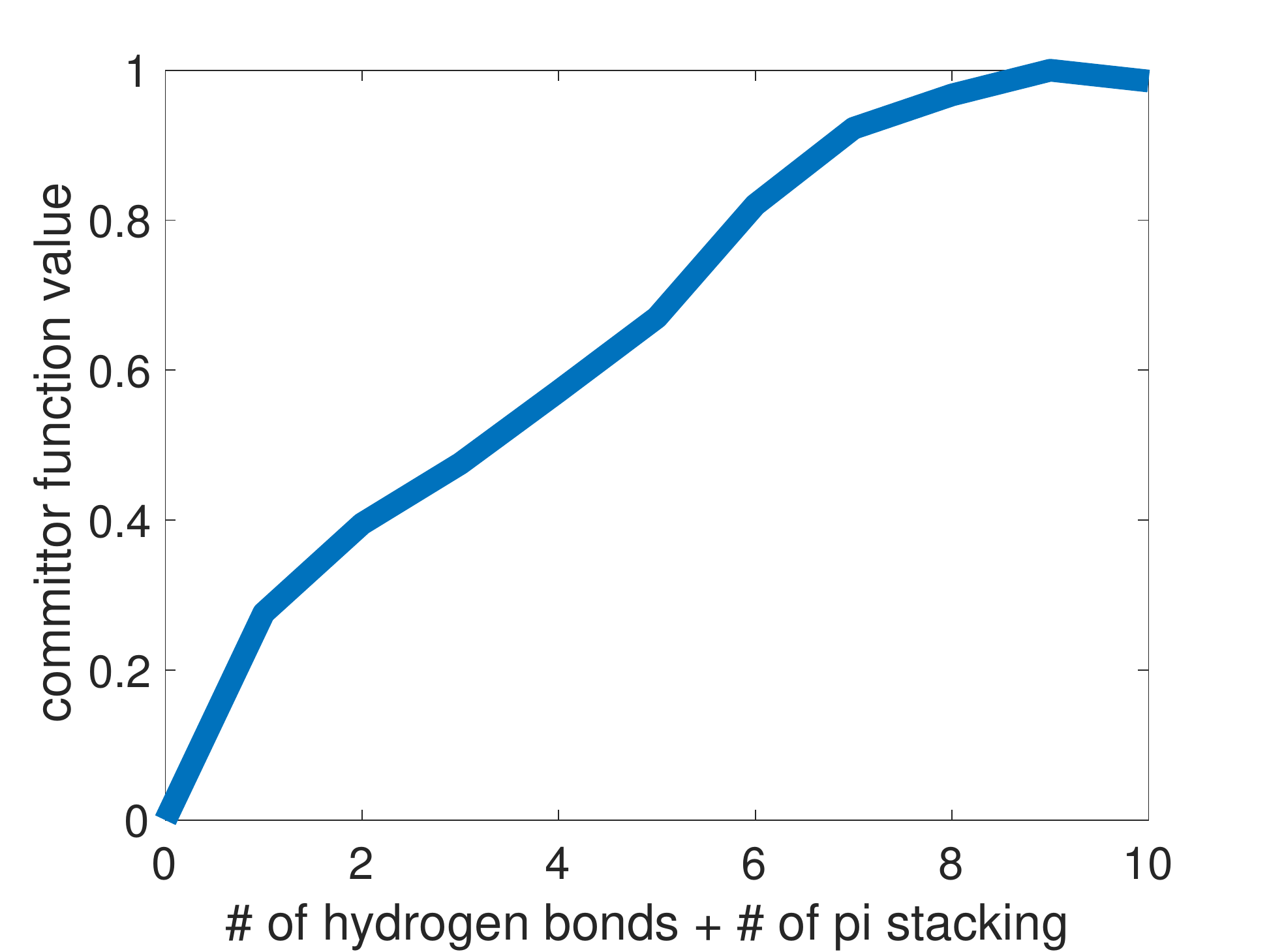} \\
(c) Free energy landscape (\emph{trans}). & (d) Committor function (\emph{trans}). \\[6pt]
\end{tabular}
\caption{\label{fig:hn_trimers_free_energy} Same as Fig.~\ref{fig:original_trimers_free_energy} but for a dimer of triazine trimers (all \emph{cis} and all \emph{trans}) with amino-ethyl side chains and amino backbone.}
\end{figure}

\begin{figure}[p]
\centering
\begin{tabular}{cc}
\includegraphics[width=80mm]{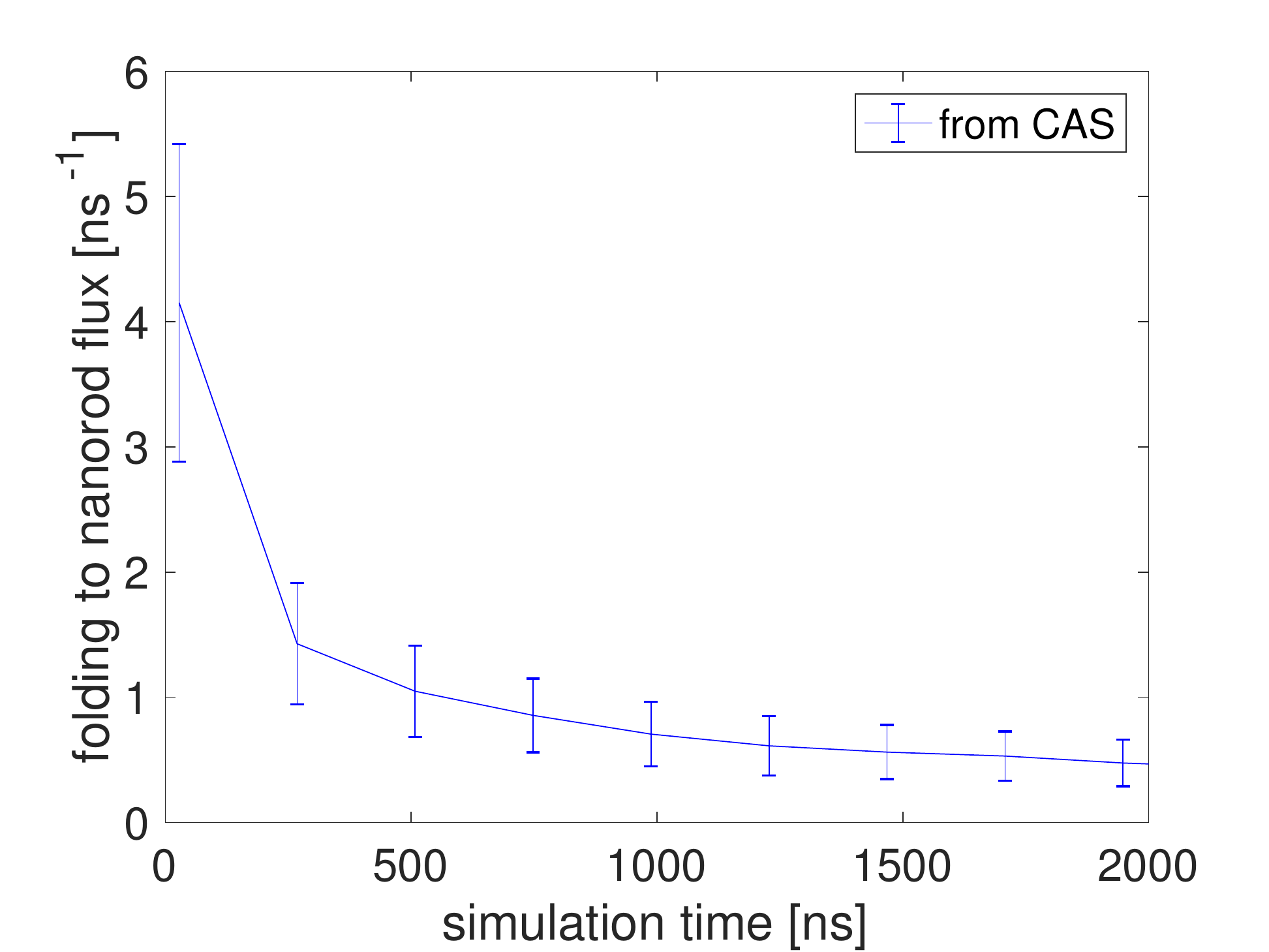} & \includegraphics[width=80mm]{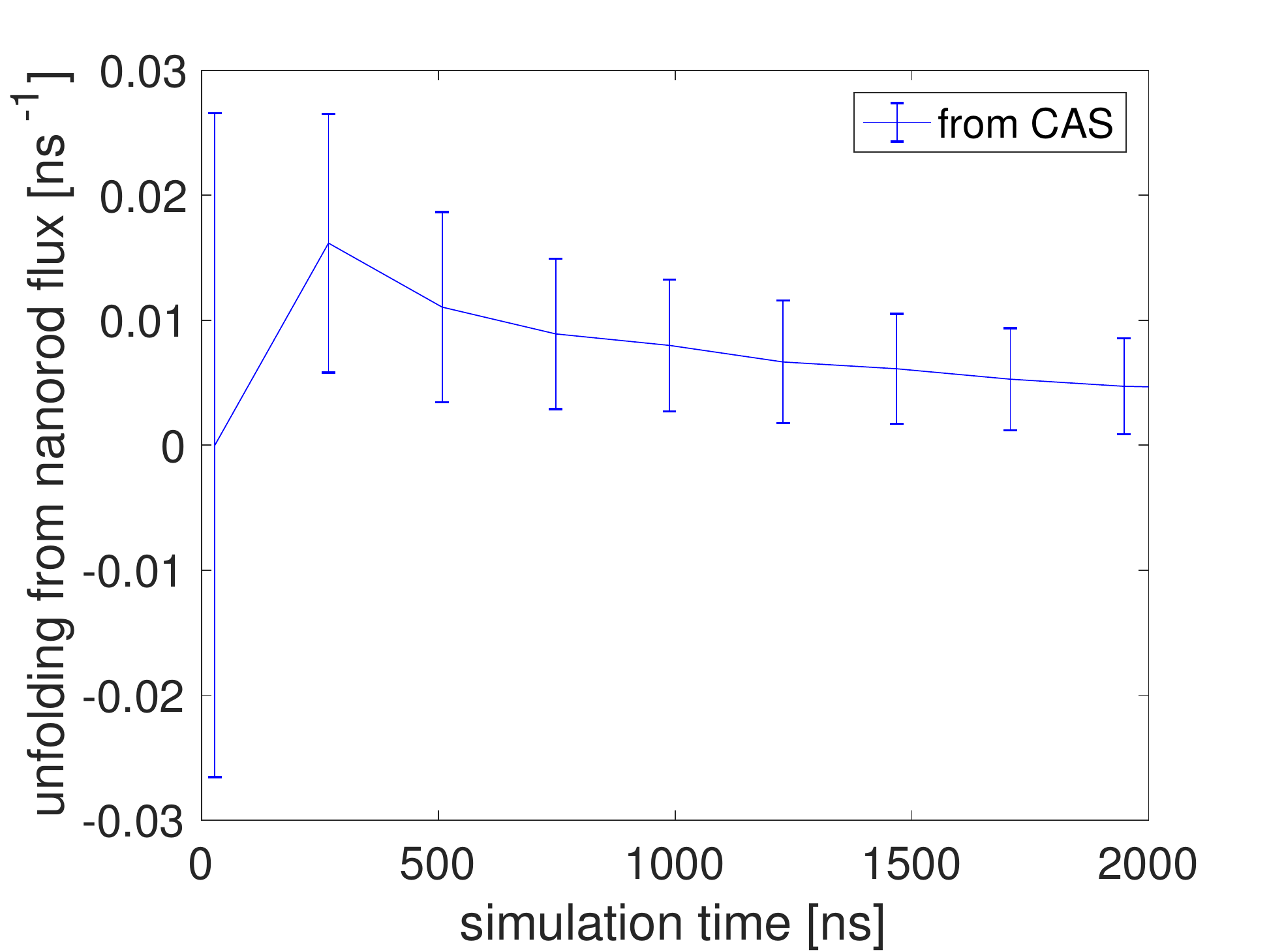} \\
(a) Forward flux (\emph{cis}). & (b) Backward flux (\emph{cis}). \\[6pt]
\includegraphics[width=80mm]{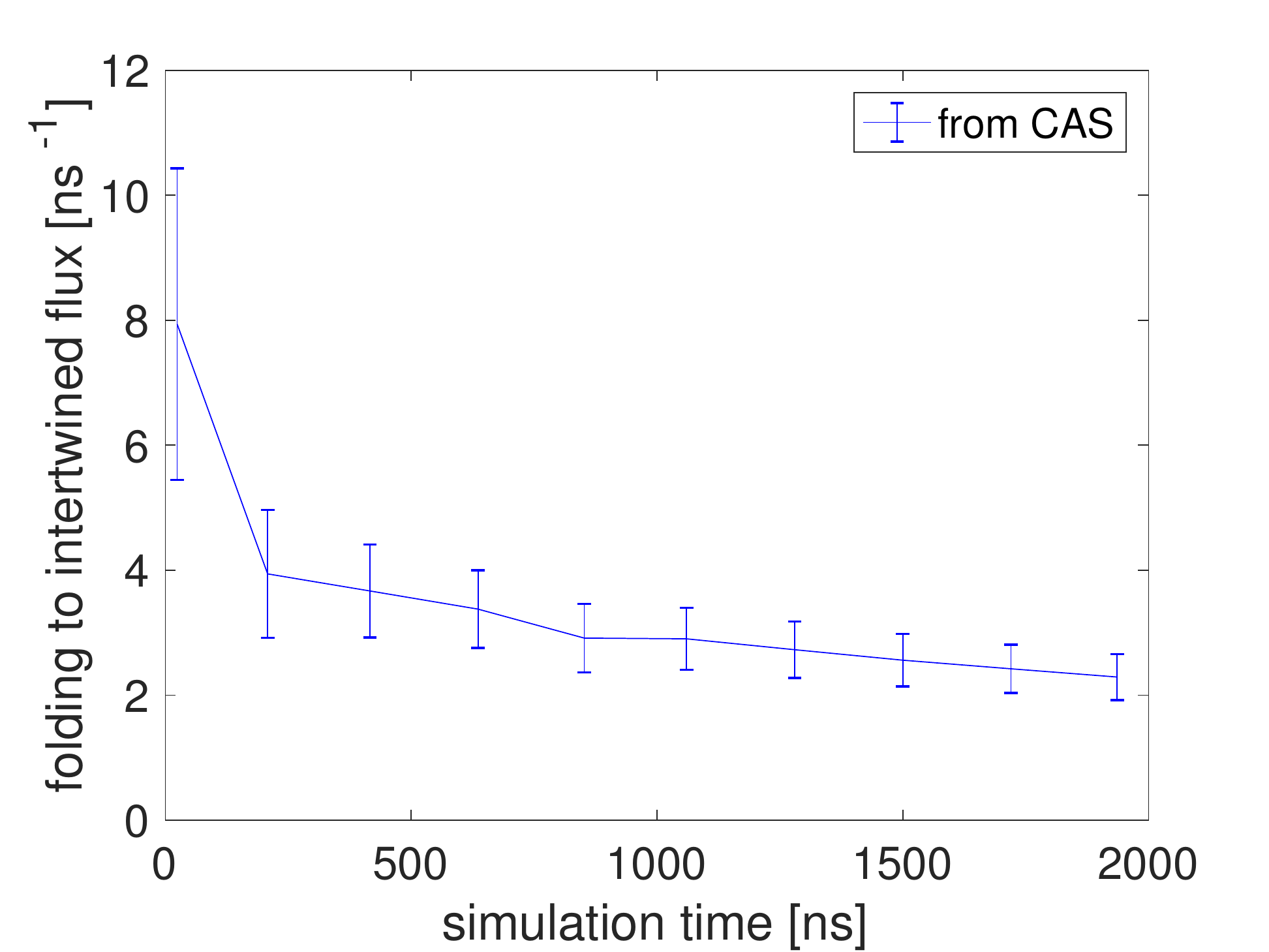} & \includegraphics[width=80mm]{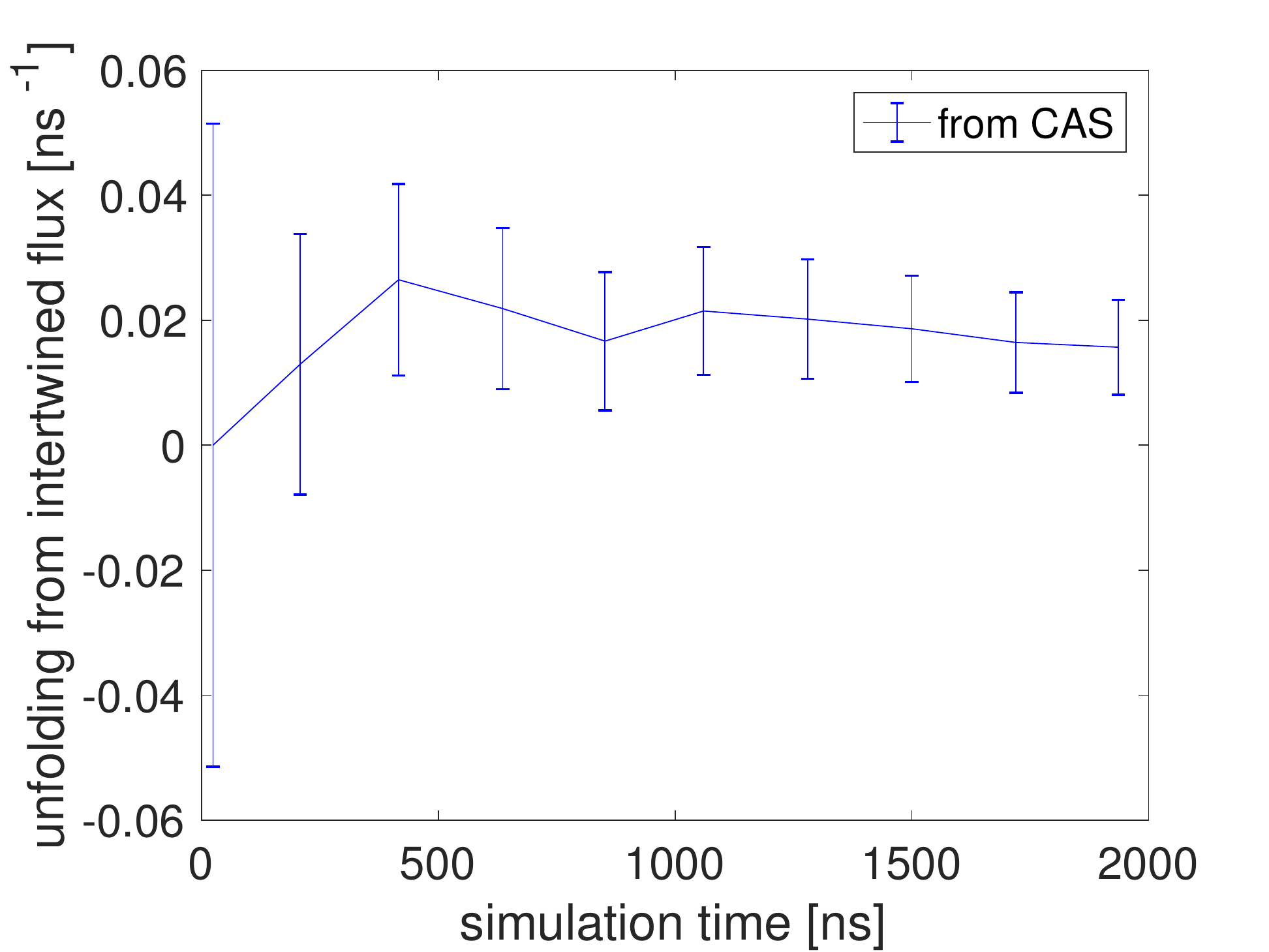} \\
(c) Forward flux (\emph{trans}). & (d) Backward flux (\emph{trans}). \\[6pt]
\end{tabular}
\caption{\label{fig:hn_trimers_fluxes} Same as Fig.~\ref{fig:original_trimers_fluxes} but for a dimer of triazine trimers (all \emph{cis} and all \emph{trans}) with amino-ethyl side chains and amino backbone.}
\end{figure}

\begin{figure}[p]
\centering
\begin{tabular}{cc}
\includegraphics[width=80mm]{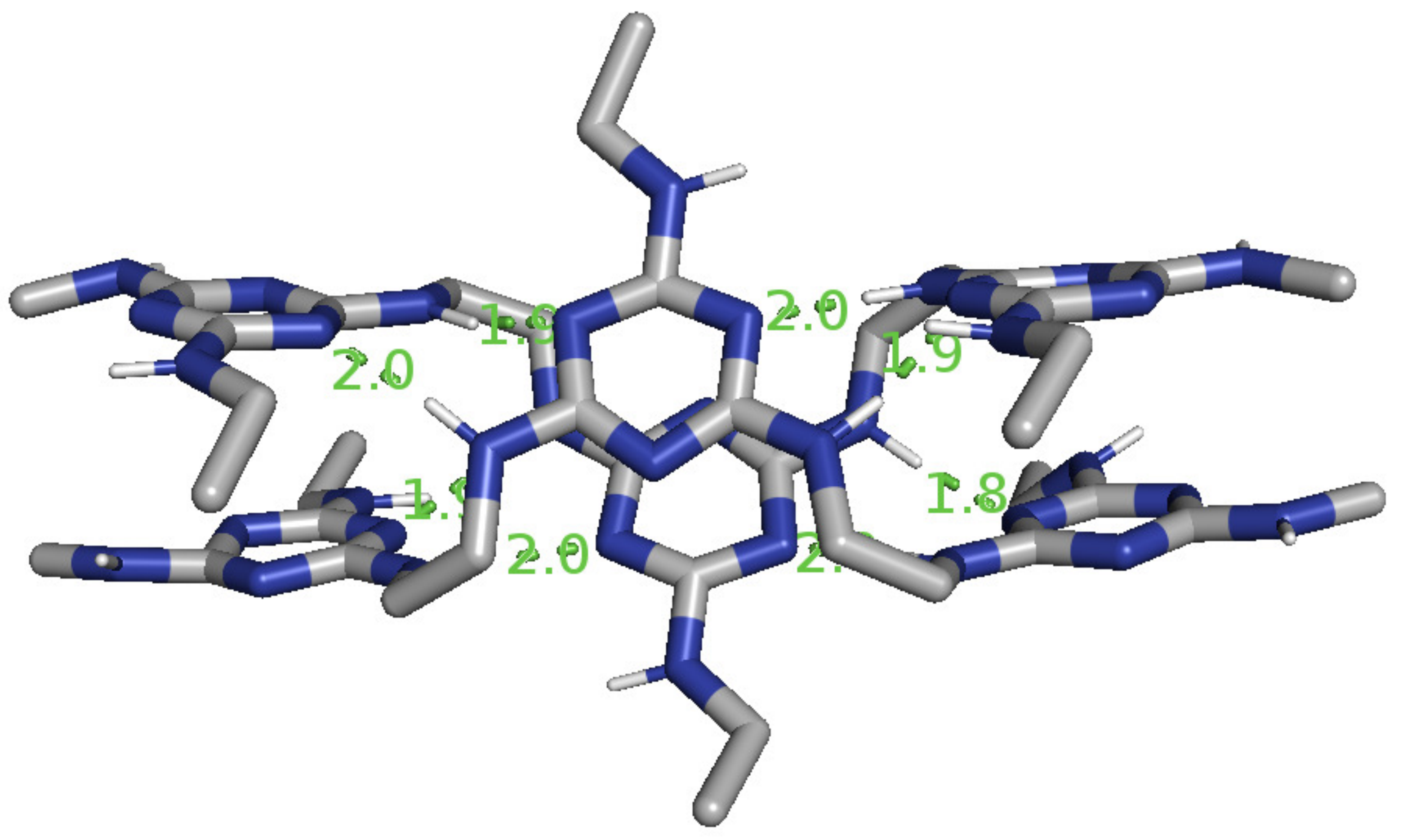} & \includegraphics[width=60mm]{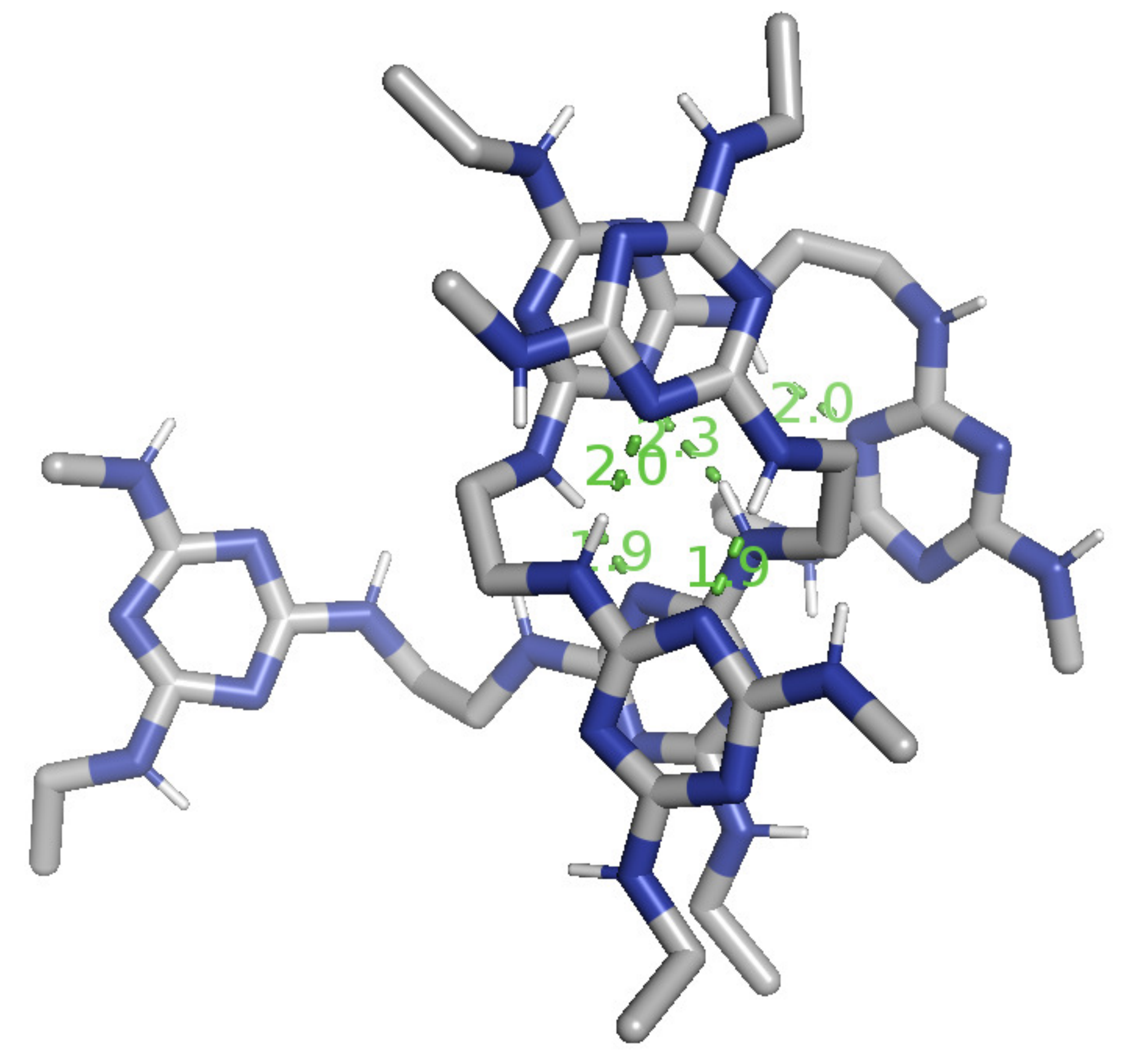} \\
(a) Nanorod structure (\emph{cis}). & (b) Intertwined structure (\emph{trans}). \\[6pt]
\includegraphics[width=80mm]{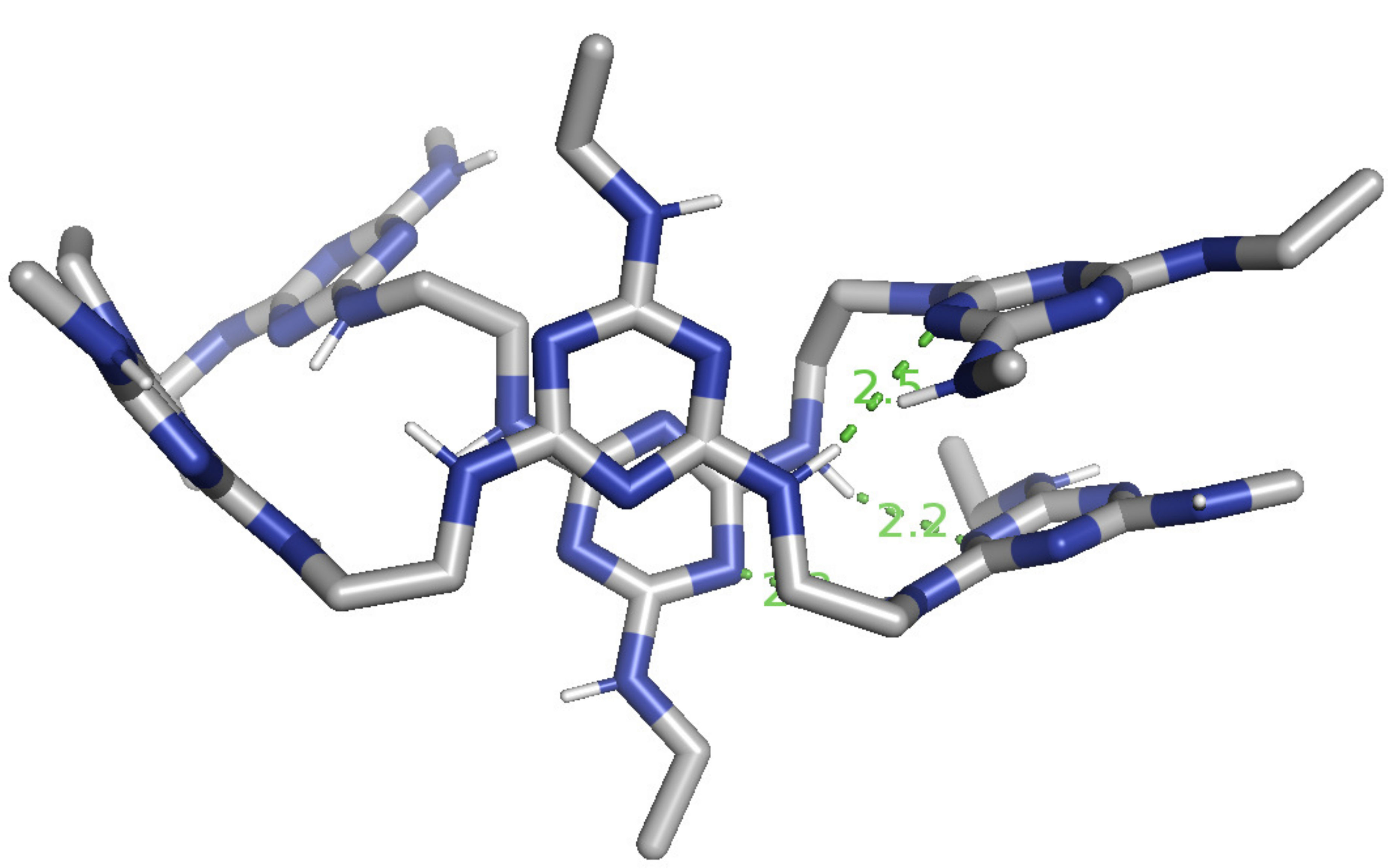} & \includegraphics[width=60mm]{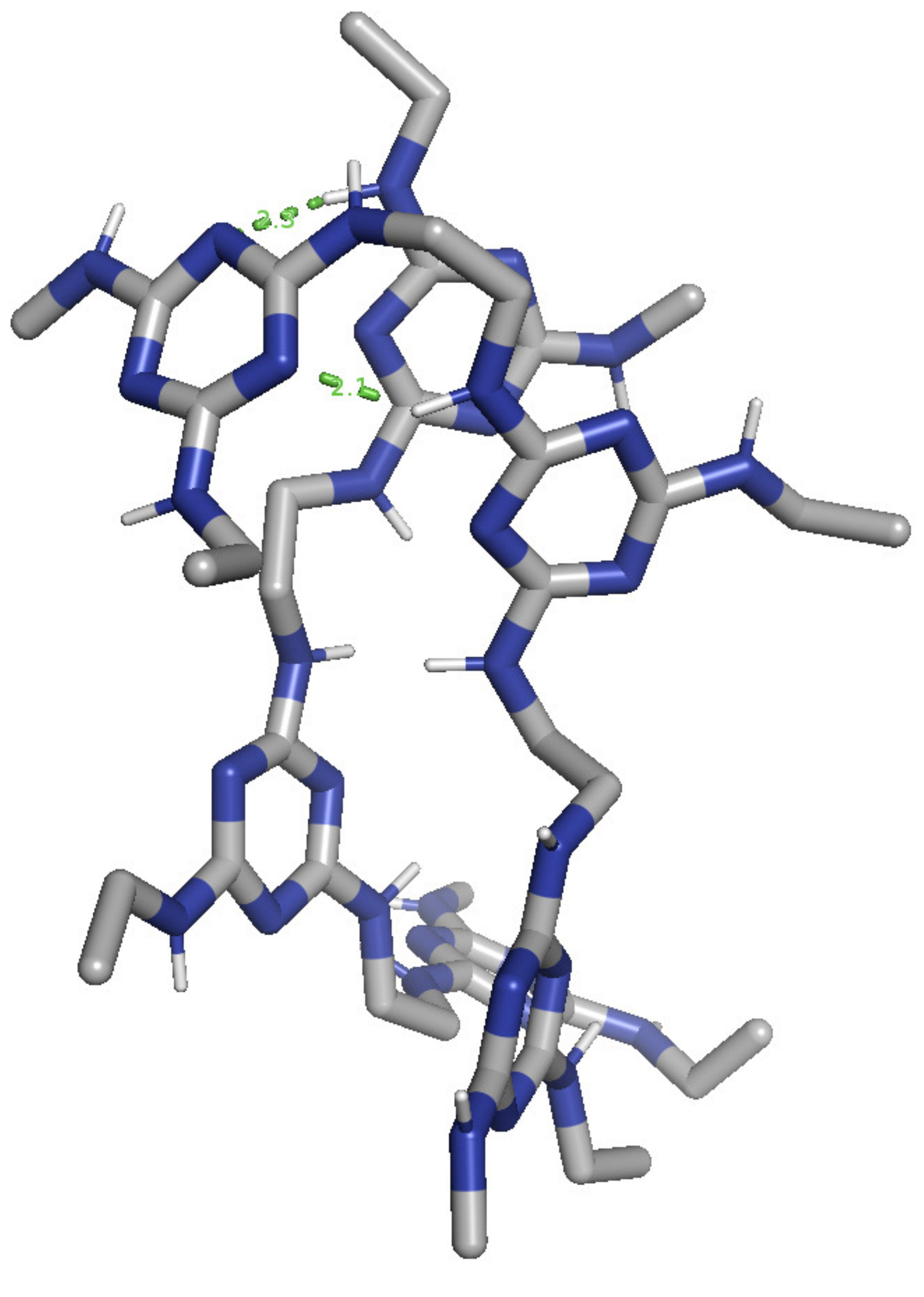} \\
(c) Intermediate conformation (\emph{cis}). & (d) Intermediate conformation (\emph{trans}). \\[6pt]
\end{tabular}
\caption{\label{fig:hn_trimers_conformations} Same as Fig.~\ref{fig:original_trimers_conformations} but for a dimer of triazine trimers (all \emph{cis} and all \emph{trans}) with amino-ethyl side chains and amino backbone. Figure (a) shows the nanorod structure that has 11 non-covalent interactions in total (8 hydrogen bonds and 3 $\pi$-$\pi$ interactions). Figure (b) shows the intertwined structure has 7 non-covalent interactions in total (5 hydrogen bonds and 2 $\pi$-$\pi$ interactions). Figures (c) and (d) show the intermediate conformations for all \emph{cis} (5 non-covalent interactions) and all \emph{trans} (3 non-covalent interactions), respectively.}
\end{figure}

\begin{figure}[p]
\centering
\begin{tabular}{cc}
\includegraphics[width=80mm]{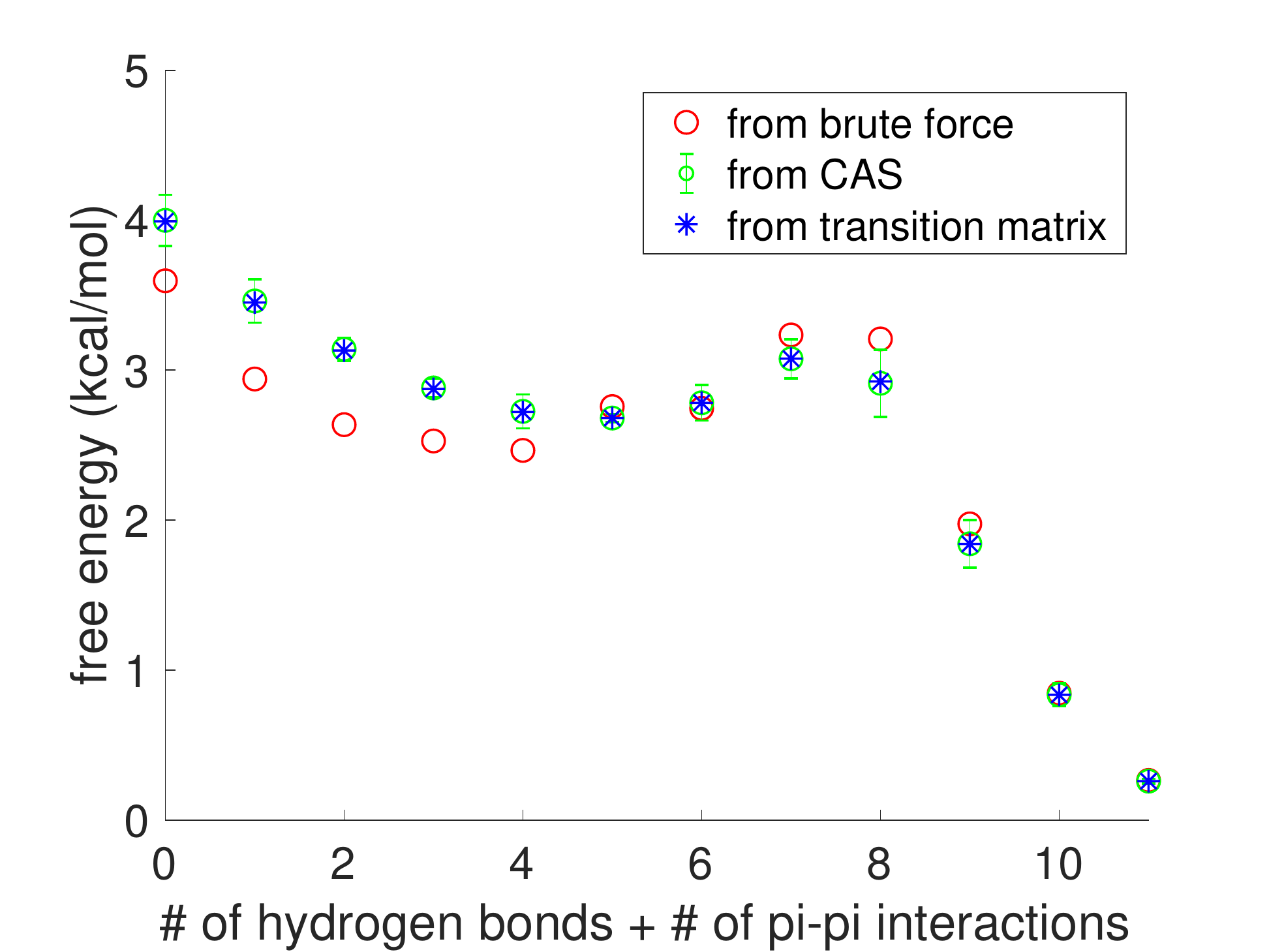} & \includegraphics[width=80mm]{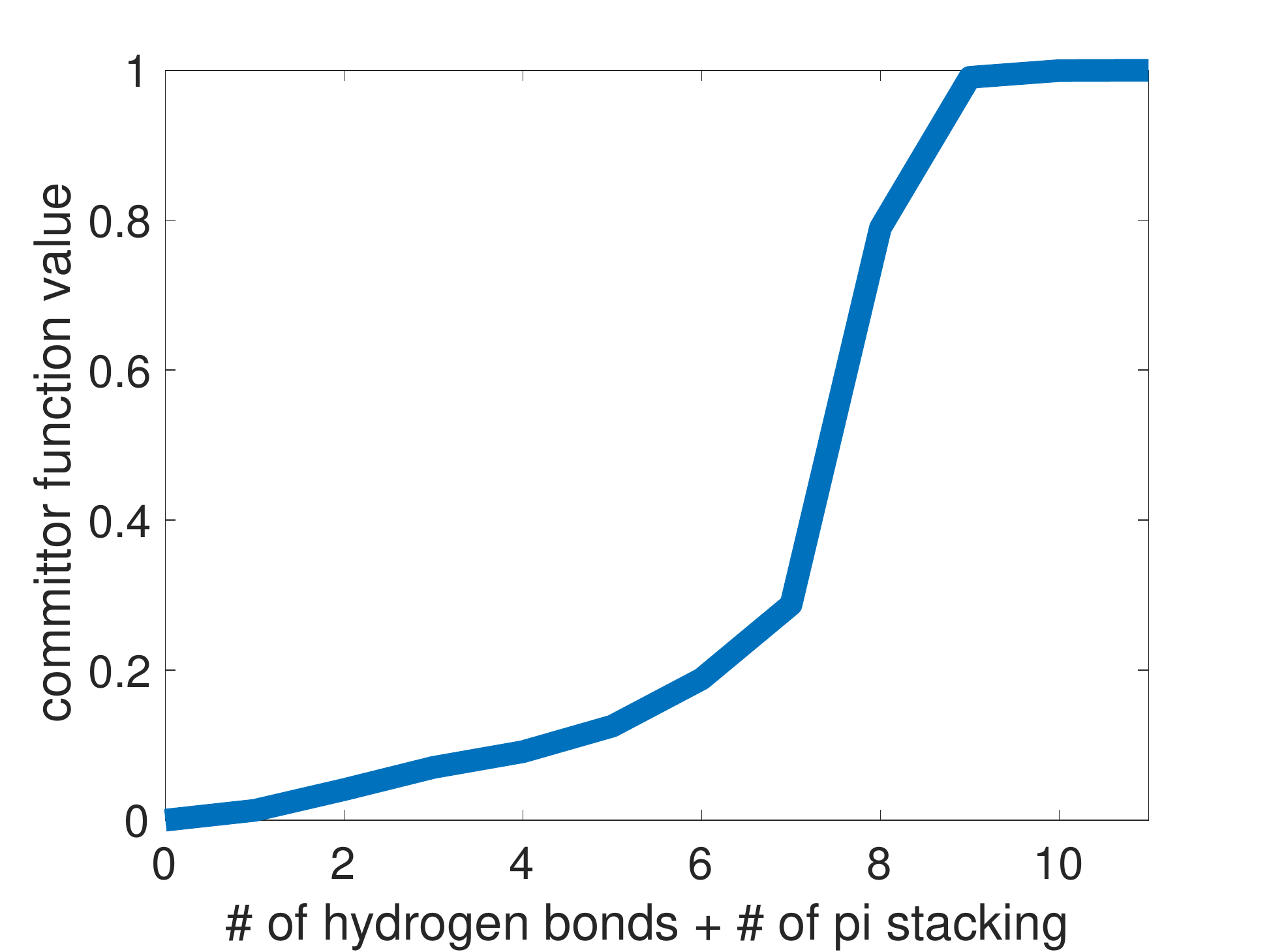} \\
(a) Free energy landscape (\emph{cis}). & (b) Committor function (\emph{cis}). \\[6pt]
\includegraphics[width=80mm]{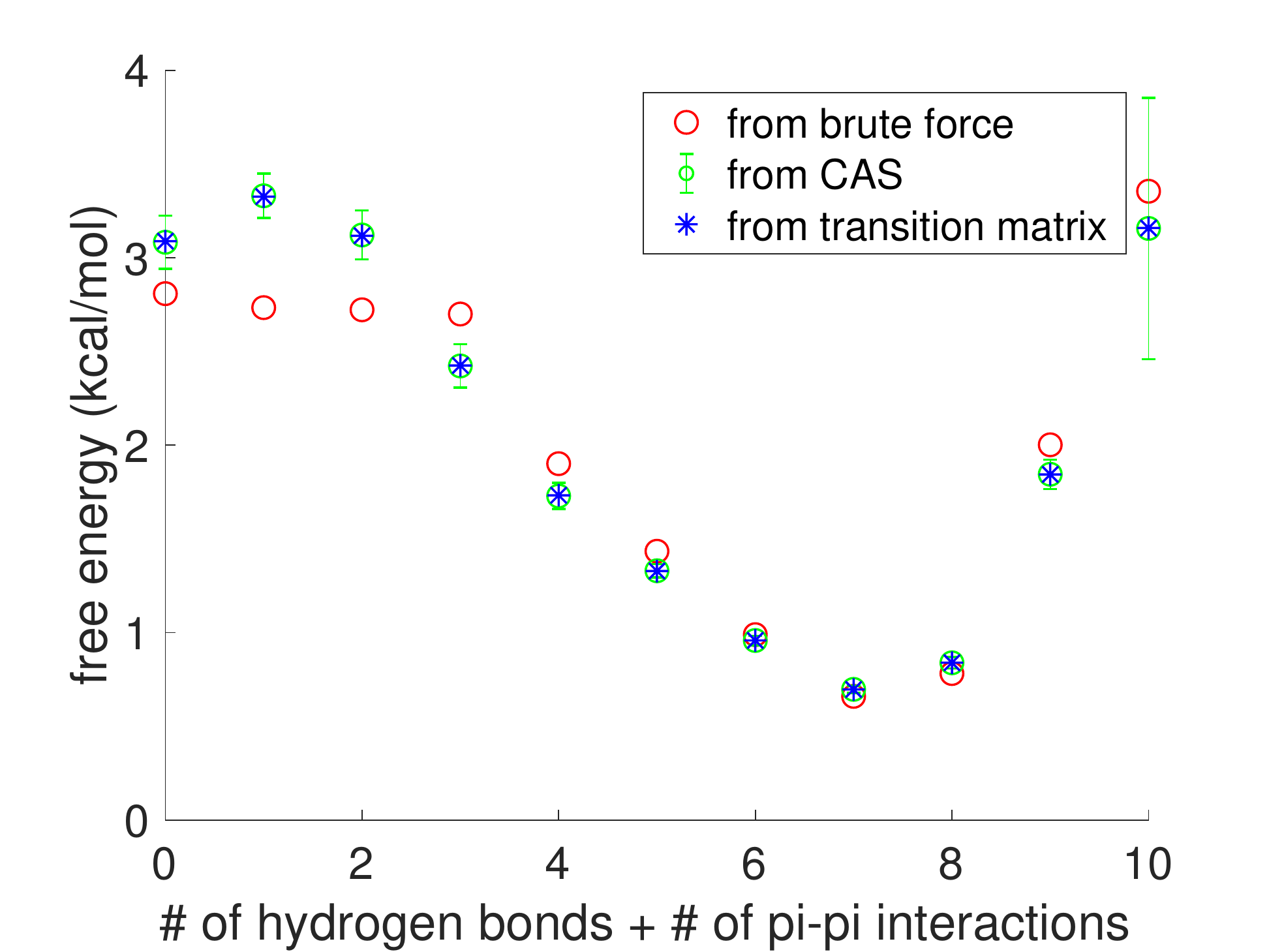} & \includegraphics[width=80mm]{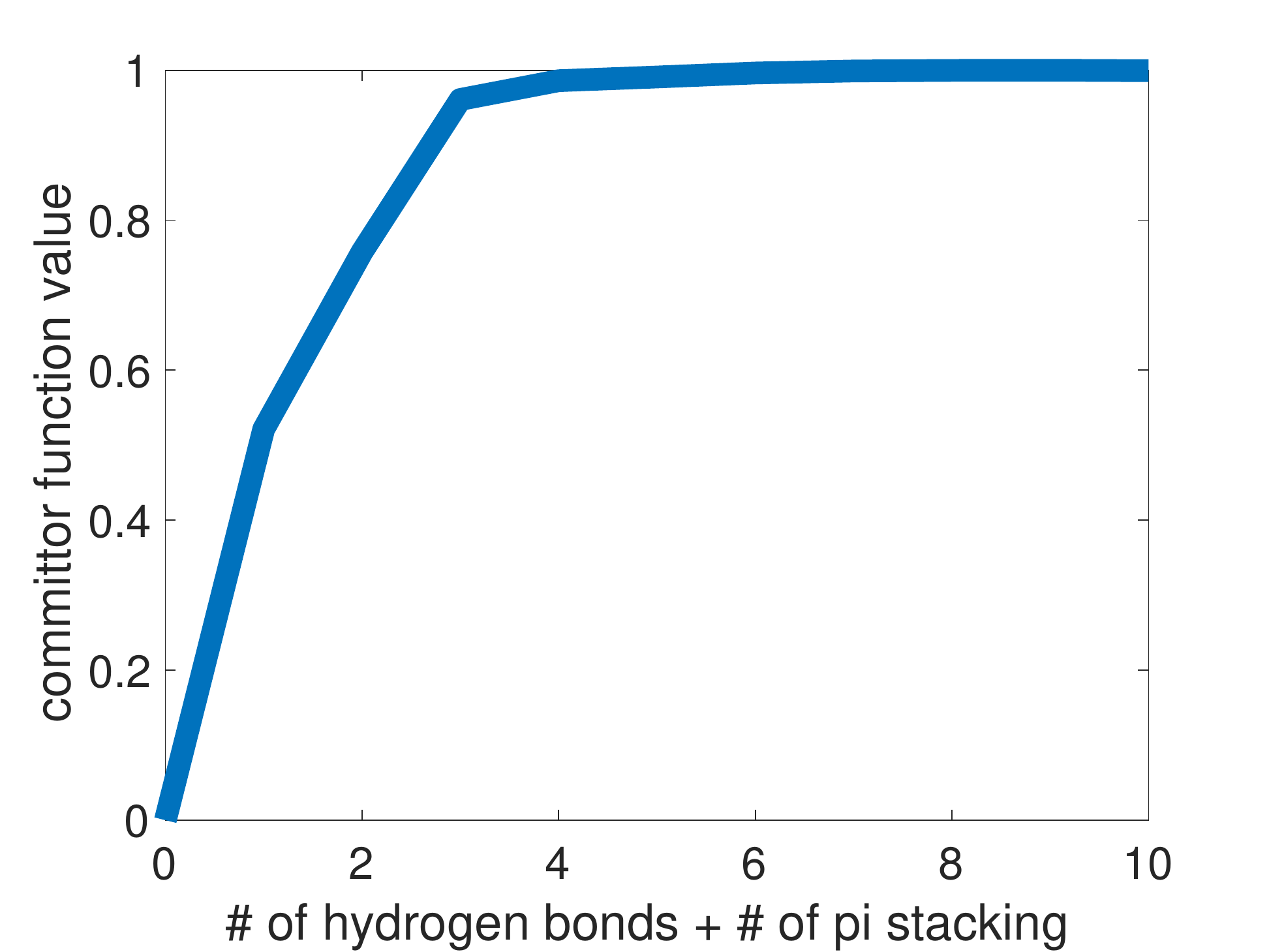} \\
(c) Free energy landscape (\emph{trans}). & (d) Committor function (\emph{trans}). \\[6pt]
\end{tabular}
\caption{\label{fig:amino_trimers_free_energy} Same as Fig.~\ref{fig:original_trimers_free_energy} but for a dimer of triazine trimers (all \emph{cis} and all \emph{trans}) with amino side chains and amino backbone.}
\end{figure}

\begin{figure}[p]
\centering
\begin{tabular}{cc}
\includegraphics[width=80mm]{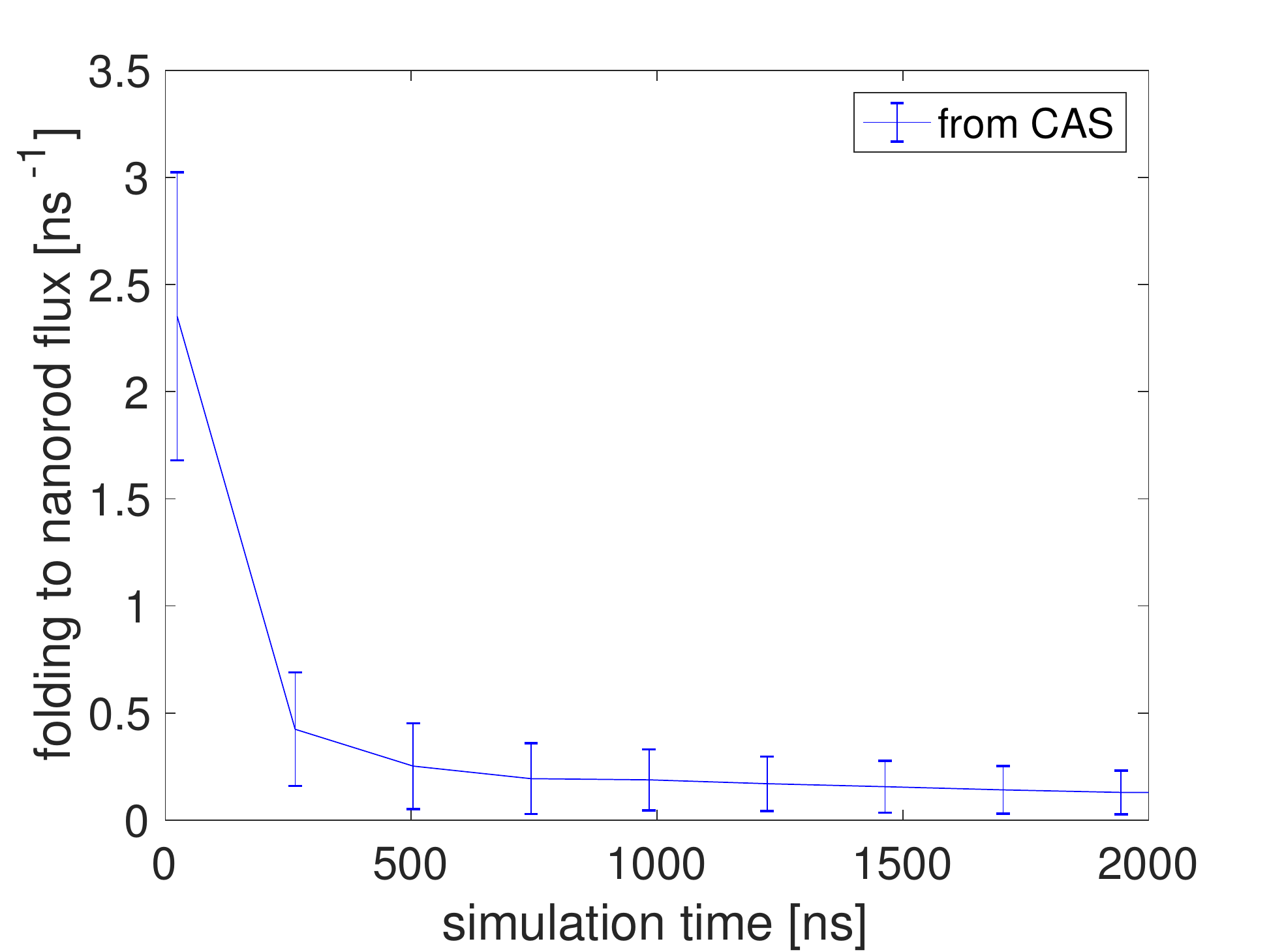} & \includegraphics[width=80mm]{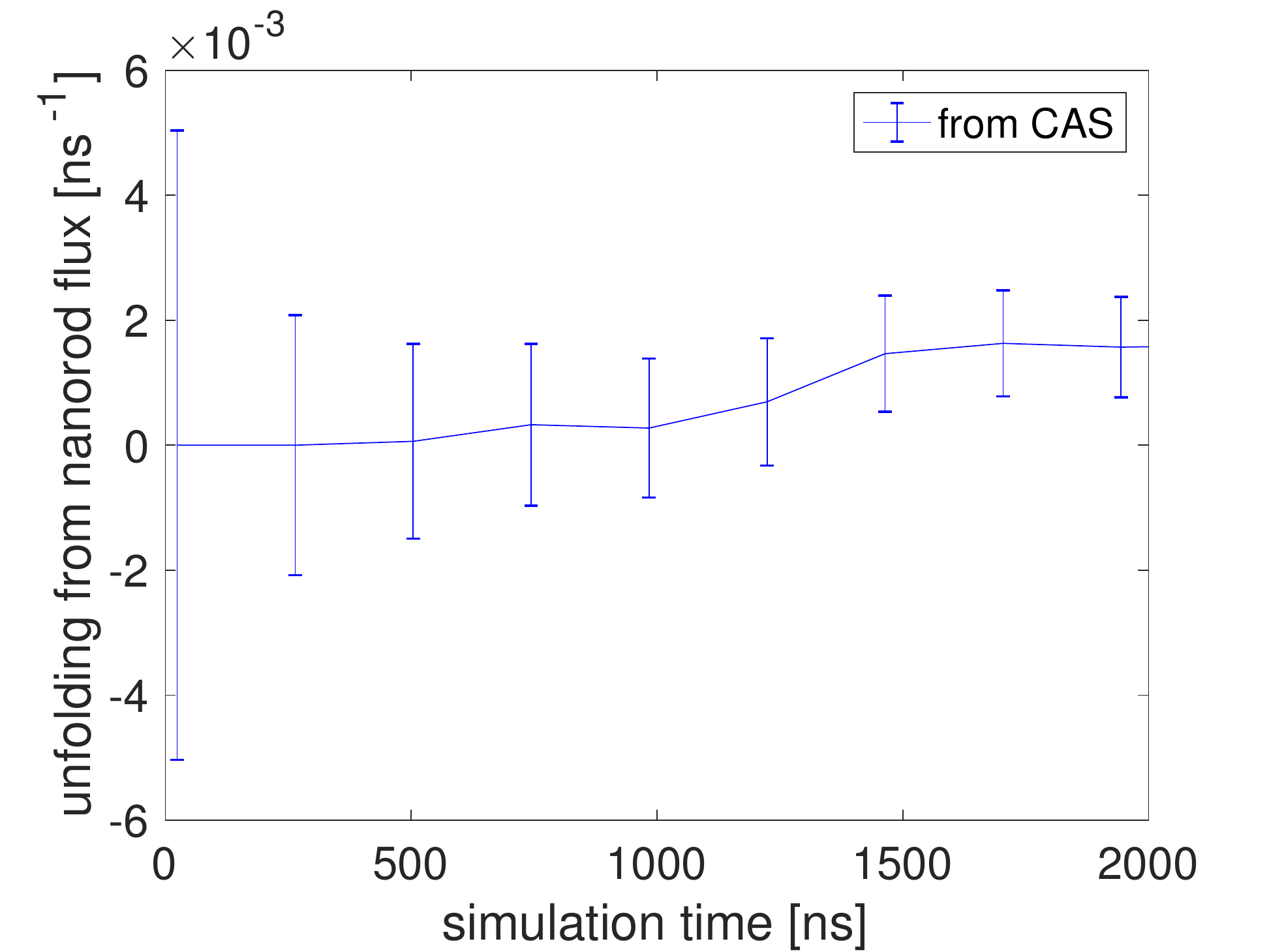} \\
(a) Forward flux (\emph{cis}). & (b) Backward flux (\emph{cis}). \\[6pt]
\includegraphics[width=80mm]{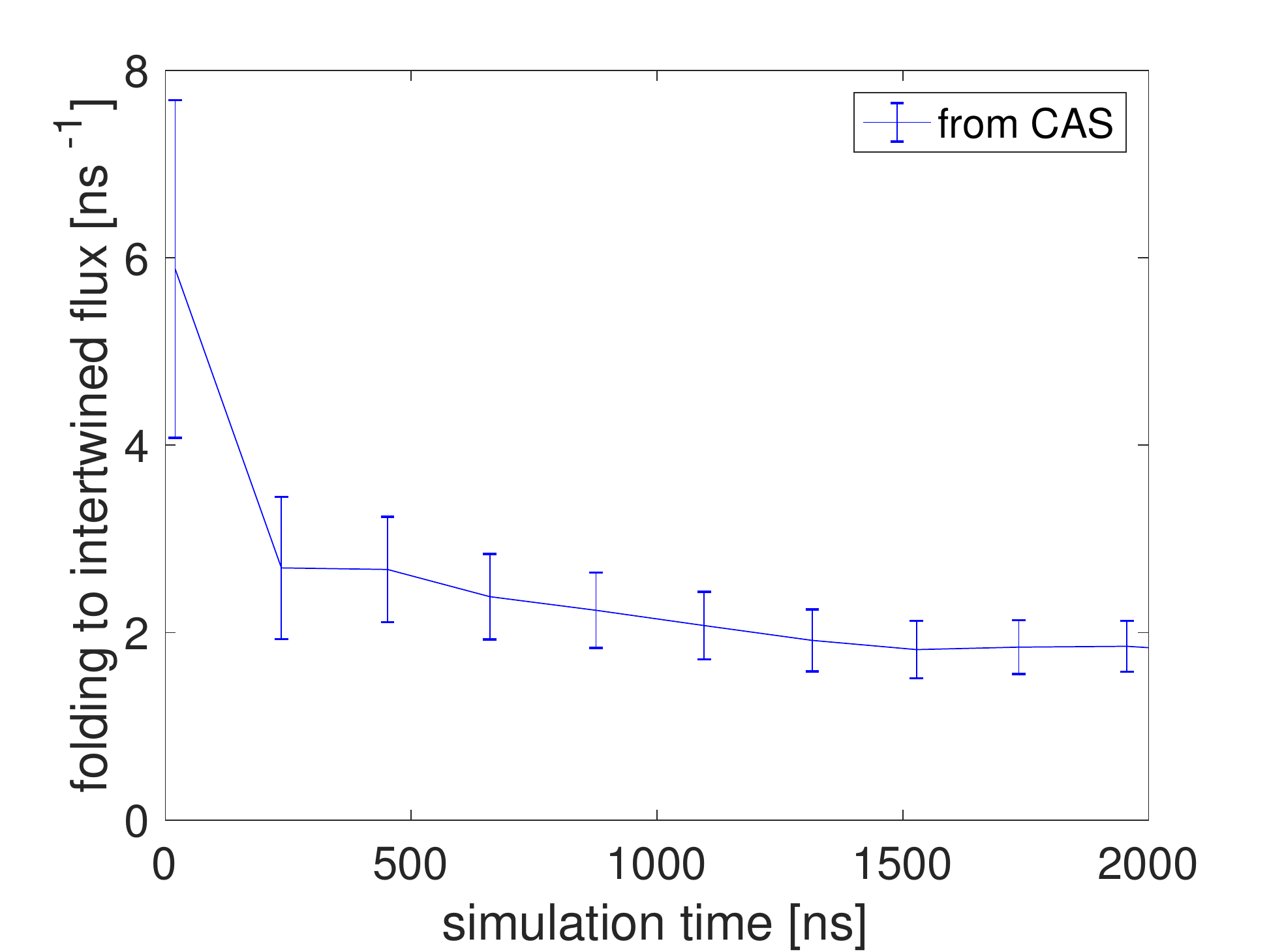} & \includegraphics[width=80mm]{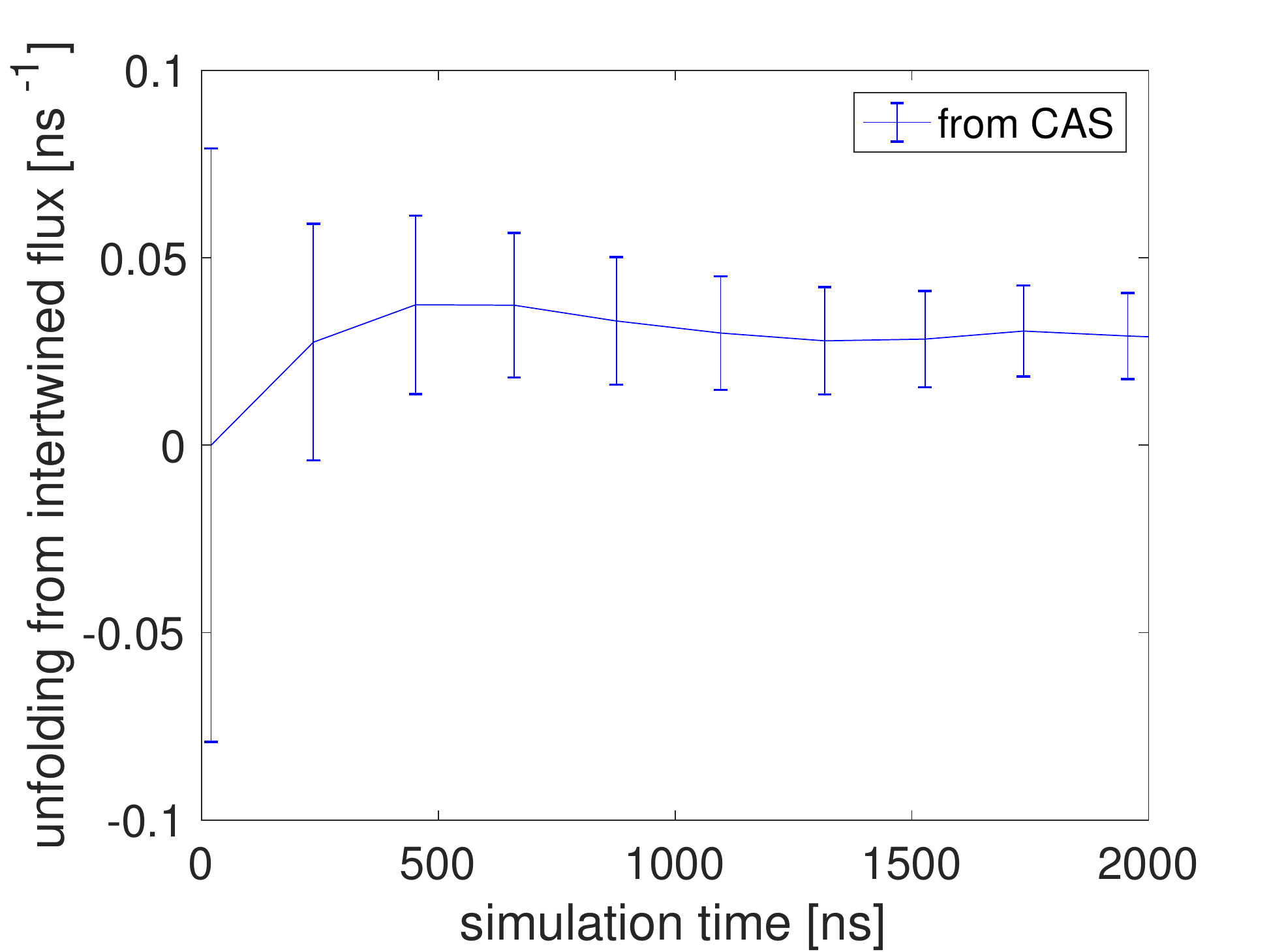} \\
(c) Forward flux (\emph{trans}). & (d) Backward flux (\emph{trans}). \\[6pt]
\end{tabular}
\caption{\label{fig:amino_trimers_fluxes} Same as Fig.~\ref{fig:original_trimers_fluxes} but for a dimer of triazine trimers (all \emph{cis} and all \emph{trans}) with amino side chains and amino backbone.}
\end{figure}

\begin{figure}[p]
\centering
\begin{tabular}{cc}
\includegraphics[width=80mm]{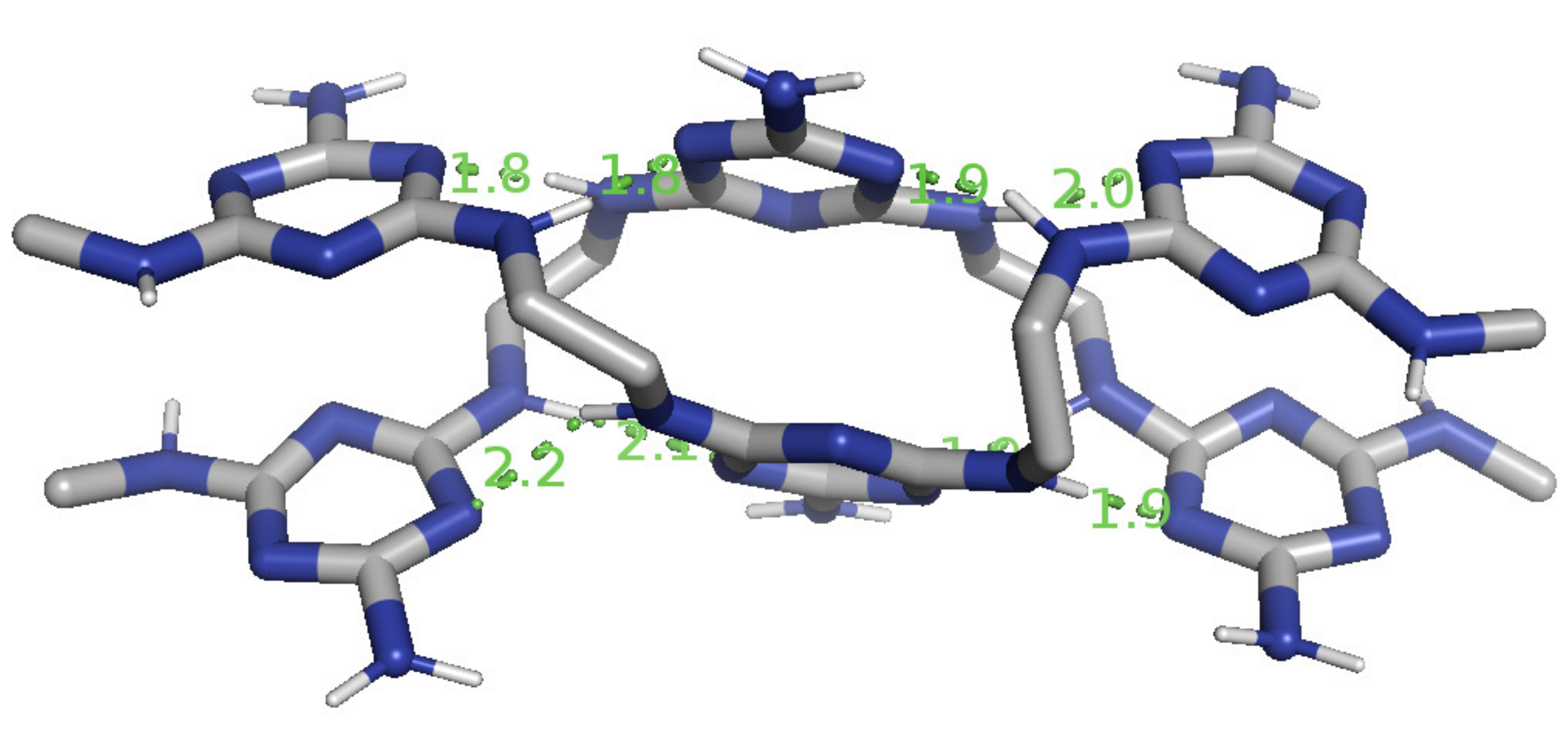} & \includegraphics[width=60mm]{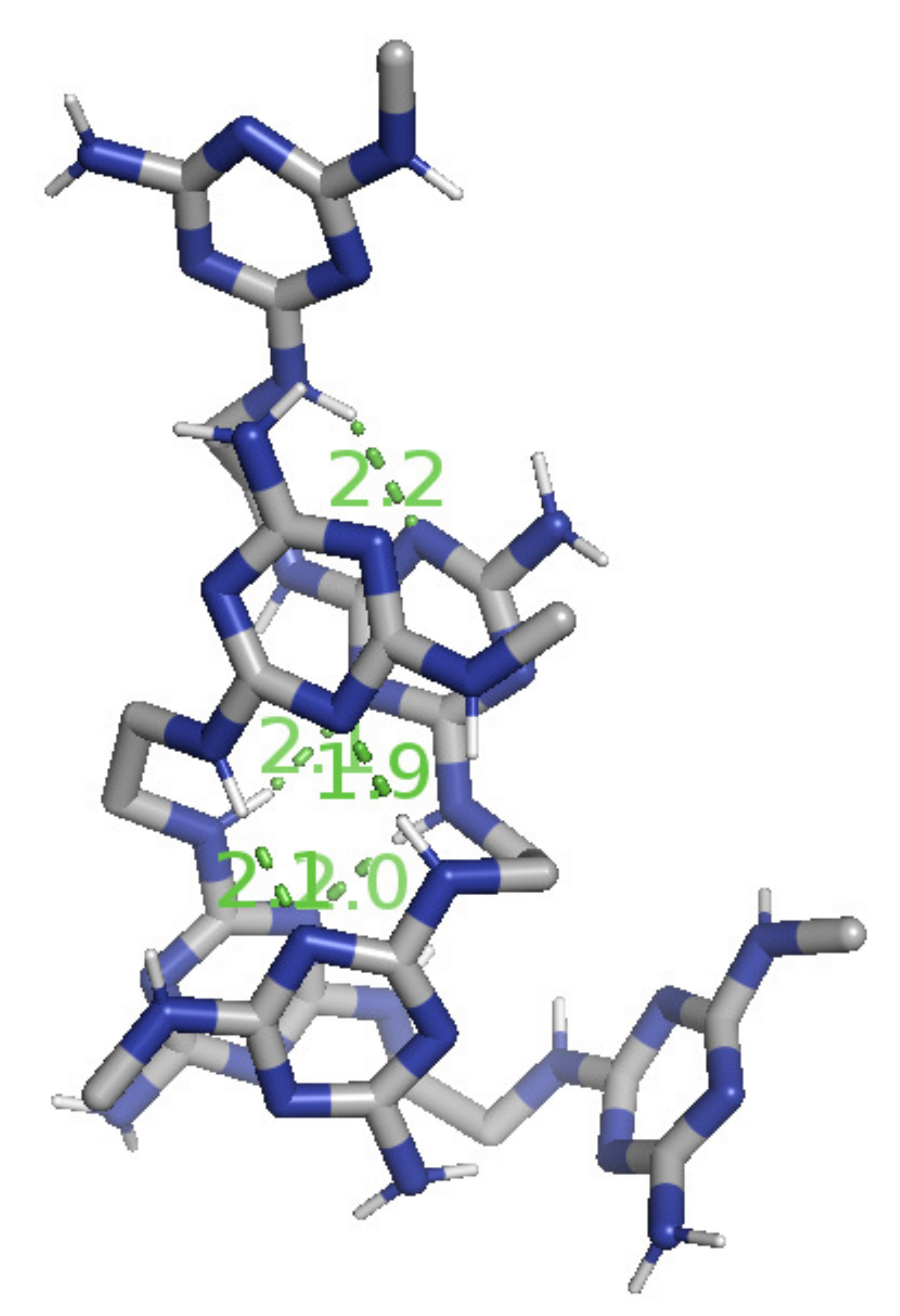} \\
(a) Nanorod structure (\emph{cis}). & (b) Intertwined structure (\emph{trans}). \\[6pt]
\includegraphics[width=80mm]{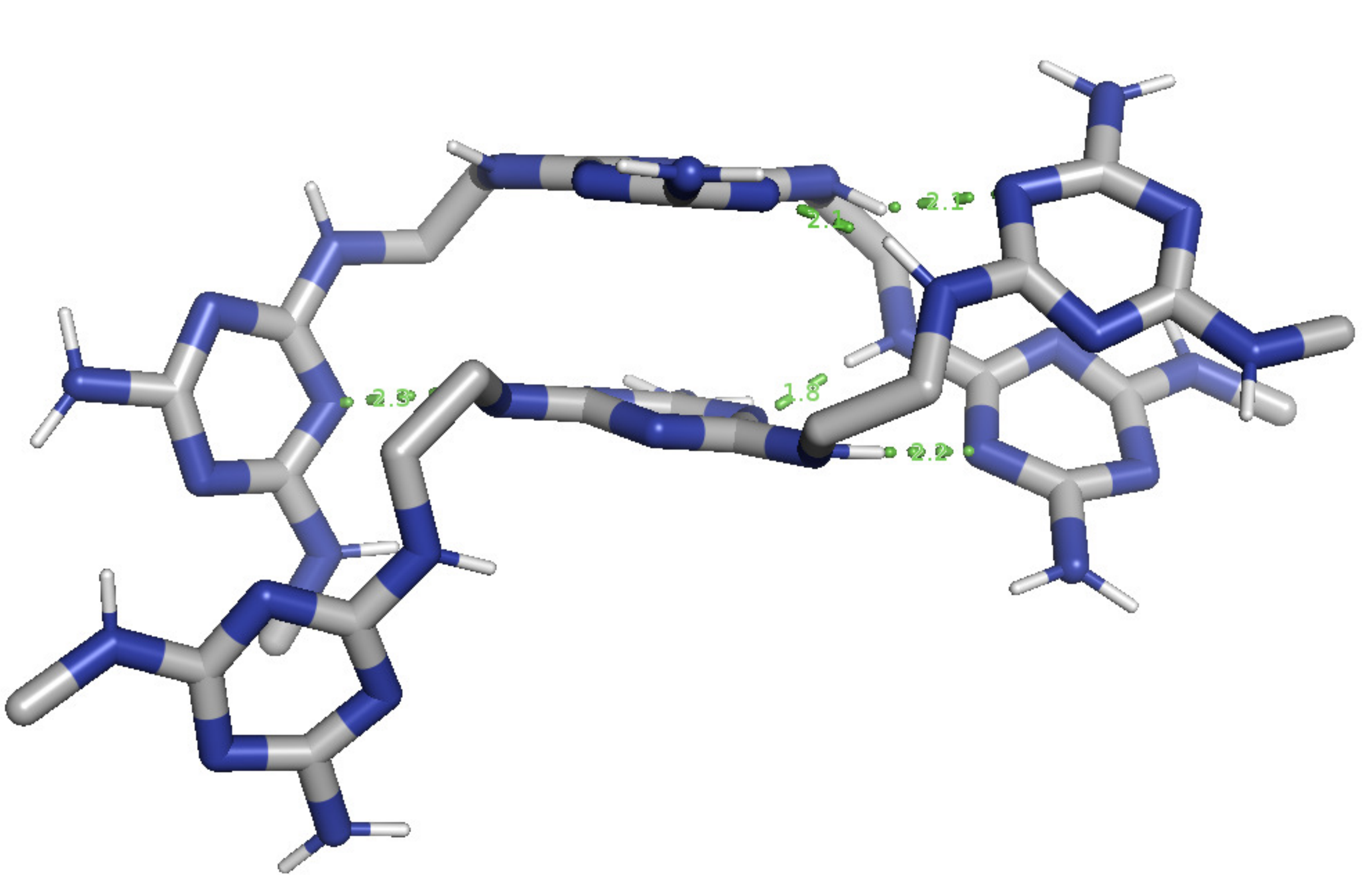} & \includegraphics[width=80mm]{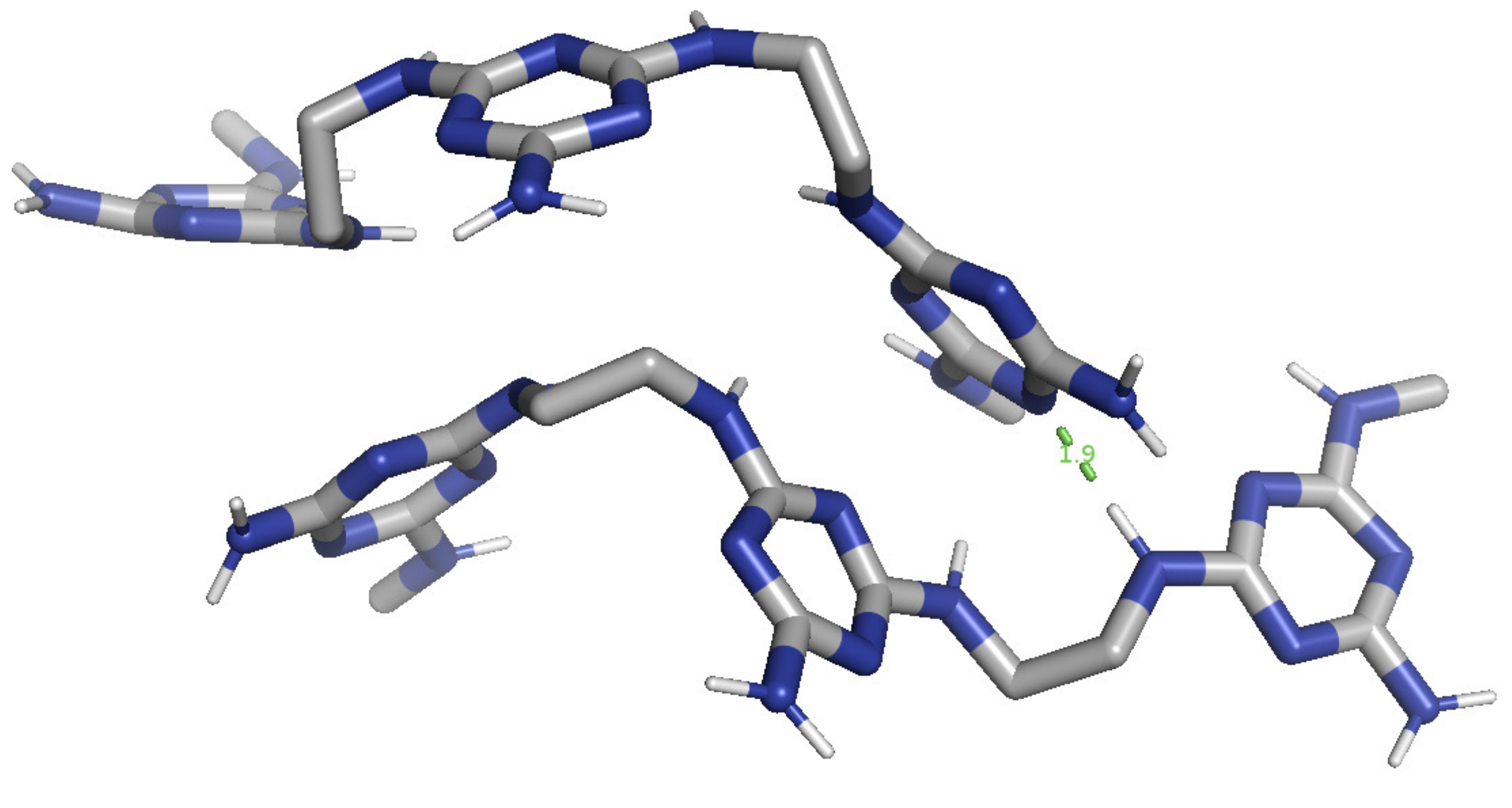} \\
(c) Intermediate conformation (\emph{cis}). & (d) Intermediate conformation (\emph{trans}). \\[6pt]
\end{tabular}
\caption{\label{fig:amino_trimers_conformations} Same as Fig.~\ref{fig:original_trimers_conformations} but for a dimer of triazine trimers (all \emph{cis} and all \emph{trans}) with amino side chains and amino backbone. Figure (a) shows the nanorod structure that has 11 non-covalent interactions in total (8 hydrogen bonds and 3 $\pi$-$\pi$ interactions). Figure (b) shows the intertwined structure has 7 non-covalent interactions in total (5 hydrogen bonds and 2 $\pi$-$\pi$ interactions). Figures (c) and (d) show the intermediate conformations for all \emph{cis} (7 non-covalent interactions) and all \emph{trans} (1 non-covalent interaction), respectively.}
\end{figure}

\begin{figure}[p]
\centering
\begin{tabular}{cc}
\includegraphics[width=60mm]{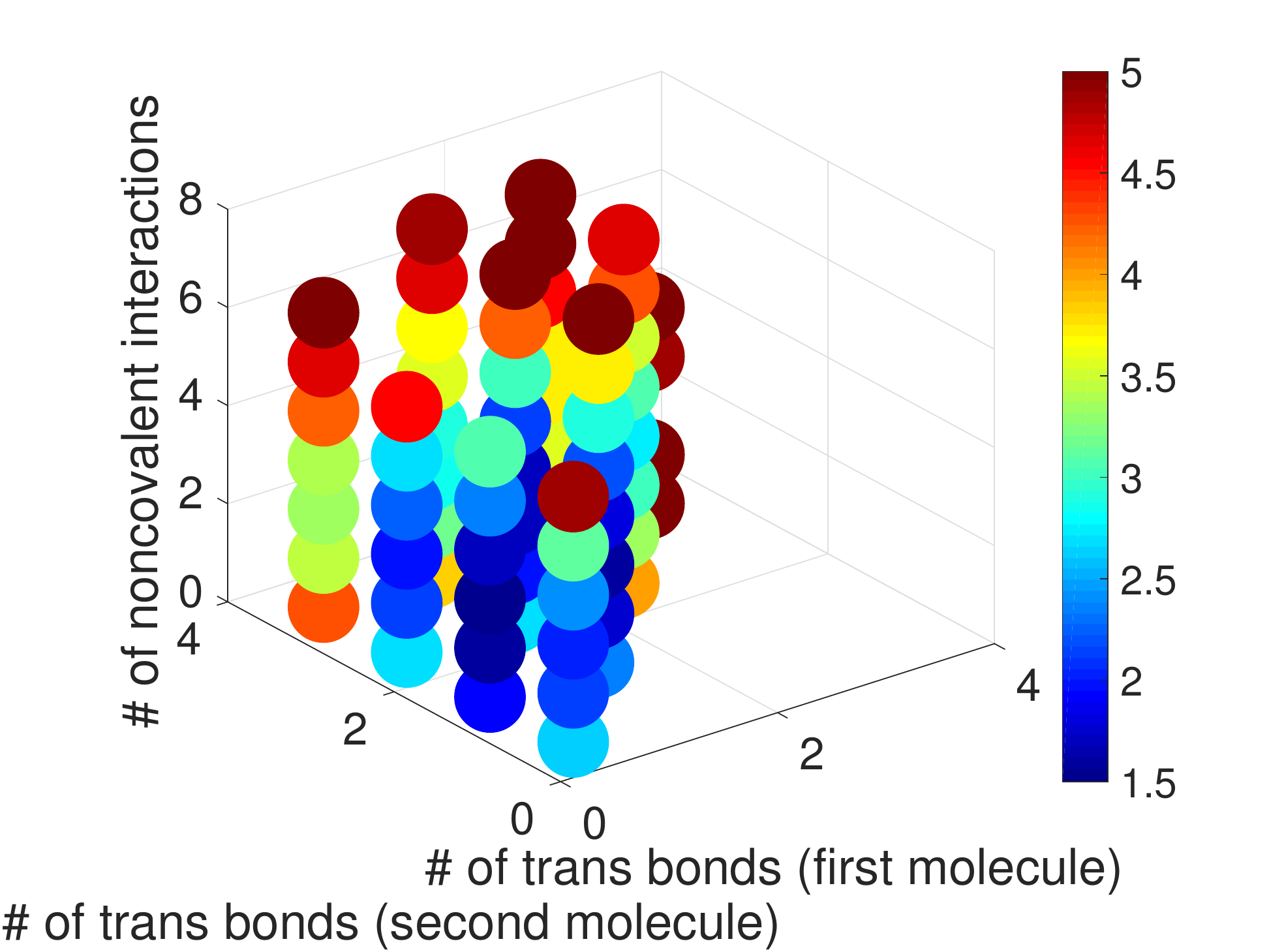} & \includegraphics[width=60mm]{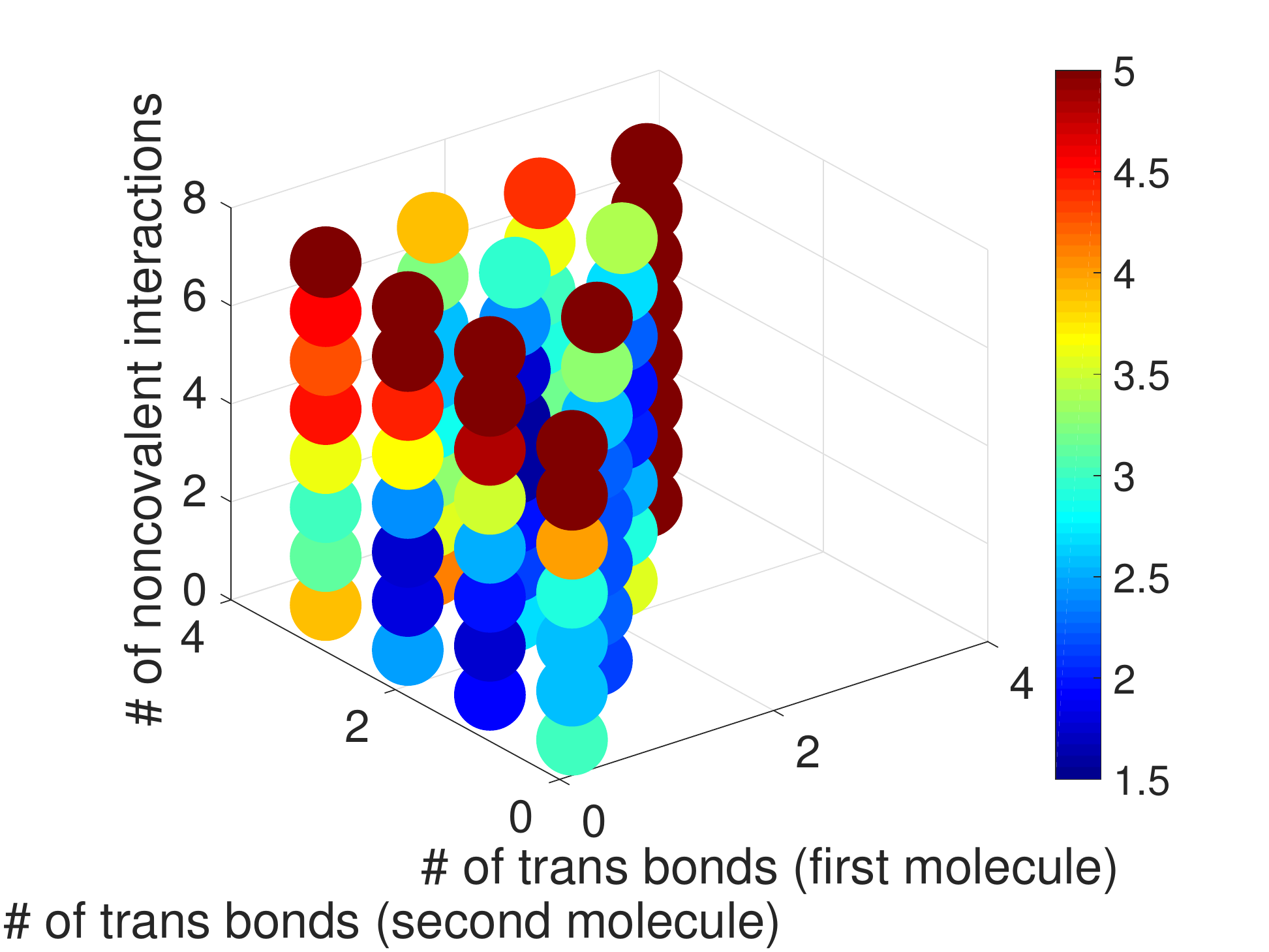} \\
(a) Free energy landscape from brute force. & (b) Free energy landscape from CAS. \\[6pt]
\includegraphics[width=60mm]{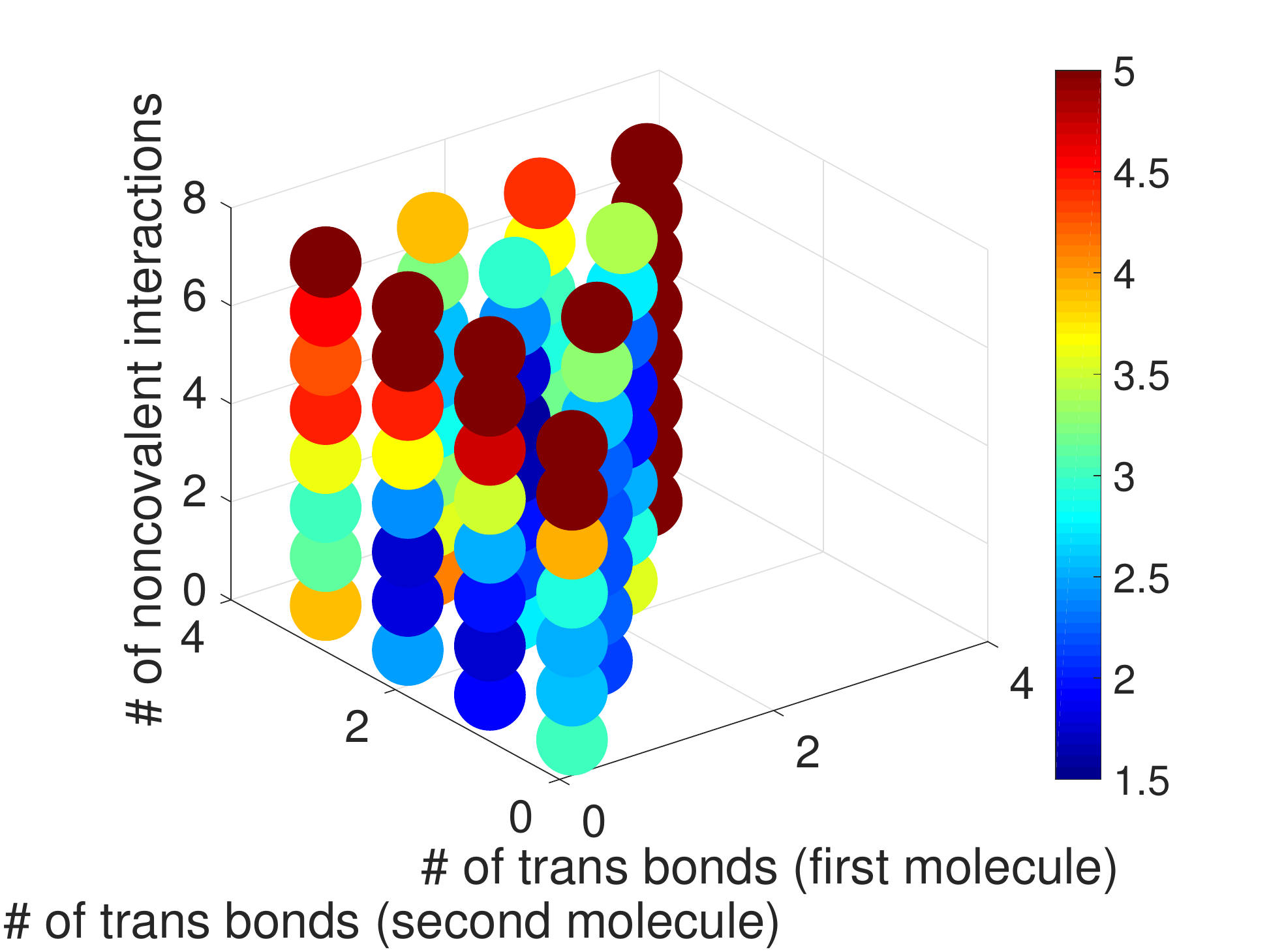} & \includegraphics[width=60mm]{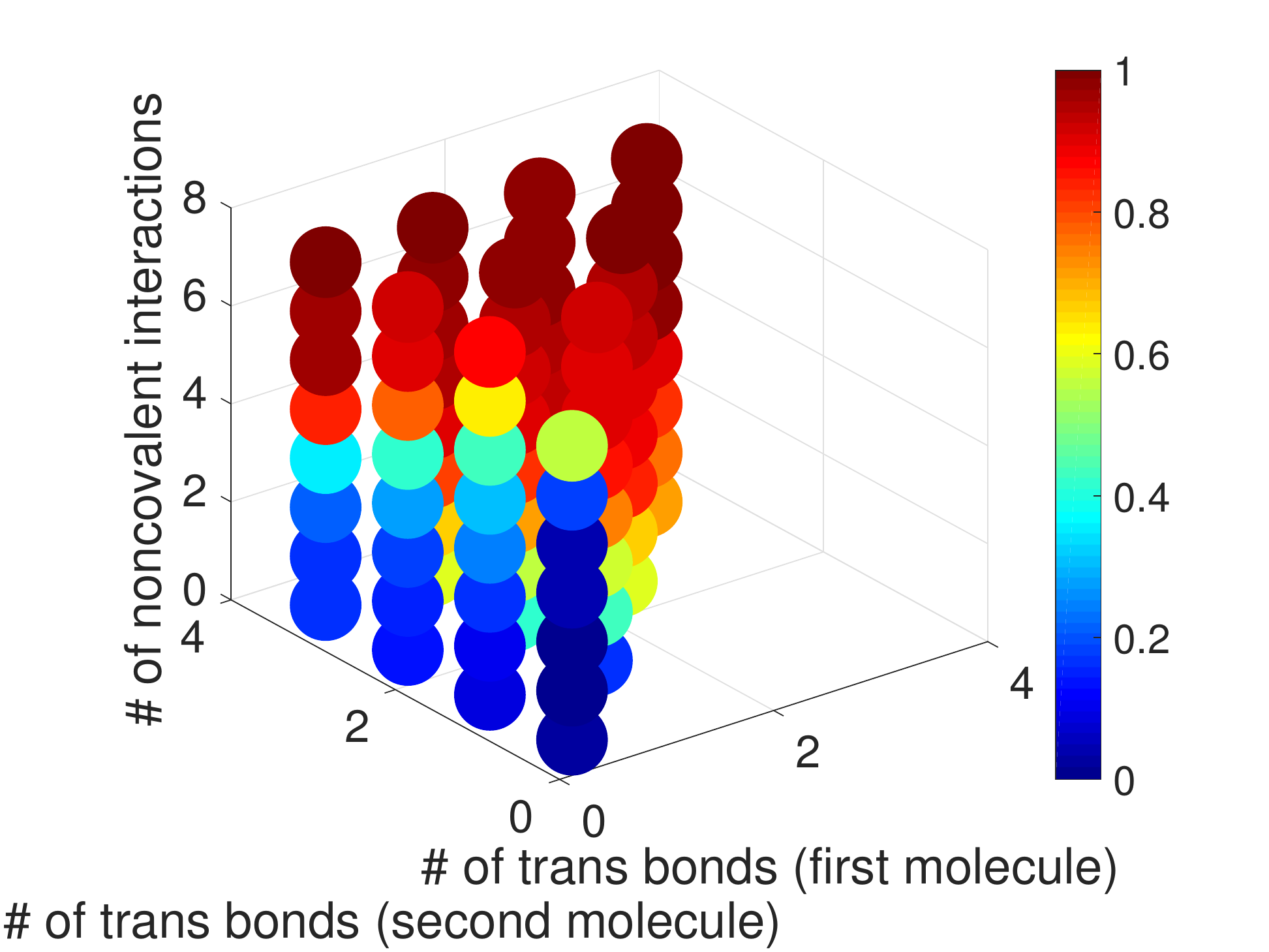} \\
(c) Free energy landscape from transition matrix. & (d) Committor function. \\[6pt]
\includegraphics[width=50mm]{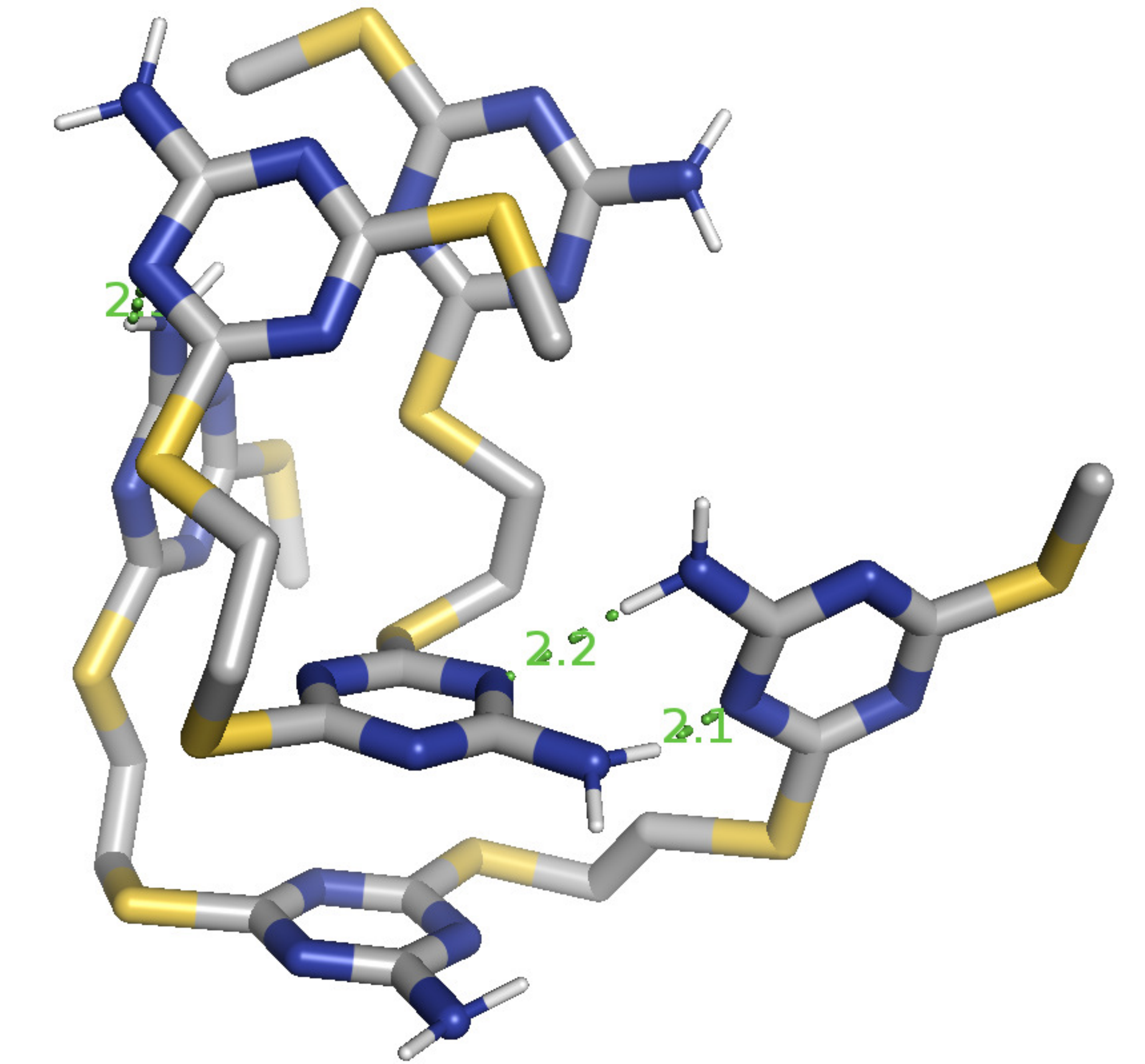} & \includegraphics[width=50mm]{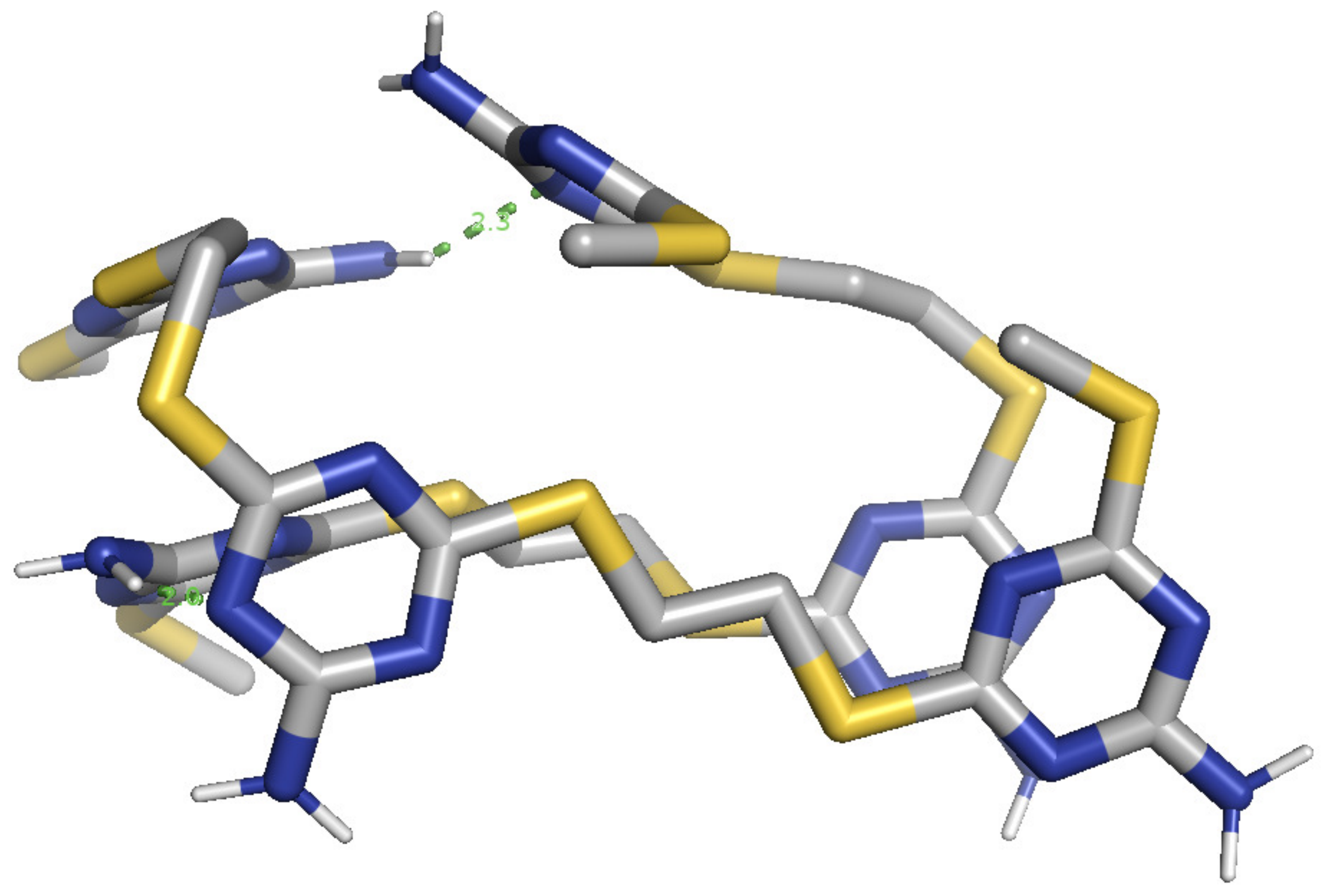} \\
(e) Most stable conformation. & (f) Intermediate conformation. \\[6pt]
\end{tabular}
\caption{\label{fig:amino_s_trimers_free_energy_conformations} Free energy landscapes, committor function, and most stable and intermediate conformations for a dimer of triazine trimers with amino side chains and sulfur backbone. Figures (a), (b), and (c) show the free energy landscapes colored in log scale or $-k_BT\ln P$ (kcal/mol), where $P$ denotes the weight, and the color bar indicates which colors correspond to which free energies in log scale (kcal/mol). Figure (d) shows the committor function. Figure (e) shows the most stable conformation that has 4 non-covalent interactions, 1 trans bond for one trimer, and 2 trans bonds for the other trimer. Figure (f) shows the intermediate conformation that has 4 non-covalent interactions, 0 trans bond for one trimer, and 2 trans bonds for the other trimer.}
\end{figure}

\begin{figure}[p]
\centering
\begin{tabular}{cc}
\includegraphics[width=60mm]{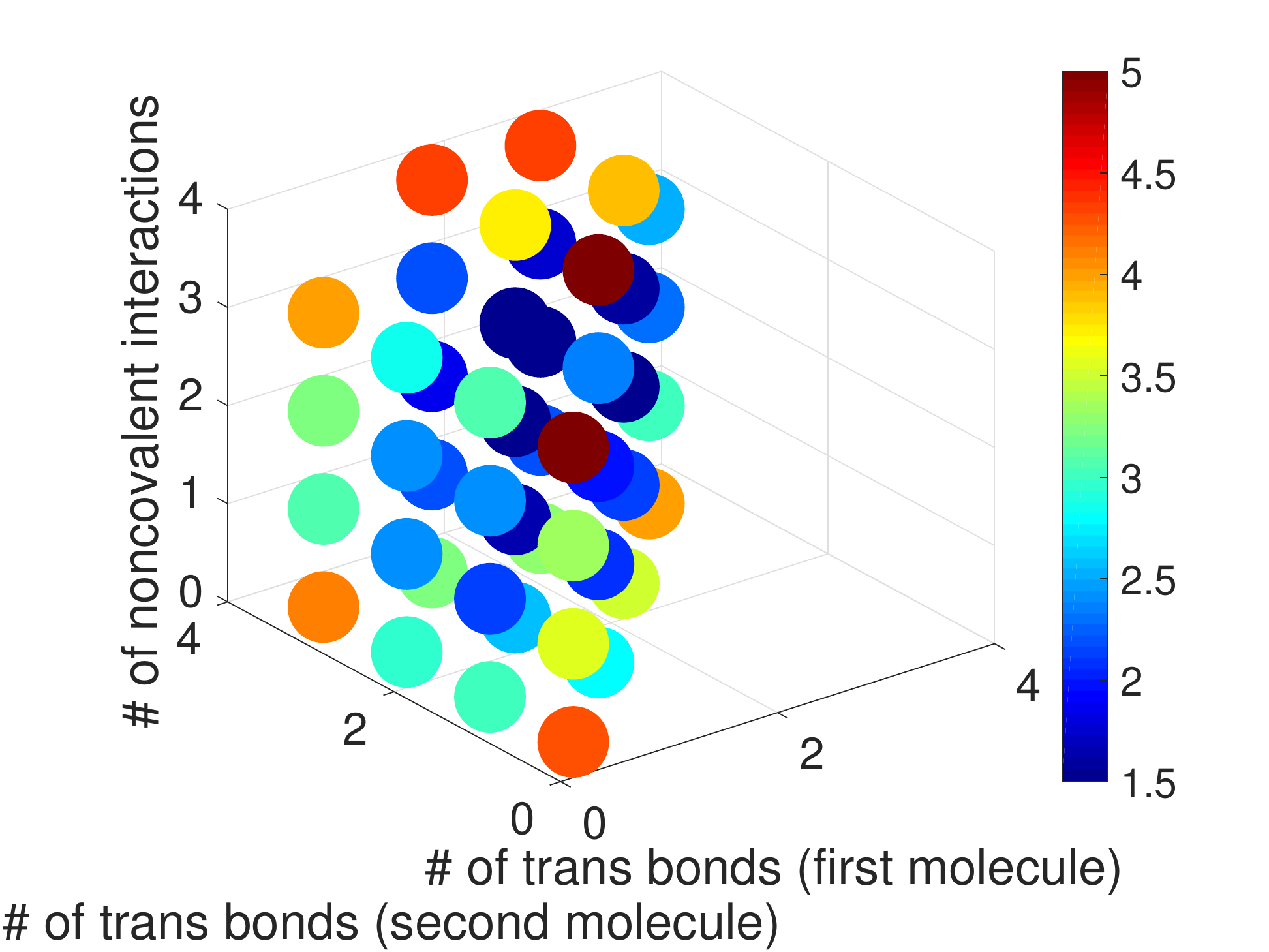} & \includegraphics[width=60mm]{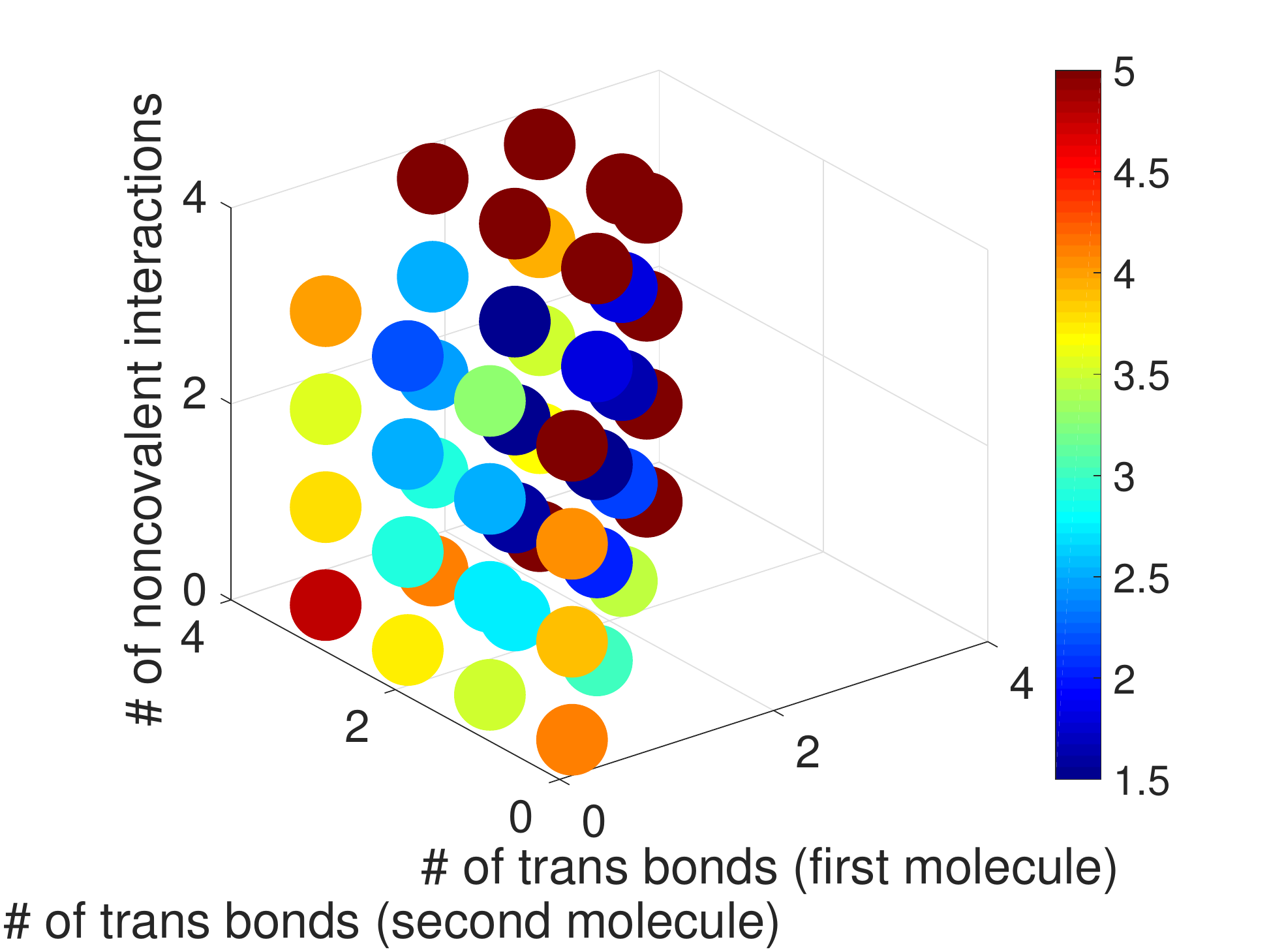} \\
(a) Free energy landscape from brute force. & (b) Free energy landscape from CAS. \\[6pt]
\includegraphics[width=60mm]{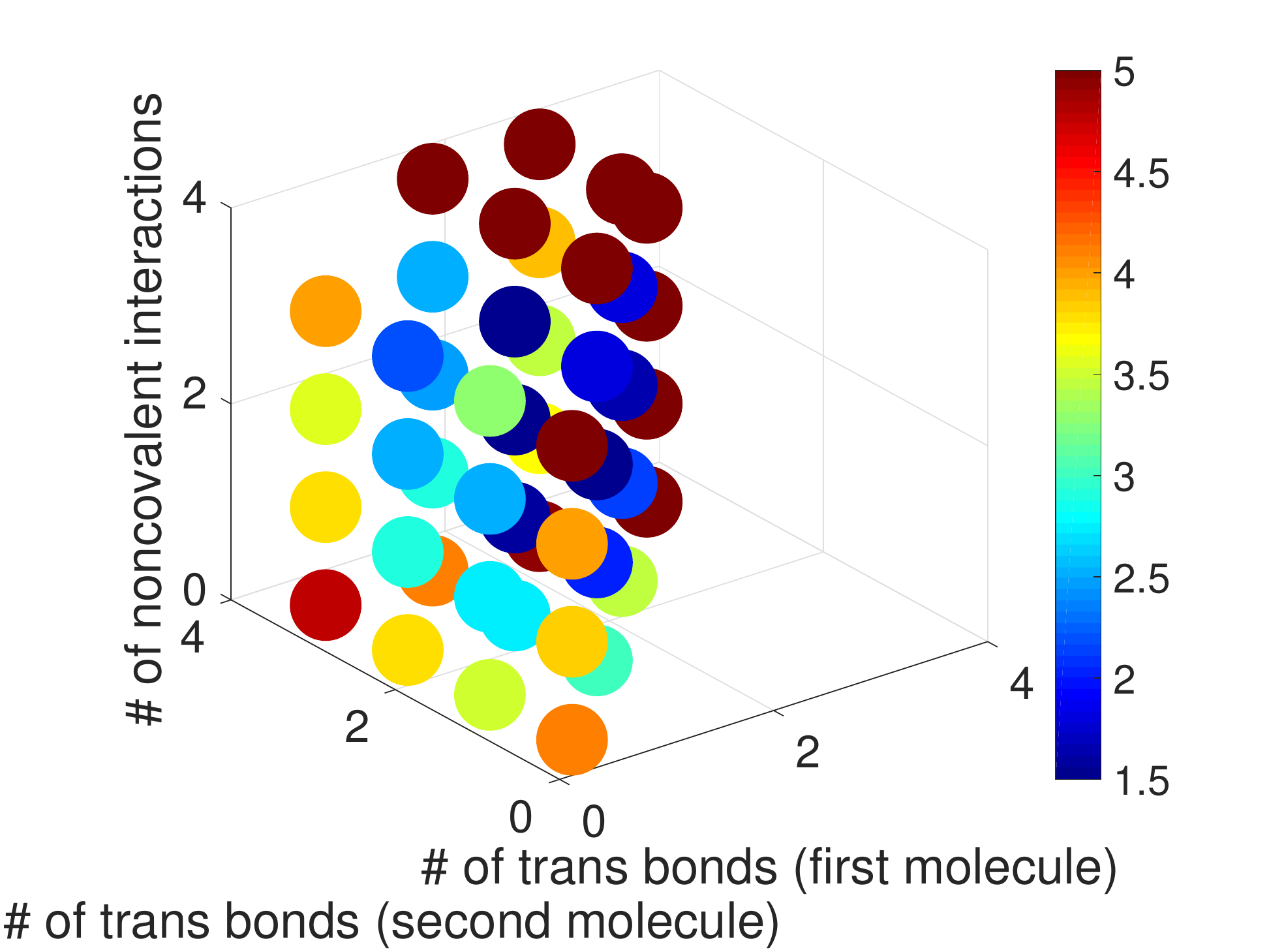} & \includegraphics[width=60mm]{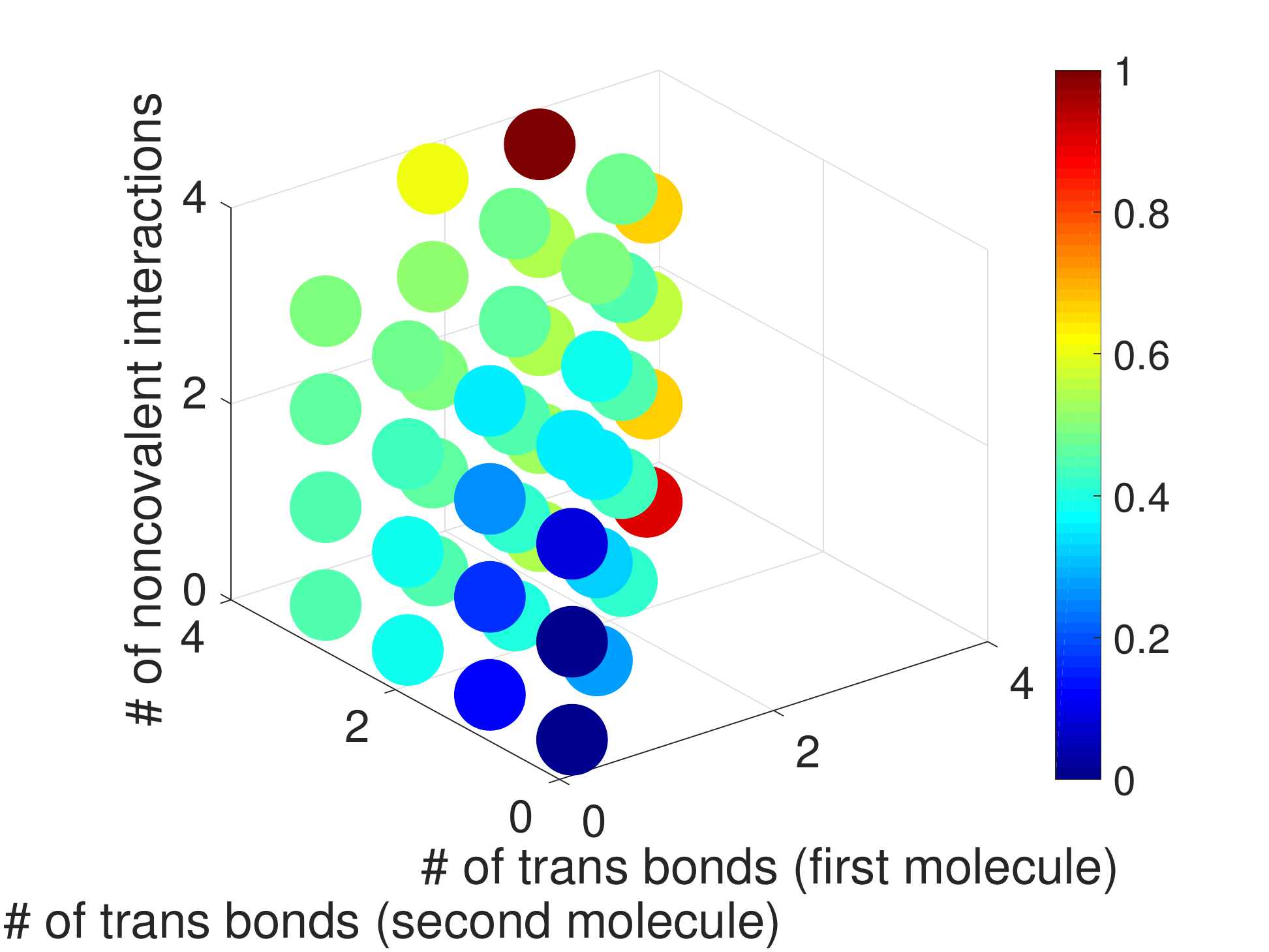} \\
(c) Free energy landscape from transition matrix. & (d) Committor function. \\[6pt]
\includegraphics[width=50mm]{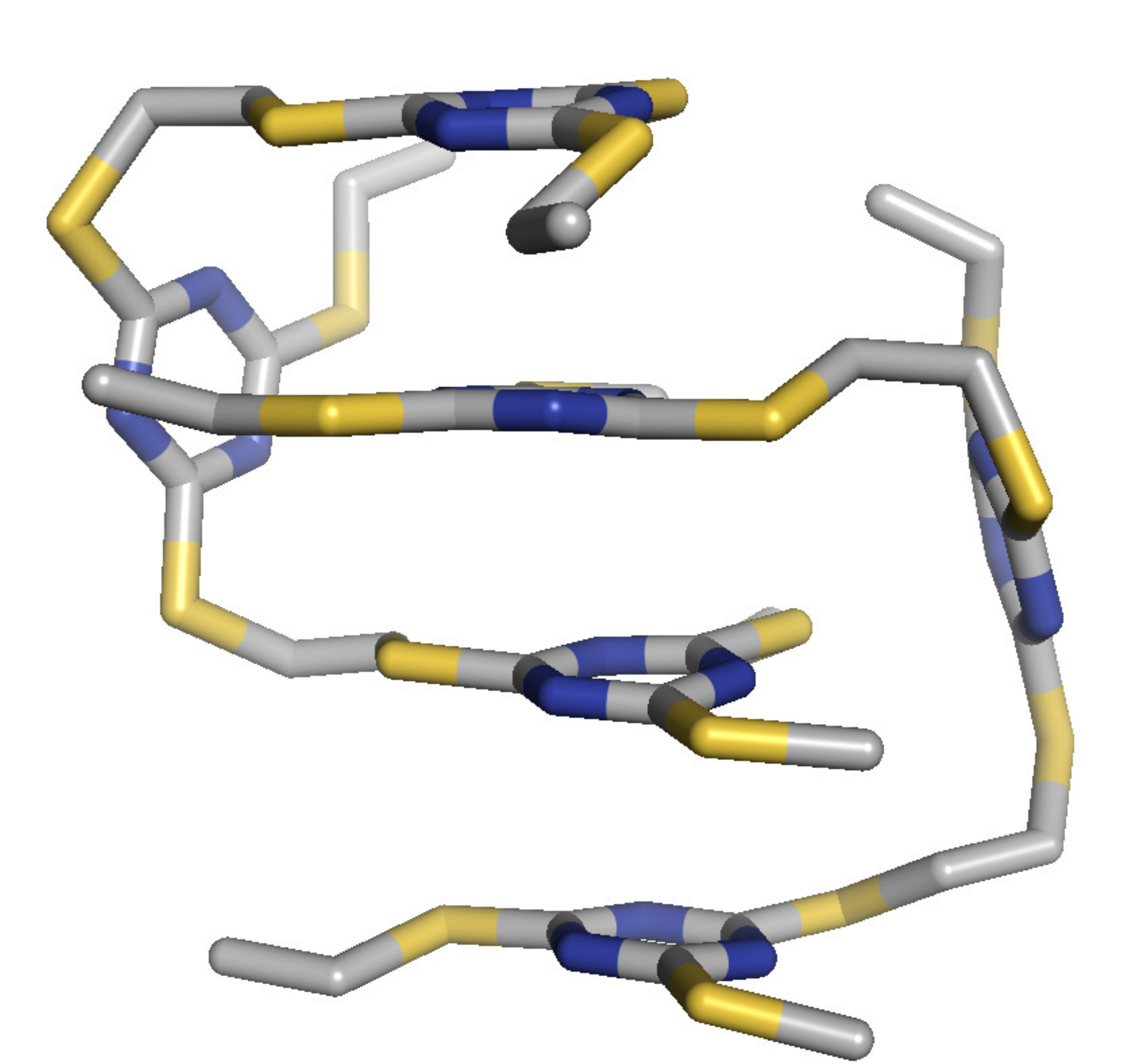} & \includegraphics[width=50mm]{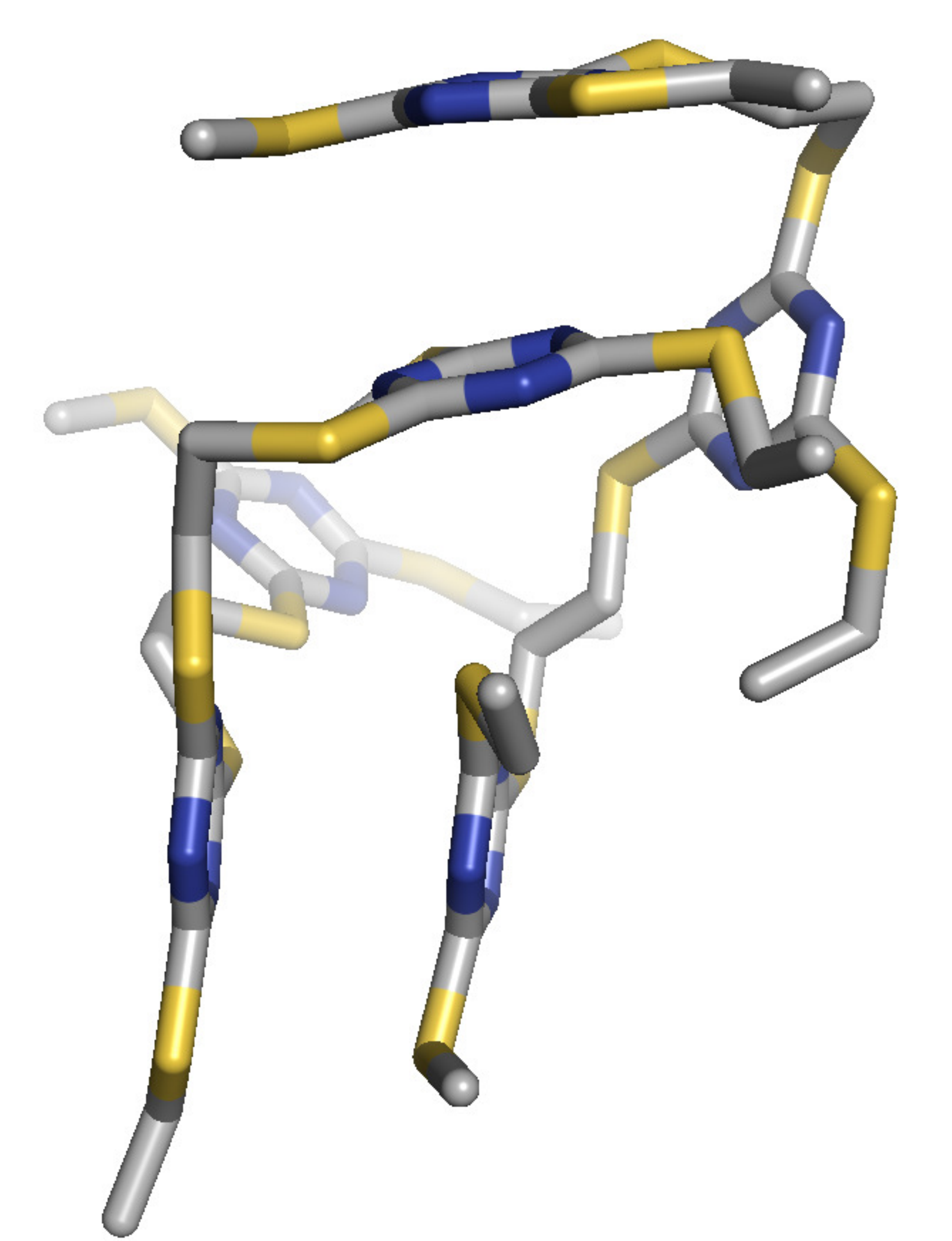} \\
(e) Most stable conformation. & (f) Intermediate conformation. \\[6pt]
\end{tabular}
\caption{\label{fig:original_s_trimers_free_energy_conformations} Free energy landscapes, committor function, and most stable and intermediate conformations for a dimer of triazine trimers with S-ethyl side chains and sulfur backbone. Figures (a), (b), and (c) show the free energy landscapes colored in log scale or $-k_BT\ln P$ (kcal/mol), where $P$ denotes the weight, and the color bar indicates which colors correspond to which free energies in log scale (kcal/mol). Figure (d) shows the committor function. Figure (e) shows the most stable conformation that has 3 non-covalent interactions, 1 trans bond for one trimer, and 2 trans bonds for the other trimer. Figure (f) shows the intermediate conformation that has 2 non-covalent interactions, 1 trans bond for one trimer, and 3 trans bonds for the other trimer.}
\end{figure}

To obtain the free energy landscape, committor function, forward and backward fluxes, and most stable and intermediate conformations for each all \emph{cis} and all \emph{trans} dimer of triazine trimers listed in Table~\ref{tab:trimers_summary}, we simulated each dimer with concurrent adaptive sampling (CAS) algorithm for 2 $\mu$s. Figs.~\ref{fig:original_trimers_free_energy} --~\ref{fig:original_trimers_conformations}, Figs.~\ref{fig:hn_trimers_free_energy} --~\ref{fig:hn_trimers_conformations}, and Figs.~\ref{fig:amino_trimers_free_energy} --~\ref{fig:amino_trimers_conformations} show results for the dimer with S-ethyl side chains and amino backbone, the dimer with amino-ethyl side chains and amino backbone, and the dimer with amino side chains and amino backbone, respectively. Table~\ref{tab:trimers_fluxes} lists the forward and backward fluxes for the dimers with amino backbone listed in Table~\ref{tab:trimers_summary}. Finally, Fig.~\ref{fig:amino_s_trimers_free_energy_conformations} and Fig.~\ref{fig:original_s_trimers_free_energy_conformations} show results for the dimer with amino side chains and sulfur backbone and for the dimer with S-ethyl side chains and sulfur backbone, respectively. 

As previously mentioned in Sec.~\ref{sec:simulation_protocol}, we used 500 ns of brute force MD simulation data for each system to initialize and speed up convergence of the CAS algorithm simulations. When simulating the dimer of triazine trimers with brute force MD simulations, both triazine trimers with amino backbone maintained their \emph{cis/trans} conformations throughout the simulation, whereas the triazine trimers with sulfur backbone converted from \emph{cis} to \emph{trans} and vice versa easily. This is due to the \emph{cis/trans} bond having partial double bond character for the triazine trimers with amino backbone and not having partial double bond character for the triazine trimers with sulfur backbone. Hence, as previously mentioned in Sec.~\ref{sec:reaction_coordinates}, for the dimers with amino backbone, we ran the CAS algorithm simulations by keeping track of the total number of non-covalent interactions (hydrogen bonds and $\pi$-$\pi$ interactions) for both all \emph{cis} and all \emph{trans} cases. For the dimer of triazine trimers with sulfur backbone, we ran the CAS algorithm simulations by keeping track of the total number of non-covalent interactions, the number of trans bonds for one triazine trimer, and the number of trans bonds for the other triazine trimer.

These various triazine trimers were tested to find out whether the nanorod structure will still be present for the all \emph{cis} triazine trimers if their backbones lose hydrogen bonding ability and/or if their side chains have hydrogen bonding abilities in different strengths. In the results, the initial brute force MD simulation data is shown for comparison to the final free energy landscape obtained from the CAS algorithm simulation. The brute force points are plotted in log scale or $-k_BT\ln P$ (kcal/mol), where $P$ denotes the overall weight obtained during the 500 ns simulation. The CAS algorithm points are also plotted as $-k_BT\ln P$ (kcal/mol), where $P$ denotes the average weight obtained during the 2 $\mu$s simulation. The standard deviation of the free energy was multiplied by 2, which approximately represents 95\% confidence interval, for error bars. To plot the transition matrix points, the transition matrix, where each entry $T_{ij}$ equals the average of the weights going from macrostate $i$ to macrostate $j$ and vice versa, was calculated at every resampling step during the 2 $\mu$s CAS algorithm simulation. The transition matrices were calculated this way so that they fulfilled detailed balance, as done in Ref.~\onlinecite{ahn2017}, which followed Ref.~\onlinecite{prinz2011}. The equilibrium weights, or the eigenvector corresponding to eigenvalue $\lambda_1 = 1$, were obtained from the averaged transition matrix. The transition matrix points are plotted as $-k_BT\ln P$ (kcal/mol), where $P$ denotes the equilibrium weights. We found that the free energy landscapes from the transition matrix and from the CAS algorithm closely matched with each other and that the CAS algorithm produced free energies with low error bars, indicating that the CAS algorithm simulation has well converged to steady-state.

For the dimers of triazine trimers with amino backbone, the forward (reactant to product) and backward (product to reactant) fluxes over simulation time for all \emph{cis} and all \emph{trans} are also shown, and their final values are listed in Table~\ref{tab:trimers_fluxes}. Specifically, for all \emph{cis} simulations, the reactant states had the total number of non-covalent interactions range from 0 to 4, whereas the product states had the number range from 9 to 11. For all \emph{trans} simulations, the reactant states had the total number of non-covalent interactions range from 0 to 2, whereas the product states had the number range from 6 to 8. The fluxes were calculated by labeling walkers with colors, which indicated whether they last came from the reactant or the product. Hence, the walkers changed color whenever they reached the other state, which effectively kept track of their history. In order to calculate the uncertainties in the flux, we used the bootstrapping procedure, which draws first passage times randomly with replacement for a number of times that is proportional to the total simulation time. The standard deviation of the flux was then multiplied by 2, which approximately represents 95\% confidence interval, for error bars. 

The brute force MD simulations were not suitable for calculating fluxes since the dimer mostly stayed in its most stable all \emph{cis} or all \emph{trans} conformation once it reached it. Note that the backward fluxes converge slower than forward fluxes since the backward reaction has to go over a much higher energy barrier than that of the forward reaction. With the CAS algorithm, we were able to obtain both forward and backward fluxes with small error bars. For all cases, the forward flux was much higher than the backward flux, indicating that it is unlikely for the dimer to unfold from the product or the most stable all \emph{cis} or all \emph{trans} conformation. In addition, the forward and backward fluxes did not linearly increase or decrease as the hydrogen bonding ability of the side chains increased as seen in Table~\ref{tab:trimers_fluxes}. This indicated that having hydrogen bonding side chains does not affect the kinetics in a straightforward, predictable manner and more studies need to be done to truly understand how different hydrogen bonding side chains affect kinetics.   

Finally, the most stable all \emph{cis} or all \emph{trans} conformation, which also corresponds to the lowest free energy point in the all \emph{cis} or all \emph{trans} isomer landscape, and the intermediate conformation, which approximately has a committor function value of 0.5, are shown for each all \emph{cis} or all \emph{trans} dimer of triazine trimers. The committor function represents the probability going from reactant to product before reaching the reactant again. The committor function was calculated by dividing the transition matrix eigenvector corresponding to the second biggest eigenvalue $\lambda_2$ by the transition matrix eigenvector corresponding to the biggest eigenvalue $\lambda_1 = 1$, or $\rho_2/\rho_1$, as done in Ref.~\onlinecite{ahn2017}. The committor function was then normalized to range from 0 to 1 to represent probability. We found that the most stable all \emph{cis} conformation is the nanorod structure and the most stable all \emph{trans} conformation is the intertwined structure for every dimer of triazine trimers with amino backbone. But note that the nanorod structure or the intertwined structure might not be the most stable conformation if we take into account all of the various \emph{cis/trans} isomers. Additionally for all cases, the conformation with a committor function value of approximately 0.5 corresponded to the energy barrier point that separated the initial state from the final, most stable all \emph{cis} or all \emph{trans} state. This makes sense since at that point, there is an equal probability of going to the final, most stable all \emph{cis} or all \emph{trans} state or going back to the initial state. On the other hand, when the dimers have sulfur backbone instead, then neither the nanorod structure nor the intertwined structure is observed and the most stable conformations become entirely different structures. Note that for the dimers of triazine trimers with sulfur backbone, only half of the cubic free energy landscape space is filled since the trimers are equivalent. Moreover, for the dimer of triazine trimers that have S-ethyl side chains and sulfur backbone, the triazine trimers have no hydrogen bonding ability so the only non-covalent interaction that binds the two triazine trimers is the $\pi$-$\pi$ interaction.

Overall, the CAS algorithm is effective at obtaining a converged one-dimensional free energy landscape and fluxes between the initial state and the final, most stable all \emph{cis} or all \emph{trans} state for the all \emph{cis} and all \emph{trans} dimers of triazine trimers with amino backbone. It is also effective at identifying intermediate conformations by calculating the committor function. Unfortunately, the dihedral angles were not sufficiently good reaction coordinates for the CAS algorithm to efficiently sample \emph{cis}-to-\emph{trans} isomerizations and vice versa and obtain steady-state weights for the dimers that had amino backbone. More work needs to be done to identify better reaction coordinates that would allow us to sample \emph{cis}-to-\emph{trans} isomerizations and vice versa for triazine polymers with amino backbone using the CAS algorithm so that we can identify the most stable \emph{cis/trans} isomer.

\begin{table}[p]
\caption{\label{tab:trimers_fluxes} Fluxes for dimers of trimers with amino backbone. The forward flux indicates the flux from the initial state to the final, most stable all \emph{cis} or all \emph{trans} state and vice versa for the backward flux.}
\begin{ruledtabular}
\begin{tabular}{ccc}
Dimer & Forward flux (ns$^{-1}$)& Backward flux (ns$^{-1}$) \\
\hline
All \emph{cis} with S-ethyl side chains & $0.164 \pm\ 0.0561$ & $0.00231 \pm\ 0.00179$\\
All \emph{trans} with S-ethyl side chains & $0.388 \pm\ 0.171$ & $0.00656 \pm\ 0.00308$\\
All \emph{cis} with amino-ethyl side chains & $0.455 \pm\ 0.182$ & $0.00455 \pm\ 0.00370$\\
All \emph{trans} with amino-ethyl side chains & $2.25 \pm\ 0.365$ & $0.0150 \pm\ 0.00747$\\
All \emph{cis} with amino side chains & $0.135 \pm\ 0.0997$ & $0.00159 \pm\ 0.000783$\\
All \emph{trans} with amino side chains & $1.86 \pm\ 0.269$ & $0.0289 \pm\ 0.0114$
\end{tabular}
\end{ruledtabular}
\end{table}

\section{\label{sec:discussion} Discussion}
The triazine-based sequence-defined polymers have immense potential to be useful building blocks for new materials. By having bonds that are not susceptible to proteases, the triazine polymers are innately robust. By being able to have various side chains, the triazine polymers can form various macromolecules with desired structure and function. However, in order for the triazine polymers to be used for production, MD simulations need to be used to first uncover their properties and mechanism. 

Ref.~\onlinecite{grate2016}, which is the first and (so far) the only publication available regarding the triazine polymers, details the experimental and computational studies done on the triazine polymers. In preliminary experiments, Grate and Mo have investigated disubstituted triazine hexamers that had two pyrene labels along the chain. The ratio of excimer to monomer fluorescence intensities is expected to be related to the physical distances between pyrene moieties. The observed trend among the disubstituted hexamers with regard to excimer to monomer ratios, based on fluorescence, is consistent with folding and inconsistent with extended linear conformations. For instance, the highest ratio, indicating the closest pyrenes, was observed for 1,6-labeled hexamers, where pyrenes are farthest in distance along the chain but would be closest if the hexamer is folded. Hence, both MD simulations and preliminary experimental work support that triazine hexamers prefer being folded. However, since there is still more to uncover regarding the triazine polymers and computational studies can be more easily done compared to experimental studies, we investigated the effects of side chains and backbone structure on the conformation and assembly of triazine polymers using MD simulations.  

Unfortunately, MD simulation by itself is limiting in timescale, so enhanced sampling methods are necessary to overcome the timescale barrier between simulations and biological processes and efficiently obtain thermodynamic and kinetic properties. By using REMD, we were able to obtain the entire free energy landscapes and identify the most stable conformation for a variety of triazine hexamers. By using the CAS algorithm\cite{CAS_Code}, we were able to obtain converged one-dimensional free energy landscapes, fluxes between the initial state and the final, most stable all \emph{cis} or all \emph{trans} state, and most stable and intermediate conformations for various all \emph{cis} and all \emph{trans} dimers of triazine trimers with amino backbone. 

With regard to forming the nanorod structure seen previously, we found, using the CAS algorithm, that if the backbone has amino groups in the linker sections, then the nanorod structure is formed and is the most stable conformation for the dimer of all \emph{cis} triazine trimers, regardless of whether the side chains can hydrogen bond or not. For the dimer of all \emph{trans} triazine trimers, we found that the intertwined structure is the most stable conformation if the backbone has amino groups in the linker section. If the backbone has sulfur instead, then the backbone loses hydrogen bonding ability and can easily isomerize from \emph{cis} to \emph{trans} and vice versa. The most stable conformations for dimers with sulfur backbones are neither the nanorod structure nor the intertwined structure and the two conformations are not observed at all. If neither the trimers' side chains nor the trimers' backbones can hydrogen bond, then there is no opportunity for hydrogen bonding and $\pi$-$\pi$ interactions dominate the conformations observed. Finally, we found that the forward and backward fluxes are different for different hydrogen bonding side chains but could not find a clear trend in how the fluxes change as the hydrogen bonding ability of the side chains increases. Similarly, we found, using REMD, that the nanorod structure appears for all cases, since all of the hexamers had hydrogen bonding amino backbones. \textbf{Taken together, we found that having hydrogen bonding backbones, rather than hydrogen bonding side chains, are critical for the triazine polymers to self-assemble into the nanorod structure.}

For future work, better reaction coordinates other than the dihedral angles need to be identified to efficiently sample \emph{cis}-to-\emph{trans} isomerizations and vice versa using the CAS algorithm. Then we will be able to identify which isomer is the most stable one for the dimer of triazine trimers. If the all \emph{cis} dimer is found to be the most stable isomer, then the nanorod structure that they form will be robust and have potential to be used as building blocks for new materials. Additional work needs to be done on longer triazine polymers and with different experimental conditions (solvent, temperature, etc.) to get a more general understanding of the self-assembly of triazine polymers.

\section{\label{sec:supplementary} Supplementary Information}
PSE files for of all of the triazine polymer conformations are available. The files can be viewed using Pymol.

\begin{acknowledgments}
This work is supported by the Applied Mathematics Program within the Department of Energy (DOE) Office of Advanced Scientific Computing Research (ASCR) as part of the Collaboratory on Mathematics for Mesoscopic Modeling of Materials (CM4). This work used the XStream computational resource, supported by the National Science Foundation Major Research Instrumentation program (ACI-1429830), Sherlock cluster at Stanford, Certainty cluster at Stanford, and PNNL's Institutional Computing (PIC) cluster. We thank Marcel Baer for helping us with making the variants of the triazine polymers and Chris Mundy and Greg Schenter for discussing the outline of the paper. We also thank the reviewers for giving us helpful comments that improved the content of the paper.
\end{acknowledgments}

\bibliography{CAS_Triazine}

\end{document}